\documentclass[a4paper,11pt]{article}
\pdfoutput=1 

\usepackage{jcappub} 

\usepackage[T1]{fontenc} 
\usepackage[dvipsnames]{xcolor}
\usepackage{tikz}

\usepackage{amssymb, amsmath, epsfig, natbib}
\usepackage{mathtools}

\renewcommand{\vec}[1]{\mathbf{#1}}

\newcommand{\vq}{\vec{q}}
\newcommand{\vx}{\vec{x}}
\newcommand{\vk}{\vec{k}}
\newcommand{\vPsi}{\vec{\Psi}}

\newcommand{\pp}{\parallel}

\newcommand{\df}{\delta}
\newcommand*{\non}  {\nonumber}
\newcommand*{\lb}  {\left(}
\newcommand*{\rb}  {\right)}

\newcommand*{\la}  {\left\langle}
\newcommand*{\ra}  {\right\rangle}

\newcommand{\eq}[1]{\begin{align}#1\end{align}}
\newcommand{\eeq}[1]{\begin{equation}#1\end{equation}}

\DeclarePairedDelimiter\floor{\lfloor}{\rfloor}

\author[a]{Zvonimir Vlah}
\author[b,c]{Martin White}

\affiliation[a]{Theory Department, CERN, 1 Esplanade des Particules, CH-1211 Gen\`eve 23, Switzerland}
\affiliation[b]{Department of Physics, University of California,
Berkeley, CA 94720}
\affiliation[c]{Department of Astronomy, University of California,
Berkeley, CA 94720}

\emailAdd{zvonimir.vlah@cern.ch}
\emailAdd{mwhite@berkeley.edu}

\title{Exploring redshift-space distortions in large-scale structure}


\abstract{We explore and compare different ways large-scale structure
observables in redshift-space and real space can be connected.
These include direct computation in Lagrangian space, moment expansions
and two formulations of the streaming model.
We derive for the first time a Fourier space version of the streaming model,
which yields an algebraic relation between the real- and redshift-space power spectra which can be compared to earlier, phenomenological models.
By considering the redshift-space 2-point function in both configuration and
Fourier space, we show how to generalize the Gaussian streaming model to
higher orders in a systematic and computationally tractable way. 
We present a closed-form solution to the Zeldovich power spectrum in redshift
space and use this as a framework for exploring convergence properties of
different expansion approaches.
While we use the Zeldovich approximation to illustrate these results,
much of the formalism and many of the relations we derive hold beyond
perturbation theory, and could be used with ingredients measured from
N-body simulations or in other areas requiring decomposition of
Cartesian tensors times plane waves.
We finish with a discussion of the redshift-space bispectrum, bias and
stochasticity and terms in Lagrangian perturbation theory up to 1-loop order.}
\arxivnumber{1812.02775}

\begin{document}
\maketitle
\flushbottom

\section{Introduction}

The large-scale structure of the Universe contains valuable information
about cosmology and fundamental physics, and a number of ambitious
observational campaigns to extract this information are underway or in
the planning stages \cite{USDE13,Euclid}.
These new observations will provide increasingly precise measurements of the
clustering of astrophysical objects on large scales, which are relatively
simple to model and where predictions are under theoretical control.
Much as for anisotropies in the cosmic microwave background, it is hoped
that the combination of robust theoretical predictions and exquisitely
sensitive observations will yield strong constraints on cosmological models
(e.g.~Ref.~\cite{Planck18-I}).

In this paper we are interested in developing analytic models for the large
scale clustering of objects in redshift space \cite{Pea99}, i.e.~as measured
in galaxy redshift surveys \cite{eBOSS,DESI,FourMost,Euclid},
the Ly$\alpha$ forest \cite{Meiksin09,McQuinn16} or line intensity mapping
experiments \cite{Kovetz17}.
In these situations the line-of-sight peculiar motions of objects contribute to their observed redshift so that they are placed at the incorrect (line-of-sight) distance \cite{Kai87,Ham92,Ham98,Pea99}.  This velocity-induced mapping from real- to redshift-space introduces an anisotropy in the clustering pattern, which can be used to test the theory and probe the growth of large-scale structure \cite{Wei13,PDG18}.

While many of our results will be valid in general, for explicit calculations
and to better bring out the physics implied by our formulae we will use
Lagrangian perturbation theory
(see e.g.~Ref.~\cite{VlaCasWhi16}, building upon the work of
 Refs.~\cite{Zel70,Buc89,Mou91,Hiv95,TayHam96,Mat08a,Mat08b,CLPT,Whi14,
  PorSenZal14,ZheFri14,Mat15,VlaSelBal15,VlaWhiAvi15,McQWhi16}).
Our focus will be on the study the low-order statistics of redshift-space
fields in both configuration and Fourier space.

The outline of the paper is as follows.  In  \S\ref{sec:background} we review
some background material on our model and establish our notation.
In \S\ref{sec:redshift} we introduce the main results of this paper,
namely a comparison of different formalims for computing the redshift-space
2-point functions.
This section also includes the development of a new variant of the streaming
model, which has some advantages over previous treatments.
We apply the general formalism in \S\ref{sec:comparison} and
\S\ref{sec:config_space}, where we compare the different approaches for
computing the redshift-space power spectrum and correlation function
(respectively) to the direct calculation, allowing a detailed study of
the convergence properties of each method.
While most of the comparisons are done for the matter field within the
Zeldovich approximation, our formalism is much more general and
\S\ref{sec:bias} explicitly develops the bias and loop expansions.
We show that the moment expansion and Fourier-space version of the streaming
model lead to relatively simple expressions for the 3-point function in
Fourier space (i.e.~the bispectrum) in \S\ref{sec:bispectrum}.
We conclude in \S\ref{sec:conclusions}.
Some technical details are relegated to a series of appendices.

Throughout the paper, we will use the index-summation convention where possible, 
the subscript $L$ will denote the linear (Eulerian) quantities,
and we will be assuming the $\Lambda$CDM cosmology with $\Omega_m=0.295$, 
$\Omega_b=0.047$, $n_s=0.968$, $\sigma_8 = 0.835$ and $h=0.688$.

\section{Background}
\label{sec:background}

In this section we give a brief review of Lagrangian perturbation theory to
fix our notation.  We refer the reader to the references below for the
development of the theory.
Lagrangian perturbation theory and effective field theory, coupled with a
flexbile bias model, offer a systematic and accurate means of predicting
the clustering of biased tracers in both configuration and Fourier space
(e.g.~Ref.~\cite{VlaCasWhi16}).

The Lagrangian approach to cosmological structure formation was developed
in \cite{Zel70,Buc89,Mou91,Hiv95,TayHam96,Mat08a,Mat08b,CLPT,Whi14,
         PorSenZal14,ZheFri14,Mat15,VlaSelBal15,VlaWhiAvi15,McQWhi16}
and traces the trajectory of an individual fluid element through space and
time.  A fluid element located at position $\vq$ at some initial time $t_0$
moves as $\vx(\vq,t) = \vq + \vPsi(\vq,t)$ with
$\ddot{\vPsi} + \mathcal{H}\dot{\vPsi} = -\nabla\Phi(\vq+\vPsi)$ where
an overdot represents a derivative with respect to conformal time and
$\mathcal{H}=aH$ is the conformal Hubble parameter.
Every element of the fluid is uniquely labeled by $\vq$ and $\vPsi(\vq,t)$
fully specifies the evolution.  We shall solve for $\vPsi$ perturbatively.
The first order solution, linear in the density field, is the Zeldovich
approximation \cite{Zel70}, which will play an important role in this paper.
Given $\vPsi$, the real-space density field at any time is simply
\begin{equation}
  1+\delta(\vx)=\int d^3q\ \delta_D\big[\vx-\vq-\vPsi(\vq)\big]
  \quad\Rightarrow\quad
  \delta(\vk)=\int d^3q\ e^{i\vk\cdot\vq}\big(e^{i\vk\cdot\vPsi(\vq)}-1\big) .
\label{eqn:deltadefn}
\end{equation}
The density of biased tracers can be modeled, assuming Lagrangian bias,
by multiplying the $\delta_D$ in the above by a function,
$F[\delta_L(\vq),\nabla^2\delta_L(\vq),\cdots]$, depending upon the linear
theory density and its derivatives \cite{Mat08b,Whi14,VlaCasWhi16}.
In the absence of explicit knowledge of $F$, the expectation values of
derivatives of $F$ take the place of unknown bias coefficients describing
the tracer under consideration.
Evaluation of the power spectrum then involves the expectation value of an
exponential, which can be evaluated using the cumulant theorem -- we refer
the reader to the above references for further details and explicit
calculations.

In what follows we will pay particular attention to the $1^{\rm st}$ order
solution to Lagrangian dynamics, i.e.~the Zeldovich approximation \cite{Zel70}.
Since the displacement field is given in terms of the linear overdensity as
$\vPsi (\vec p)=(i\vec p/p^2)\ \delta_L(\vec p)$ (where $\vec p$ is the
momentum variable corresponding to the Lagrangian coordinate $\vec q$),
it follows that the Zeldovich matter power spectrum is given by
\cite{TayHam96,FisNus96,Bha96,Mat08a,Mat08b,Whi14,Tas14a,Tas14b}
\begin{equation}
  P(\vec k) = \int d^3q
  \ e^{i \vec k . \vec q} \exp\left[-\frac{1}{2}k_i k_j A_{i j}(\vq) \right]
  \label{eq:ps_noshift}
\end{equation}
with $A_{\ell m}(\vq) = \left\langle\Delta_\ell\Delta_m\right\rangle$,
and $\Delta_i=\Psi_i(\vec{q})-\Psi_i(\vec{0})$ (see \S\ref{sec:redshift} and
Eq.~\eqref{eq:LagPS_chi}, setting $\Delta u=0$ and $\delta_a=\delta_b=0$).
The argument of the exponential can  be expressed in terms of integrals over
the linear theory power spectrum.
Writing $A_{ij}(\vq)=X(q)\delta_{ij}+Y(q)\hat{q}_i\hat{q}_j
 =\tfrac{2}{3}\delta^K_{ij}\left(\mathcal{J}_{0}(0)-\mathcal{J}_{0}(q)\right)+
  2\lb\hat{q}_i\hat{q}_j-\tfrac{1}{3}\rb\mathcal{J}_2(q)$ we have
\begin{equation}
  \mathcal{J}_0(q) = \int_0^\infty\frac{dk}{2\pi^2}\ P_L(k) j_0(kq)
  \quad {\rm and} \quad
  \mathcal{J}_2(q) = \int_0^\infty\frac{dk}{2\pi^2}\ P_L(k) j_2(kq),
\end{equation}
where $j_\ell$ are the spherical Bessel functions.
We shall return to an evaluation of Eq.~(\ref{eq:ps_noshift}) in \S\ref{sec:comparison}.

\section{Redshift space}
\label{sec:redshift}

The line-of-sight component of the peculiar motion of each object or fluid
element affects its measured redshift, and thus the radial distance at
which it is inferred to lie using the distance-redshift (Hubble) relation
\cite{Kai87,Ham92,Ham98,Pea99}.
Specifically, an object with peculiar velocity $\vec{v}$ which truly lies
at $\vx$ will be assigned a ``redshift-space position''
$\vec{s}=\vx+\hat{n}\,(\vec{v}\cdot\hat{n}/\mathcal{H})$ if $\hat{n}$ is
the line of sight.
In our Lagrangian formalism, the shift to redshift space is easily accomplished
by adding $\hat{n}\lb\hat{n}\cdot\dot{\vPsi}\rb$ to $\vPsi$.  We shall use the
shorthand notation $\dot{\vPsi}_{\hat n}$ for $\hat{n}\lb \hat{n}\cdot\dot{\vPsi} \rb$.

The Fourier-space density contrast in redshift space is thus
\begin{eqnarray}
  (2\pi)^3\delta^D(\vec k) +  \delta_s(\vec{k})
  &=& \int d^3q\ F\exp\left[ i\vk\cdot\left\{\vq+\vPsi(\vq)
  +\dot{\vPsi}_{\hat n}(\vq)\right\} \right] \\
  &=& \int d^3x\ \left[1+\delta(\vec{x})\right]
  \exp\left[i\vk\cdot\left\{\vec{x}+\vec{u}(\vec{x})\right\}\right],
\label{eqn:delk_def}
\end{eqnarray}
where we have introduced a dimensionless velocity, $\vec{u}$.
Note that this says the Fourier transform of the shifted field differs from
that of the unshifted field by a phase, $\exp[i\vec{k}\cdot\vec{u}]$, as
might have been expected.
By writing this expression in Lagrangian coordinates we have not needed
to make the single-stream approximation, or that the mapping
$\vec{x}\to\vec{x}+\vec{u}(\vec{x})$ is one-to-one.
The 2-point function of the shifted fields can now be expressed in terms
of a `moment' generating function\footnote{Expanding the exponential in
powers of $\vec{J}$, the $n^{\rm th}$ order moment is the coefficient of
the $J^n$ term. This can be trivially generalized for higher point functions,
and we consider the bispectrum in \S\ref{sec:bispectrum}.}  
(see e.g.\ Ref.~\cite{Sco04}\footnote{
Our definition differs slightly from the one introduced in e.g.\ Ref.~\cite{Sco04}
in that we allow consideration of two different tracers, as well as 3D free vector $\vec J$. 
The latter allows one to consider, in principle, RSD effects beyond the plane parallel approximation.
There is also a difference in a density weighting factor $(1+\xi)$ where in \ Ref.~\cite{Sco04}
the pairwise velocity generating function is defined as 
$ (1+\xi(\vec r))M(\lambda, \vec r)\equiv  \left\langle
  \big[ 1 + \df(\vec{x}_1) \big]
  \big[ 1 + \df(\vec{x}_2) \big]
  e^{ i \lambda \Delta u_{\pp} } \right\rangle$. In that respect, our definition of the 
moment generating function is more in line with the $\mathcal Z(\lambda, \vec r)$ quantity defined in Eq.(15) of this reference.
All the physical considerations will, of course, not depend on any of these definition differences.
})
\begin{equation}
  1 + \mathcal{M}^{ab}(\vec{J},\vec{r}) = \left\langle
  \big[ 1 + \df_a(\vec{x}_1) \big]
  \big[ 1 + \df_b(\vec{x}_2) \big]
  e^{ i \vec{J}\cdot\Delta u_{ab} } \right\rangle ,
\label{eqn:M_def}
\end{equation}
where $\Delta u_{ab} = u_b(\vec x_2) - u_a(\vec x_1)$ and
$\vec{r} = \vec{x}_2 - \vec{x}_1$, and $a$ and $b$ are the  labels of
different tracers.  Note that the translational invariance of the generating
function, $\mathcal{M}$, is explicit, i.e.~$\mathcal{M}$ depends only on
$\vec{r}$.  This can be seen clearly by expanding the exponent and noticing
that after ensemble averaging each term in the sum depends only on $\vec{r}$.

We can Fourier transform the generating function to obtain
\begin{equation}
  \widetilde{\mathcal{M}}^{ab}(\vec J,\vec k) =
  \frac{k^3}{2\pi^2}\int d^3r\ e^{i\vec{k}\cdot\vec{r}}
  \left\langle(1+\delta_a(\vec{x}))(1+\delta_b(\vec{x}'))
  e^{i\vec{J}\cdot\Delta u_{ab}}\right\rangle.
\label{eqn:M_def_ink}
\end{equation}
where the prefactor of $k^3/(2\pi^2)$ is inserted to make $\widetilde{\mathcal{M}}$ dimensionless and for later convenience.
Directly from Eqs.~(\ref{eqn:delk_def},\ref{eqn:M_def}, \ref{eqn:M_def_ink}) we see that the generating function, $\mathcal{M}$, has a simple relation to the power spectrum,
\begin{equation}
  \frac{k^3}{2\pi^2}P^{ab}_s(\vec k) =\widetilde{\mathcal{M}}^{ab}(\vec{J}=\vk,\vec{k}) = \frac{k^3}{2\pi^2}\int d^3r\ e^{i\vk\cdot\vec{r}}
  \mathcal{M}^{ab}(\vec{J}=\vk,\vec{r})
  \label{eq:power_master}
\end{equation}
and that this relation holds beyond perturbation theory.  We note as well that the
correlation function can be obtained here by one more Fourier transform
\begin{equation}
  \xi^{ab}_s(\vec r)  = \int \frac{d^3k}{(2\pi)^3} \ e^{- i\vk \cdot \vec{r}} P^{ab}_s(\vec k).
\end{equation}
We point out that to obtain the correlation function we have to perform one
additional transform compared to the power spectrum regardless of whether we
start from $\mathcal{M}$ or $\widetilde{\mathcal{M}}$.
The reason for this, of course, is that in obtaining the power spectrum
$\vec{J}$ could be specified as the last step, while for the correlation
function the sum over the modes needs to be performed after specifying
$\vec{J}$.  This observation will lead to some interesting consequences below.

We now consider several general methods for computing the 2-point statistics
of these redshift-space fields, by manipulating Eq.~\eqref{eqn:M_def}.
There has been lots of theoretical activity in the RSD literature over the
past several decades, with different approaches leading to seemingly very different models and different results.  We will see that the differences in all of these
approaches are simply due to different levels of approximation and techniques
used to obtain the constituents. In fact, all of these approaches can be
categorized based on the manner in which they approach the generating function:
$\mathcal{M}$
\begin{enumerate}
\item Direct Lagrangian approach: the moment generating function,
$\mathcal{M}$, is transformed into Lagrangian coordinates and then
contributions are estimated using the cumulant theorem.
Examples include Refs.~\cite{Mat08a,Mat08b,CLPT,Whi14,VlaWhiAvi15}.
\item Moment expansion approach: the exponential in the generating
function, $\mathcal{M}$, is expanded and the moments individually evaluated.
Examples of this approach are the distribution function approach
\cite{SelMcD11,Oku12,Oku12II,Vla12,Vla13,Oku14,Oku15,Hand2017} 
as well as the direct SPT-loop expansion \cite{Ber02,Makino92,Sco99,Mat08a}.
Various Eulerian EFT based approaches also fit into this class (see
e.g.\ Refs.~\cite{Senatore2014,Perko2016,Bella2017,Ivanov2018,Desjacques2018}).
\item Streaming model: the cumulant theorem is used in Eulerian space after
transforming the moment generating function $\mathcal{M}$ into the cumulant
generating function $\mathcal{Z}$ (defined below in Eq.~\ref{eq:cumgen_Z}).
Examples include Refs.~\cite{Pee80,Fis95,ReiWhi11,Rei12,WanReiWhi14,
Uhl15,Kop16,VlaCasWhi16,Kuruvilla2018}.
\item
Fourth approach, widely developed in the literature, is the `smoothing kernel' approach.  
Here the cumulant expansion is used on each of the 4 contributions that arise upon expanding the product
of $(1+\df)$ terms and the exponential in Eq.~\eqref{eqn:M_def}, giving a
result widened by a ``smoothing kernel''.
This approach was first presented in Ref.~\cite{Sco04} and further developed
in Ref.~\cite{TNS10}, which further expands the exponential contributions and approximates
the smoothing kernel as a one parameter Fourier space Gaussian or Lorentzian.
\end{enumerate}
As mentioned above, the differentiation into four classes and the labels are
primarily historical.  It is important to stress that all these methods would
be mathematically identical if carried to the same order and with the same
approximations.  The differences are thus primarily of convenience: some
aspects of the problem are easier to handle in some approaches than others.
It is also the case that the ingredients to each method can be supplied by
a perturbative, analytic model or they could (in principle) be measured in
simulations. We now consider each method in turn.

\subsection{Direct Lagrangian approach}
\label{sec:dir_lag_app}

Directly from the definition of $\vec{u}$, and using the continuity
equation, we can write
\begin{equation}
  1 + \mathcal{M}^{ab}\big(\vec J,\vec r\big) =
  \int\frac{d^3p}{(2\pi)^3}\,d^3q
  \ e^{-i\vec{p}\cdot(\vec{r}-\vec{q})}\ \left\langle
  \big(1+\df_a\big(\vec{q}_1))\big(1+\df_b(\vec{q}_2)\big)
  e^{i\vec{p}\cdot\Delta(\vec q) + i\vec{J}\cdot\Delta u_{ab}} \right\rangle,
\label{eq:M_direct}
\end{equation}
and similarly
\begin{equation}
  (2\pi)^3\df^D(\vec k) + P^{ab}_s(\vec k)
  = \int d^3q\ e^{i\vec{k}\cdot\vec q}
  \ \left\langle \big(1+\df_{a}(\vec q_2)\big)
                 \big(1+\df_{b}(\vec q_1)\big)
  \exp\Big[i\vec{k}\cdot\Delta + i\vec{k}\cdot\Delta u_{ab}\Big]
  \right\rangle,
  \label{eq:LagPS_chi}
\end{equation}
where $\Delta(\vq)=\vPsi(\vq_2)-\vPsi(\vq_1)$ and all quantities are functions
of Lagrangian coordinates. For the velocity in particular we have
$\vec{u}(\vec x)  = \dot{\vPsi}_{\hat n}(\vec q)$ for properly normalized time units.
As before the redshift-space power spectrum, $P^{ab}_s(\vec{k})$, is simply the Fourier
transform of $\mathcal{M}^{ab}(\vec{J}=\vec{k},\vec{r})$, but now
expressed entirely in Lagrangian coordinates.
We also note that there is no difference between using $\mathcal M$ or
$\widetilde{\mathcal M}$ as the starting point for the derivation.

There has been significant attention paid to this Lagrangian framework in
recent years
\cite{Mat08a,Mat08b,CLPT,Whi14,PorSenZal14,ZheFri14,Mat15,
      VlaSelBal15,VlaWhiAvi15,McQWhi16,VlaCasWhi16},
as it lends itself naturally to the implementation of redshift-space
distortions.
In addition it has been employed as the basis of an effective field theory
expansion \cite{PorSenZal14,VlaWhiAvi15,VlaCasWhi16}
and to model higher order correlation functions \cite{Tas14b}.
We shall consider the evaluation of this expression further later, and for
now turn to the second expansion.

\subsection{Moment expansion approach}

The moment expansion approach is equivalent to the distribution function
approach \cite{SelMcD11,Vla13} and proceeds by expanding the exponential
term in $\mathcal{M}$ to obtain the (density weighted) moments of the velocity
field (for the explicit connection to the distribution function approach
we refer the reader to App \ref{sec:and_decomp_ofvelocities}):
\begin{equation}
  \Xi_{i_1,\ldots,i_n}(\vec{r}) =
  \left\langle \big(1 + \df_a(\vec{x})\big)\big(1+\df_b(\vec{x}')\big)
  \Delta u_{ab,i_1}\ldots\Delta u_{ab,i_n}\right\rangle .
\label{eqn:def_moments}
\end{equation}
We can write the power spectrum as the sum of the Fourier transforms of
these moments, viz
\begin{eqnarray}
  (2\pi)^3 \delta^D(\vec{k}) + P^{ab}_s(\vec{k})
  &=& \sum_{n=0}^\infty \frac{i^n}{n!} k_{i_1} \ldots k_{i_n}
  \widetilde \Xi_{i_1,\ldots,i_n}(\vec{k}) \\
  &=& \sum_{n=0}^\infty \frac{i^n}{n!} k_{i_1}\ldots k_{i_n}
  \int d^3r\ \Xi_{i_1,\ldots,i_n}(\vec{r}) e^{i\vec{k}\cdot\vec{r}},
\label{eq:PS_momentsum}
\end{eqnarray}
where $\widehat{\Xi}_{i_1\ldots i_n}^{(n)}(\vec{k}) =(-i)^n
\left.\partial \widetilde{ \mathcal{M}}^{ab}(\vec{J},\vec{k})/
\partial J_{i_1} \ldots \partial J_{i_n} \right|_{\vec{J}=0}$
with $\widehat{\Xi}\equiv k^3\widetilde{\Xi}/(2\pi^2)$.
The moment expansion approach is the most straightforward of the 3 methods,
although some of the effects that can be captured nonlinearly in other
approaches (e.g.~finger-of-god terms in the case of RSD) might be missed
here if one truncates the expansion at low order.
On the other hand, such nonlinear terms should always be resummable
afterwards to obtain results equivalent to the other methods\footnote{
It is challenging to robustly define an expansion parameter
in many of these approaches. Field variance $\sigma_8$ could be considered as 
one potential candidate although many of the methods do not consistently resum
in this parameter. 
}.
This was the strategy adopted in the distribution function approach
\cite{SelMcD11,Vla13}.

The leading order results are straightforwardly obtained within this approach.
Since $u$ is a quantity of order $\delta$,  we can compute the low order contributions as
\begin{eqnarray}
  P^{ab}_s(\vec{k})
  &=& \widetilde \Xi_0(\vec{k}) + ik_i\widetilde \Xi_{1,i}(\vec{k})
  -\frac{1}{2}k_ik_j\widetilde \Xi_{2,ij} (\vec{k}) +\cdots  \\
  &=& \int d^3r\ e^{i\vec{k}\cdot\vec{r}}
  \left\langle\df_a(\vec{x})\df_b(\vec{x}')\right\rangle
  \nonumber \\
  && +ik_i\int d^3r\ e^{i\vec{k}\cdot\vec{r}}
  \Big(\left\langle\df_a(\vec{x} )u_{b,i}(\vec{x}')\right\rangle
     - \left\langle\df_b(\vec{x}')u_{a,i}(\vec{x} )\right\rangle\Big)
  \nonumber \\
  && + k_ik_j\int d^3r\ e^{i\vec{k}\cdot\vec{r}}
    \left\langle u_{a,i}(\vec x)u_{b, j}(\vec x')\right\rangle + \cdots
\label{eq:Kaiser_general}
\end{eqnarray}
where we have dropped terms proportional to $\df^D(\vec{k})$.
This leads to the well known Kaiser formula \cite{Kai87}.

\subsection{Streaming approach}

The streaming approach is sometimes regarded as a phenomenological model,
but in fact can be derived as an expansion of $\mathcal{M}$ or
$\widetilde{\mathcal{M}}$ using the cumulant theorem (see below).
This can be done in either configuration or Fourier space.
The two forms are not equivalent, because at any finite order the
cumulant expansion and the Fourier transform do not commute.
The configuration space streaming model has been extensively explored
in the literature
\cite{Pee80,Fis95,ReiWhi11,WanReiWhi14,Uhl15,Kop16,VlaCasWhi16}
and applied to data \cite{Rei12,Samushia13,Samushia14,Alam17,Zarrouk18}.
The Fourier space expansion has not been explored in the literature to date
and is new to this paper.  As we will see, it has some nice properties
compared to the more common, configuration-space approach, but also some
subtleties.

\subsubsection{Configuration space}
\label{sec:SM_config_space}

We can perform the cumulant expansion by taking the logarithm
$\mathcal{Z}^{ab}(\vec{J}, \vec{r}) =
   \ln \big[ 1+ \mathcal{M}^{ab}
   \big( \vec J, \vec r \big) \big]$
and expanding in $\vec{J}$:
\begin{equation}
  \mathcal{Z}^{ab}(\vec{J}, \vec{r}) =
  \sum_{n=0}^\infty \frac{i^n}{n!}J_{i_1}\ldots J_{i_n}
  \mathcal{C}_{i_1\ldots i_n}^{(n)}(\vec{r}),
\label{eq:cumgen_Z}
\end{equation}
where
$\mathcal{C}_{i_1\ldots i_n}^{(n)}(\vec{r}) =(-i)^n
\left.\partial\mathcal{Z}^{ab}(\vec{J},\vec{r})/
\partial J_{i_1} \ldots \partial J_{i_n} \right|_{\vec{J}=0}$
are the cumulants of the (density weighted) velocities, $\Delta u$.
The first few cumulants are
\begin{eqnarray}
\mathcal C^{(0)}(\vec{r}) &=& \ln \big[ 1+ \xi_{ab}(\vec r) \big] , \non\\
\mathcal C^{(1)}_{i}(\vec{r}) &=& \Xi_{i}(\vec r)/
  \left(1+\xi_{ab}(\vec r)\right), \non\\
\mathcal C^{(2)}_{ij}(\vec{r}) &=& \Xi_{ij}(\vec r)/
   \left(1+\xi_{ab}(\vec r)\right)
  -\mathcal C^{(1)}_{i}\mathcal C^{(1)}_{j}, \non\\
\mathcal C^{(3)}_{ijk}(\vec{r}) &=& \Xi_{ijk}(\vec r)/
   \left(1+\xi_{ab}(\vec r)\right)
  -\mathcal C^{(2)}_{\{ ij }\mathcal C^{(1)}_{k\}}
  -\mathcal C^{(1)}_{i}\mathcal C^{(1)}_{j}\mathcal C^{(1)}_{k},
\label{eq:cums}
\end{eqnarray}
where $\Xi$'s are the shift field moments given by Eq.~\eqref{eqn:def_moments}
and the $\{\cdots\}$ indicates all the nontrivial permutations of the indices.
Note that physically the denominators in the above are positive definite,
however this is not guaranteed for all perturbation theory schemes and scales.
The first and second cumulants are the pairwise velocity,
$v_{12}$, and the dispersion, $\sigma_{12}$.

If we introduce the kernel, $\mathcal{K}^{ab}$, defined as
\begin{equation}
  \ln\mathcal{K}^{ab}(\vec{k},\vec{r}) =
  \sum_{n=1}^\infty \frac{i^n}{n!} k_{i_1}\ldots k_{i_n}
  \mathcal C^{(n)}_{i_1\ldots i_n} (\vec{r})
  \label{eq:SM_Kfnc}
\end{equation}
we can write
\begin{equation}
  (2\pi)^3\delta^D(\vec k) + P^{ab}_s(\vec k)  =
   \int d^3r\ e^{i \vec k \cdot \vec r} \big[ 1 + \xi^{ab}(\vec r) \big]
  \mathcal{K}^{ab}(\vec{k},\vec{r})
  \label{eq:PS_stream_mom}
\end{equation}
and
\begin{equation}
  1 +  \xi^{ab}_s(\vec{s}) =
  \int d^3r\ \big[ 1 + \xi^{ab}(\vec{r}) \big]
  \int\frac{d^3k}{(2\pi)^3}\  e^{-i\vec{k}\cdot(\vec{s}-\vec{r})}
  \mathcal{K}^{ab}(\vec{k},\vec{r}).
\end{equation}
The Gaussian streaming model follows immediately by truncating the cumulant
expansion in $\mathcal{K}^{ab}$ at second order,
\begin{equation}
  1 +  \xi^{ab}_s(\vec{s}) \simeq
  \int d^3r\ \big[ 1 + \xi^{ab}(\vec{r}) \big]
  \int\frac{d^3k}{(2\pi)^3}\  e^{-ik_j(s_j-r_j-\mathcal{C}^{(1)}_j)}
  e^{-(1/2)k_ik_j \mathcal{C}^{(2)}_{ij}} \quad ,
\end{equation}
and doing the Gaussian integral over $d^3k$:
\begin{equation}
  1 +  \xi^{ab}_s(\vec{s}) =
  \int \frac{d^3r}{\sqrt{ (2\pi)^3 \det [\mathcal{C}^{(2)}] }}
  \ \big[ 1 + \xi^{ab}(\vec{r}) \big]\  \exp\left[
  -\frac{1}{2}(\vec s-\vec r-\mathcal{C}^{(1)})
  [\mathcal{C}^{(2)}]^{-1}(\vec s-\vec r-\mathcal{C}^{(1)})\right] \quad .
\end{equation}
This expression can be further simplified due to the simple matrix structure
of $\mathcal{C}^{(2)}$ (see later).

\subsubsection{Fourier space}
\label{subsec:FS_model}

We can also work in Fourier space and perform the cumulant expansion by writing
$\widetilde{\mathcal{Z}}^{ab}(\vec{J}, \vec{k}) =
   \ln \big[1 + \widetilde{\mathcal{M}}^{ab}(\vec{J}, \vec{k}) \big]$
and expanding in $\vec{J}$ as above. The first few cumulants are
\begin{eqnarray}
\widetilde{\mathcal C}^{(0)}( \vec k ) &=& \ln \big[ 1+\Delta^2_{ab}(k) \big], \\
\widetilde{\mathcal C}^{(1)}_{i}( \vec k ) &=& \widehat{\Xi}_i(\vec k)/\big[ 1+\Delta^2_{ab} \big], \non\\
\widetilde{\mathcal C}^{(2)}_{ij}( \vec k ) &=& \widehat{\Xi}_{ij}(\vec k)/\big[ 1+\Delta^2_{ab} \big] - \widetilde{C}^{(1)}_{i}\widetilde{C}^{(1)}_{j} , \non\\
\widetilde{\mathcal C}^{(3)}_{ijk}( \vec k ) &=& \widehat{\Xi}_{ijk}(\vec k)/\big[ 1+\Delta^2_{ab} \big]
- \widetilde{C}^{(2)}_{\{ ij } \widetilde C^{(1)}_{k\}} - \widetilde C^{(1)}_{i}\widetilde C^{(1)}_{j}\widetilde C^{(1)}_{k}. \non
\label{eq:k-cums}
\end{eqnarray}
where we have used the common notation $\Delta^2=k^3P(k)/(2\pi^2)$ and defined
$\widehat{\Xi}=k^3\widetilde{\Xi}/(2\pi^2)$ as before.
Since the power spectrum is always a positive quantity, both the ratios
and the $\log$ in the first cumulant are well defined for all $\vec{k}$
and we shall assume that this property is satisfied on the relevant scales
by the perturbation theories of relevance here.
In analogy to what we had before, we can introduce the kernel
\begin{equation}
  \ln \widetilde{\mathcal{K}}^{ab}(\vec{k}) =
  \sum_{n=1}^\infty \frac{i^n}{n!} k_{i_1}\ldots k_{i_n}
  \widetilde{\mathcal C}^{(n)}_{i_1\ldots i_n} (\vec{k}).
\end{equation}
We note that the translation kernel in this case depends on $\vec k$ only
and not on $\vec r$ as was the case above.
This is significant since in order to compute the power spectrum no additional
Fourier transform is needed.
It should be clear that the kernel $\widetilde{\mathcal{K}}^{ab}$
is not simply the Fourier transform of $\mathcal{K}^{ab}$, and
neither do $\widetilde{\mathcal{Z}}^{ab}$ and
$\mathcal{Z}^{ab}$ form a Fourier transform pair.

The redshift-space power spectrum can now be written
\begin{align}
  \frac{k^3}{2\pi^2}P^{ab}_s(\vec{k}) &=
  \exp\left[ \widetilde{\mathcal{Z}}^{ab}(\vec{k}, \vec{k}) \right] -1
  \non \\
  &= \big[1+\Delta^2_{ab}\big]\widetilde{\mathcal{K}}^{ab}(\vec{k}) -1
  \non \\
  &=  \big[1+\Delta^2_{ab}\big]
  \exp \Bigg[ \sum_{n=1}^\infty \frac{i^n}{n!} k_{i_1}\ldots k_{i_n}
  \widetilde{\mathcal C}^{(n)}_{i_1\ldots i_n} (\vec{k}) \Bigg] - 1 .
  \label{eq:streaming_ps_ink}
\end{align}
We note again that if we had expanded the moment generating function,
$\mathcal{M}$, then the Fourier and configuration space expressions would have
been conjugate.  However the non-linearity inherent in the cumulant expansion
(the fact that we are expanding the log of $\mathcal{M}$ and not $\mathcal{M}$
itself) means that these two ``streaming models'' make different predictions
for both $\xi$ and $P$.

Interestingly, the configuration-space streaming model has a long history
(dating back to \cite{Pee80,Fis95}),
but its Fourier counterpart does not seem to have been developed previously
even though it yields a much simpler power spectrum structure
(Eq.~\ref{eq:streaming_ps_ink}).

\subsection{Smoothing kernel approach}
\label{subsec:smoothing_kernel}

The final approach uses the cumulant expansion on each of the four terms
after expanding the $(1+\df_a)(1+\df_b)$ piece of the generating function
$\mathcal{M}$ in Eq. \eqref{eqn:M_def}. In the case of RSD this approach was first proposed in Ref.~\cite{Sco04}
and later further developed in Ref.~\cite{TNS10}, who expanded the exponential and approximated the smoothing kernel as a Gaussian. 

We re-derive this approach here in a slightly different way than presented in Ref.~\cite{Sco04}. 
For details of the standard derivation we refer the reader to the original reference. 
We can consider writing the generating function using an `inertia' operator
$\hat{\mathcal{O}} = 1-i\left(\partial_{\lambda_a}+\partial_{\lambda_b}\right)
 - \partial_{\lambda_a}\partial_{\lambda_b}$, which gives
\begin{equation}
  1 + \mathcal{M}^{ab}\big(\vec J,\vec r\big) = \hat{\mathcal{O}}
  \left\langle e^{i\lambda_a\df_a(\vec x_1)
                 +i\lambda_b\df_b(\vec x_2)
                 +i\vec J\cdot\Delta u_{ab}}
  \right\rangle \Big|_{\lambda_a=\lambda_b=0}.
\end{equation}
Defining the auxiliary quantities
\begin{eqnarray}
  \mathcal A_{ab}(\vec J, \vec x_1, \vec x_2) &\equiv& \vec{J}\cdot\Delta u_{ab}, \non\\
  \mathcal B_{ab}(\vec J, \vec x_1, \vec x_2) &\equiv& \mathcal A_{ab}(\vec J, \vec x_1, \vec x_2) +
 \lambda_a\df_a(\vec x_1) + \lambda_b\df_b(\vec x_2),
\end{eqnarray}
we can use the cumulant theorem to compute
\begin{eqnarray}
\hat{\mathcal{O}}\left\langle e^{\mathcal B_{ab}}\right\rangle
  \bigg|_{\lambda_a=\lambda_b=0}
  &=& \hat{\mathcal{O}}  \exp\Bigg[\sum_{n=2}^\infty\frac{i^n}{n!}
 \left\langle \mathcal B_{ab}^n \right\rangle_c \Bigg]
 \Bigg|_{\lambda_a=\lambda_b=0} \\
  &=& \exp\Bigg[\sum_{n=2}^\infty\frac{i^n}{n!}
      \left\langle \mathcal A_{ab}^n \right\rangle_c \Bigg]
  \Bigg(1-i\sum_{n=2}^\infty\frac{i^n}{n!}
  \Big(\la\partial_a \mathcal B_{ab}^n\ra_c + \la\partial_b \mathcal B_{ab}^n\ra_c\Big)
  \nonumber \\
  && - \lb\sum_{n=2}^\infty \frac{i^n}{n!}\la\partial_a \mathcal B_{ab}^n\ra_c\rb
  \lb\sum_{n=2}^\infty \frac{i^n}{n!}\la\partial_b \mathcal B_{ab}^n\ra_c\rb
  -\sum_{n=2}^\infty\frac{i^n}{n!}\la\partial_a\partial_b
  \mathcal B_{ab}^n\ra_c\Bigg)\Bigg|_{\lambda_a=\lambda_b=0}.
\end{eqnarray}
The generating function then becomes
\begin{align}
  1 + \mathcal{M}^{ab} \big(\vec J,\vec r\big)
  &= \exp\Bigg[ \sum_{n=2}^\infty \frac{i^n}{n!}\la \mathcal A_{ab}^n\ra_c\Bigg]
  \bigg(1 + \sum_{n=1}^\infty \frac{i^n}{n!}
  \Big(\la\df_a \mathcal A_{ab}^n\ra_c + \la\df_b \mathcal A_{ab}^n\ra_c \Big)
  \nonumber  \\
  & \hspace{1.1cm} + \lb\sum_{n=1}^\infty\frac{i^n}{n!}\la\df_a \mathcal A_{ab}^n\ra_c\rb
       \lb\sum_{n=1}^\infty\frac{i^n}{n!}\la\df_b \mathcal A_{ab}^n\ra_c\rb
  + \sum_{n=0}^\infty\frac{i^n}{n!}\la\df_a\df_b \mathcal A_{ab}^n\ra_c\bigg).
\end{align}
This can be simplified by defining a ``smoothing'' kernel
\begin{eqnarray}
  K \lb\vec J,\vec r\rb &\equiv&
  \exp\Big[\la e^{ i \vec J\cdot\Delta u_{ab}}\ra_c  -1 \Big] \non\\
  &=& \exp\Bigg[ \sum_{n=2}^\infty \frac{i^n}{n!} J_{i_1}\ldots J_{i_n}
  \la\Delta u_{ab, i_1}\ldots\Delta u_{ab,i_n}\ra_c\Bigg].
  \label{eq:smo_ker}
\end{eqnarray}
This ``smoothing'' kernel can be considered as a general and full form of what is typically called the ``finger
of god'' term in many RSD models.  Historically this term was approximated to be a
scale independent contribution, originating from the zero-lag velocity
dispersion \cite{Pea92,Par94,PD94,BPH96,Ham98,HatCol99,Whi01,TNS10}.
In the literature, such terms were frequently re-summed, more often than
not on phenomenological grounds, yielding the family of so-called dispersion RSD models.
A serious drawback of such schemes in PT calculations was that at a certain
order in PT they typically broke the Galilean invariance of the theory 
(see also the related discussion in e.g. \cite{Sco04,Vla12}).
Note that this is explicitly not the case for the full smoothing kernel in Eq. \eqref{eq:smo_ker} where,
in addition to explicit Galilean invariance, the full scale dependence
of the Eulerian velocity field $\Delta u$ cumulants is included in the exponent. 

Finally, collecting all above, we have for the generating function
\begin{align}
  1 + \mathcal{M}^{ab}\big(\vec J,\vec r\big) &=
  K \lb \vec J, \vec r \rb
  \bigg( 1 + \la \lb \df_a +\df_b \rb e^{ i \vec J \cdot \Delta u_{ab}} \ra_c \non\\
  &\hspace{2.7cm}
  +  \la \df_a e^{ i \vec J \cdot \Delta u_{ab}}\ra_c
  \la \df_b e^{ i \vec J \cdot \Delta u_{ab}} \ra_c
+ \la \df_a \df_b e^{i \vec J \cdot \Delta u_{ab}} \ra_c \bigg),
\end{align}
where the power spectrum can now be obtained via Eq.~\eqref{eq:power_master}.
This result is equivalent to Eq. (31) in Ref.~\cite{Sco04}.

Note that the last step, computing the power spectrum, involves computing
one additional Fourier transform.  In this respect this method resembles
the configuration space streaming model.  However, the kernel method does
not allow for a Fourier counterpart, such as in the streaming model case,
and so the last integration step in Eq.~\eqref{eq:power_master} is
unavoidable.  Further, the quantities which enter above are most naturally 
evaluated in the Eulerian picture, given that the ``smoothing'' kernel $K(\vec J, \vec r)$ 
depends on the Eulerian velocity field $u$. Unlike in other approaches 
discussed earlier, $u$ is now a volume-weighted quantity rather than mass-weighted. 
For this reason, it is less straightforward to evaluate these correlators using Lagrangian 
theory, and we shall not consider this approach further in the rest of the paper. 
Nonetheless, it would be interesting to perform a detailed comparative study 
including this approach, and we hope that this question will be addressed in the future.

\section{Fourier space application and comparison of methods}
\label{sec:comparison}

We now compare the performance and convergence of the different approaches
detailed in the previous section, first in Fourier space (this section) and
then in configuration space (\S\ref{sec:config_space}).
In order to bring out the essential points we shall adopt a simplified
dynamical model, since it serves to highlight some of the more interesting
aspects of the problem.

Thus, while in reality the effects of translations and nonlinear dynamics
are intertwined and can give raise to effects of comparable importance on
scales of interest, we shall assume they can be separated.
In order to focus attention on the effects of translations, below we will
restrict the dynamics to the linear displacements (i.e.~the Zeldovich
approximation \cite{Zel70}), and neglect the higher order corrections.
We will also neglect bias for now.
Of course this is purely for the purposes of presentation -- each formalism
can easily be applied to more general dynamical models and biased tracers
(see \S\ref{sec:bias}).

Moving into redshift space amounts to replacing the displacement,
$\vPsi$, by
\begin{equation}
  \vPsi\to\vPsi^s=\vPsi+\frac{\hat{n}\,(\dot{\vPsi}\cdot\hat{n})}{\mathcal{H}}
  = \mathbf{R}\vPsi
\end{equation}
where $R_{ij}=\delta_{ij}+f\hat{n}_i\hat{n}_j$ and $\hat{n}$ is the line of sight.
In the distant-observer limit\footnote{Note that one can relax the distant
observer approximation within the Zeldovich approximation
\cite{Tas14b,CasWhi18b}.}
the angular dependence is normally expanded in a Legendre series:
\begin{equation}
  P(k,\nu) = \sum_{\ell=0}^\infty P_\ell(k)\mathcal{L}_\ell(\nu)
  \quad {\rm and} \quad
  \xi(s,\mu_s) = \sum_{\ell=0}^\infty \xi_\ell(s)\mathcal{L}_\ell(\mu_s)
\end{equation}
where $\mathcal{L}_\ell$ is the Legendre polynomial of order $\ell$,
$\nu=\hat{k}\cdot\hat{n}$ and $\mu_s=\hat{s}\cdot\hat{n}$.  Note we have used
$\nu$ for what is commonly called $\mu_k$ to distinguish it from the other
cosines which appear later.
The multipole moments of $\xi$ and $P$ are related through a Hankel transform,
\begin{equation}
  \xi_\ell(s) = i^\ell \int_0^\infty\frac{k^2\,dk}{2\pi^2}\ P_\ell(k)j_\ell(ks)
\end{equation}
where $j_\ell$ is the spherical Bessel function of order $\ell$.

The linear theory predictions are the same in each approach and amount to
the well known \cite{Kai87}
\begin{equation}
  P_s(k,\nu) = \left(1 + f\nu^2\right)^2 P_L(k) \quad ,
\end{equation}
where $f$ is the growth rate $f=d\ln D/d\ln a$.  In linear theory only
$P_0$, $P_2$ and $P_4$ are non-zero and each is proportional to $P_L$.
Including the higher order terms in the Zeldovich approximation we begin
to see the differences in the approaches.

\subsection{Direct Lagrangian approach}
\label{sec:direct_lag}

We consider two different methods to directly evaluate the Zeldovich
power spectrum in redshift space.  We will label the methods MI and MII.
Both methods rely on representing the power spectrum in term of a
series of spherical Bessel functions that can be truncated and
efficiently evaluated numerically.
We will compare the two methods and their efficiency.

The redshift-space power spectrum is given by Eq.~\eqref{eq:ps_noshift}
but with the transformation
$A_{ij} \to A^s_{ij} = R_{i\ell} R_{jm} A_{\ell m}$.  Note that
\eq{
\label{eq:kkRRA}
  k_i k_j R_{i\ell} R_{jm} A_{\ell m} &=
  k^2X\left[ 1 + 2f \nu^2 + f^2\nu^2 \right] +
  k^2Y\left[ \mu^2+2f\mu \nu (\hat n \cdot \hat q)
           + f^2\nu^2 (\hat n \cdot \hat q)^2\right] \\
&= k^2 X \alpha_0(\nu) + k^2 Y \Big(  \alpha_1(\nu)
+  \alpha_2(\nu)  \gamma(\mu, \nu) \cos \phi +  \alpha_3(\nu) \gamma(\mu, \nu)^2 (\cos \phi)^2 \Big) \mu^2, \non
}
where $\mu=\hat{q}\cdot\hat{k}$, $X$ and $Y$ are linear theory displacement contributions explicitly defined in Eq.~\eqref{eq:Xi_1234} and \eqref{eq:XYVT}, 
and we introduce angle factors $\gamma(\mu, \nu)= \sqrt{1-\mu^2}\sqrt{1-\nu^2}/\mu\nu$ and 
\eeq{
\alpha_0(\nu) =  \Big( 1 + f \lb 2 + f \rb \nu^2 \Big), 
~~~ \alpha_1(\nu) = \Big( 1 + f \nu^2 \Big)^2, 
~~~ \alpha_2(\nu) = 2 f \nu^2 (1 + f \nu^2), 
~~~ \alpha_3(\nu) = f^2 \nu^4. \non
}
For method MI we first perform the integral over azimuthal angle $\phi$,
and thus we can write the power spectrum in the form
\eq{
(2\pi)^3\df^D(\vec k) + P_s(\vec k, \tau) 
&= \int d^3 q ~ e^{ i \vec k \cdot \vec q} ~ \exp \Big[ - \frac{1}{2} k_i k_j  A^s_{ij} \Big] \\
&= \int d^3 q ~ e^{ i \vec k \cdot \vec q} ~ 
e^{-\frac{1}{2}k^2 \big( \alpha_0(\nu) X^{\rm lin}(q) +   \alpha_1(\nu) \mu^2 Y^{\rm lin}(q) \big) }  
I_\phi \lb \mu, \nu, -\tfrac{1}{2}k^2 Y^{\rm lin}(q) \rb. \non
}
where in the second line we have introduced an azimuthal integral $I_\phi$.
We can expand $I_\phi$ in powers of $\mu$ to get
\eq{
I_\phi \lb \mu, \nu, C \rb = \int_0^{2\pi} \frac{d\phi}{2\pi}~e^{ C \mu^2 \big( \alpha_2(\nu)  \gamma(\mu, \nu) \cos (\phi) 
+ \alpha_3(\nu)  \gamma(\mu, \nu)^2 \cos (\phi)^2 \big) }
=  \sum_{\ell=0}^\infty F_{\ell}(\nu, \alpha_1 C) \lb \mu^{2} \alpha_1 C \rb^\ell~, \non
}
where, using the confluent hypergeometric function of the first kind (Kummer's) $M(a,b,z)$, we introduce a further function 
\eq{
F_{\ell}(\nu, x) &= \sum_{m=0}^\ell
\frac{(-1)^m 4^{\ell-m}\Gamma(m+\tfrac{1}{2})}
     {\pi^{1/2}\Gamma(m+1)\Gamma(1+2m-\ell)\Gamma(2\ell-2m+1)}
     \lb \frac{\alpha_0}{\alpha_1} \rb^m \non\\
&\hspace{2.4cm} \times  M\left(\ell-2m; \ell-m+\tfrac{1}{2};  x \right)
 M\left(m+\tfrac{1}{2};m+1; \frac{\alpha_0}{\alpha_1}\,x \right).
}
Next we do the integral over the angle $\mu$. This can be done by using well-known formulae for the integrals of powers of $\mu$ times
a Gaussian in $\mu$.
Following Ref.~\cite{VlaSelBal15} and using the integral
\eeq{
\frac{1}{2} \int_{-1}^1 d \mu ~ \mu^{2\ell} e^{i \mu A + \mu^2 B } = \frac{(-1)^{\ell} e^{B}}{B^\ell} \sum_{n=0}^{\infty} U(-\ell,n -\ell +1,-B) \lb -\frac{2B}{A} \rb^n j_n(A)~ ,
\label{eq:int_mu_n}
}
where $U(a,b,z)$  is a confluent hypergeometric function of the second kind
(Tricomi's), we can write the redshift-space, Zeldovich power spectrum as
\begin{align}
  P_s(k,\nu) &= 4\pi  \sum_{n=0}^{\infty}
  \int q^2\,dq\ \mathcal K^s_n \big(k,q,\nu\big)
  \ e^{-\frac{1}{2}k^2 \big( X(q)+Y(q) \big) }
  \lb\frac{k\,Y(q)}{q}\rb^n\, j_n(kq) \quad .
\label{eq:PS_Zel_met_I}
\end{align}
All of the RSD effects are contained in the kernel
\begin{equation}
\mathcal K_n^s \big(k, q, \nu  \big) = \Big( 1 + f \nu^2 \Big)^{2n}
  e^{-(1/2)f\nu^2k^2\big[\lb 2+f\rb X(q) + \lb 2+f\nu^2 \rb Y(q)\big]}
  K^s_n\Big(\nu,-\tfrac{1}{2}a_1k^2Y\Big)
\end{equation}
with
\eq{
K_n^s(\nu,x) &= \sum_{\ell=0}^\infty (-1)^\ell F_{\ell}(\nu, x)
  U(-\ell,n -\ell +1,-x).
}
Again, above we use standard functions $M(a,b,z)$ and $U(a,b,z)$, which are confluent hypergeometric
function of the first (Kummer's) and second (Tricomi's) kind respectively.

While this formula looks cumbersome, it lends itself to efficient numerical
evaluation.  The functions $j_n$, $X$ and $Y$ can be pre-computed on a
regular log-spaced grid then Fourier transforms can be employed to do the
$q$-integral \cite{Ham00}.  The sum over $n$ can be truncated at finite
$n$, with more terms needed for higher $k$.  We find sub-percent convergence
with between 10-20 terms for all $k$ where PT can be expected to hold
($k<1\,h\,{\rm Mpc}^{-1}$).  The infinite sum in the $K_n^s(\nu,x) $ can
also be truncated.  For the typical values of our arguments,
$\ell_{\rm max} = 10-15$ is sufficient.

For our second method (MII) we change the coordinate frame for the integration.
Rather than using $\hat{k}$ as our $z$-axis we instead introduce a new variable
\eeq{
  K_i = k_j \lb \df^K_{ij} + f \hat n_i \hat n_j \rb,
  \qquad
  K^2 = k^2 \left[ 1 + f \lb 2 + f \rb \nu^2 \right],
}
and set up the coordinate frame so that the $z$-axis is along $\vec K$.
This makes integration over the azimuthal angle trivial.
This coordinate frame was also suggested by Ref.~\cite{TayHam96}.
The redshift-space, Zeldovich power spectrum is then given as
\eeq{
  P_s(\vec k) = 2 \pi \int q^2 dq\, d \mu
  \ e^{i k q c \mu  - (1/2)K^2(X+\mu^2 Y) }
  J_0 \lb k q s \sqrt{1 - \mu^2} \rb,
\label{eq:P_zel_Ham&Tay}
}
with
\eeq{
  s = f \nu \frac{\sqrt{1-\nu^2}}{\sqrt{1 + f \lb 2 + f \rb \nu^2}}
  \quad , \quad
  c = \sqrt{1-s^2}
}
so that $(k q c)^2+(k q s)^2 = (k q)^2 (c^2 +s^2) = (kq)^2$.
We can use integral 6.677(6) from Ref.~\cite{GRint2014}
\eeq{
  \int_{-1}^1 d\mu ~e^{i \mu A}   J_0\lb C \sqrt{1-\mu^2} \rb
  = 2\frac{\sin \sqrt{ A^2 + C^2 } } {\sqrt{ A^2 + C^2 }},
}
for simplifying Eq.~(\ref{eq:P_zel_Ham&Tay}).
Ref.~\cite{TayHam96} suggested (in their appendix) taking the derivatives
with respect to the variable $A$ in order to obtain a series for handling
the additional $\mu^2$ in the exponent of Eq.~(\ref{eq:P_zel_Ham&Tay}).
We found this method converged very slowly so we take a different path that
allows us to obtain a closed-form representation of the $\mu$ integral.
Using the differential equation for the Bessel function
\begin{equation}
  x^2 J_0''(x) + x J_0'(x) + x^2 J_0(x) = 0,
\end{equation}
we can write 
\eeq{
(1-\mu^2) J_0(C \sqrt{1-\mu^2}) = - \lb \partial^2_C + \frac{1}{C} \partial_C \rb J_0(C \sqrt{1-\mu^2}),
}
and thus
\eq{
\int_{-1}^1 d\mu ~e^{i \mu A + \mu^2 B}   J_0\lb C \sqrt{1-\mu^2} \rb
&= e^{B}\sum_{n=0}^\infty \frac{(-B)^n}{n!} \int_{-1}^1 d\mu ~ e^{i \mu A } (1 - \mu^{2})^n J_0\lb C \sqrt{1-\mu^2} \rb \non\\
&= 2 e^{B}\sum_{n=0}^\infty \frac{B^n}{n!} \lb \partial^2_C + \frac{1}{C} \partial_C \rb^n \frac{\sin \sqrt{ A^2 + C^2 } } {\sqrt{ A^2 + C^2 }}.
}
Introducing the variable $\rho = \sqrt{A^2 + C^2}$, we can rewrite the integral in the form
\eq{
\int_{-1}^1 d\mu ~e^{i \mu A + \mu^2 B}   J_0\lb C \sqrt{1-\mu^2} \rb
&= 2\exp\left\{ \frac{B}{\rho} \left[ (\rho^2 - A^2) \frac{d}{d\rho} \lb \frac{1}{\rho} \frac{d}{d\rho} \rb + 2 \frac{d }{d\rho}  + \rho \right] \right\} j_0(\rho) \non\\
&= 2 e^{B} \sum_{\ell = 0}^\infty G_\ell(A, B, \rho) j_\ell (\rho),
}
where in the last line we have introduced a function
\eq{
G_m(A, B, \rho) &=
\lb - \frac{2}{\rho} \rb^m \sum_{n=m}^\infty
\lb \frac{B A^2}{\rho^2} \rb^n \frac{\Gamma \left(m+n+\frac{1}{2}\right)}{\Gamma (m+1) \Gamma \left(n+\frac{1}{2}\right) \Gamma(1-m+n)} \non\\
&\hspace{5.5cm} \times \,_2F_1\left(\frac{1}{2}-n,-n; \frac{1}{2} - m - n; \frac{\rho ^2}{A^2}\right),
}
where ${}_2F_1$ is the ordinary (Gauss) hypergeometric function.
Thus Eq.~\eqref{eq:P_zel_Ham&Tay} becomes
\eeq{
P_s (\vec k) = 4 \pi \sum_{\ell = 0}^\infty \int_0^\infty q^2 dq ~ e^{-(1/2)K^2(X+Y)} G_\ell \lb ckq, - \tfrac{1}{2}K^2 Y,  kq \rb j_\ell (kq).
\label{eq:P_zel_Ham&Tay_fin}
}
To our knowledge this is the first direct and complete redshift-space,
Zeldovich power spectrum calculation presented in the literature,
although Ref.~\cite{TayHam96} outlined in their appendix a direction
very similar to MII presented above.

\begin{figure}
\begin{center}
\resizebox{\columnwidth}{!}{\includegraphics{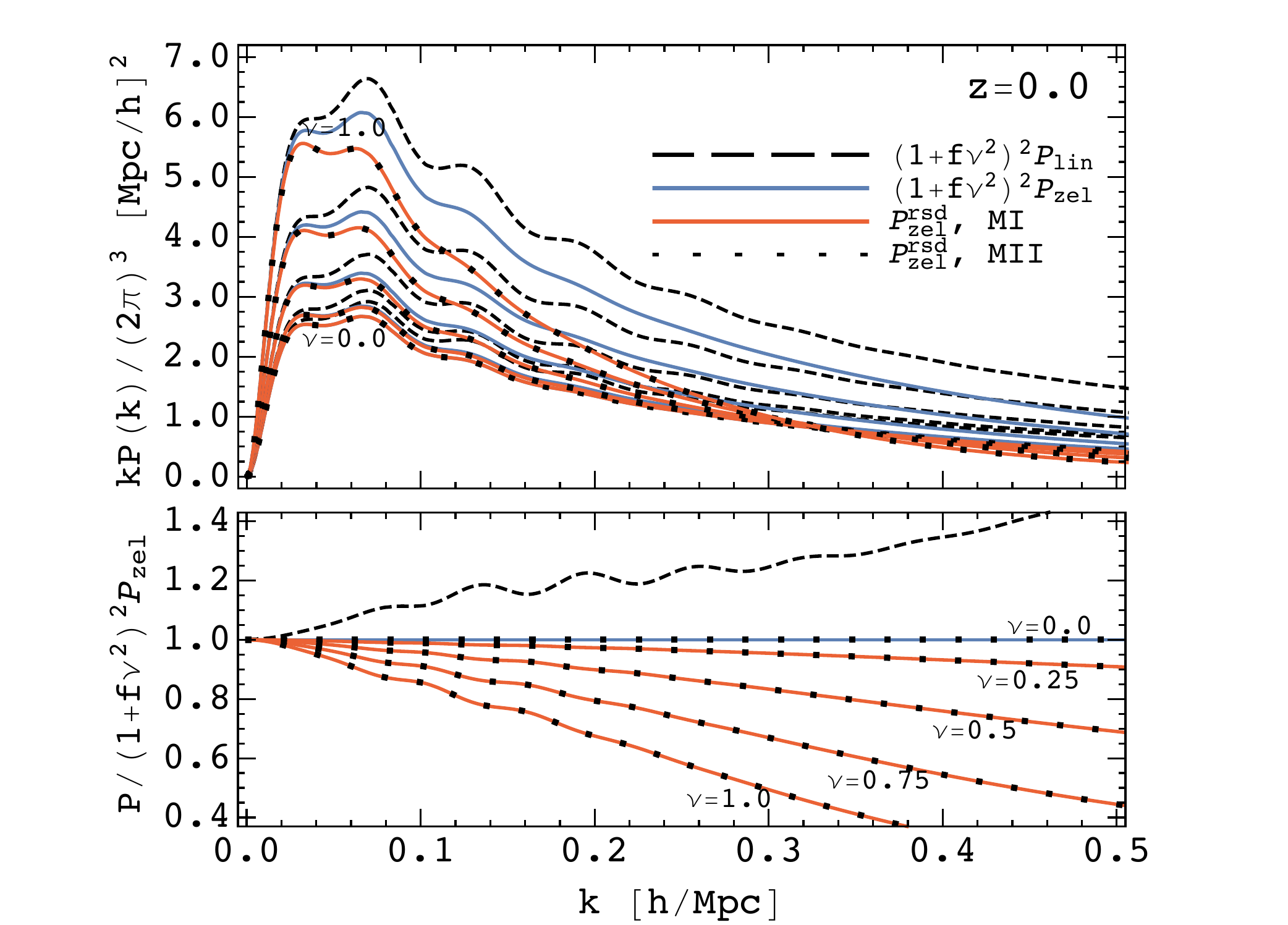}}
\end{center}
\vspace*{-5mm}
\caption{The redshift-space, Zeldovich power spectrum as a function of $k$
for select values of $\nu$ for $\Lambda$CDM at $z=0$.
We highlight the agreement between the two series expansions MI
(Eq.~\eqref{eq:PS_Zel_met_I}; red solid lines)
and MII (Eq.~\eqref{eq:P_zel_Ham&Tay_fin}; black dotted lines) described
in the text.  For comparison we show the linear theory results,
$(1+f\nu^2)^2P_L$, as the black dashed line
and the real-space, Zel'dovich power spectrum multiplied by the linear
RSD term $(1+f\nu^2)^2$ as the blue lines.  The upper panel shows the power
spectrum, times $k$ to reduce the dynamic range, while the lower panel shows
the ratio to $(1+f\nu^2)P_{\rm Zel}$ which highlights the change in the
damping as a function of $\nu$.
}
\label{fig:Pk_DirectLagrangian}
\end{figure}

\begin{figure}
\begin{center}
\resizebox{\columnwidth}{!}{\includegraphics{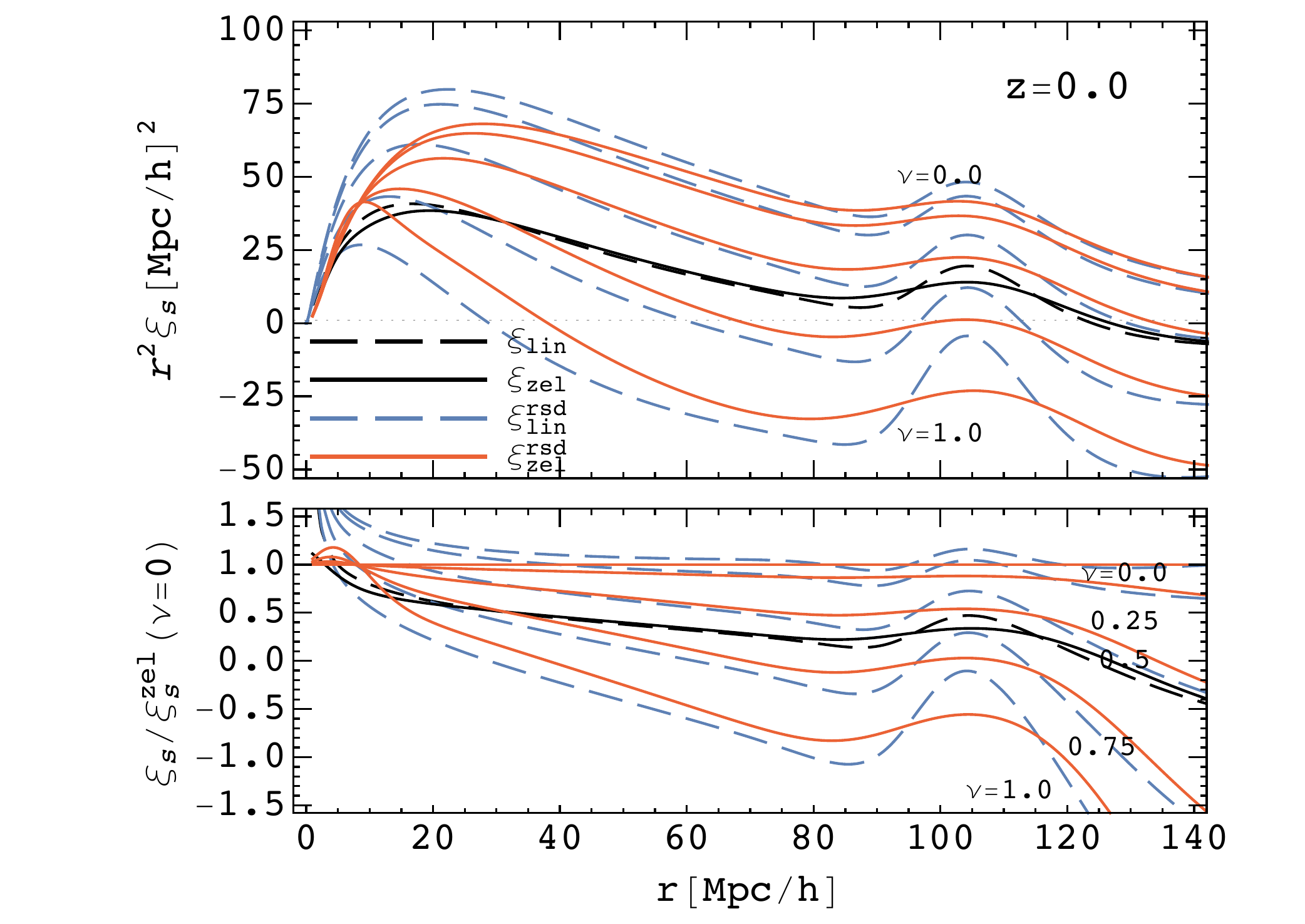}}
\end{center}
\vspace*{-5mm}
\caption{The redshift-space, Zeldovich correlation function as a function
of $s$ for select values of $\nu$ for $\Lambda$CDM at $z=0$.
}
\label{fig:xi_DirectLagrangian}
\end{figure}

We evaluate the Zeldovich power spectrum and correlation function for the 
$\Lambda$cdm cosmology using the parameters $\Omega_m=0.295$, $\Omega_b=0.047$, $n_s=0.968$, $\sigma_8 = 0.835$ and $h=0.688$.
Fig.~\ref{fig:Pk_DirectLagrangian} shows the power spectrum results, and the good agreement
between MI and MII.  At high $k$ we see the familiar damping of power from
the Zeldovich dynamics, with the damping being larger along the line of
sight than transverse to it.  This damping corresponds to the fact that
small scale structure does not form properly in the Zeldovich approximation.
This is not of concern to us, since we are using the Zeldovich approximation
for illustration.  We are more interested in checking the self-consistency
of the different approaches and the consistency of methods MI and MII.
The two methods give excellent agreement on the scales of interest
($k<1\,h\,{\rm Mpc}^{-1}$), though on smaller scales differences begin to
creep in that are due to our truncations in the sums in
Eqs.~\eqref{eq:PS_Zel_met_I} and \eqref{eq:P_zel_Ham&Tay_fin}.  These
differences could be reduced further by including more terms.
For comparison we also show the deviations from linear theory,
$(1+f\nu^2)^2P_L$, as well as a phenomenological `Kaiser-Zeldovich'
approximation: $(1+f\nu^2)^2P_{\rm Zel}$.  The latter highlights how the
damping in the Zeldovich approximation depends upon angle to the line of sight.

Since it will be useful later, we also reproduce the configuration-space
2-point function, $\xi(s,\nu)$, within the Zeldovich approximation.
Fig.~\ref{fig:xi_DirectLagrangian} shows $s^2\,\xi(s,\nu)$ for several values
of $\nu$ with the characteristic BAO peak clearly visible at
$s\simeq 100\,h^{-1}$Mpc.  The BAO peak is broadened from its linear theory
value by non-linear structure formation and that broadening is anisotropic.
The angle-dependence of the Zeldovich correlation function is different from
that of linear theory even at relatively large scales, showing that the Kaiser
limit is approached very slowly in configuration space.
We shall return to $\xi(s,\nu)$ in \S\ref{sec:config_space}.

\subsection{Moment expansion approach}
\label{sec:mom_exp}

In order to explore the moment expansion approach we return to the generating
function.  We are interested in ascertaining how well the moment expansion
approach works compared to the exact result of the previous section.
In other words, how many terms do we need to keep in the moment expansion
in order to achieve good accuracy at the scales of interest?
Since we are using leading order Lagrangian dynamics (the Zeldovich
approximation) we can answer this question robustly as we can derive
closed-form expressions for arbitrary velocity moments.  Even though the
full dynamics is not properly captured, we believe the answers will be
relevant to the more general case as well.

We start from the expressions in Sec.~\ref{sec:dir_lag_app}
(setting $\df_a(\vec q) = \df_b(\vec q) = 0$ since we are not interested in
the biased tracers at this point),
and use the cumulant expansion theorem (for Gaussian fields) to obtain
\eq{
  \widetilde{\mathcal M}(\vec J, \vec k) &= \frac{k^3}{2\pi^2}
  \int d^3q\ e^{i\vec{k}\cdot\vec q}
  \  e^{-\frac{1}{2} \lb k_i k_j + f (\vec{J}\cdot \hat n) k_{\{ i}\hat n_{j\}}
   + f^2 (\vec{J}\cdot \hat n)^2 \hat n_i \hat n_j \rb A_{ij} } \non\\
  &= \frac{k^3}{2\pi^2}\int d^3q\ e^{i\vec{k}\cdot\vec q}
  \ e^{ - \frac{1}{2} k_i k_j A_{ij}}
   e^{-\frac{1}{2}\lb fJ_{\hat n}\mathcal A + f^2 J_{\hat n}^2\mathcal B \rb },
\label{eq:moment_gen_fnc}
}
where we introduced $J_{\hat n} = \vec{J}\cdot \hat n $,
$\mathcal A = k_{\{ i} \hat n_{j\}} A_{ij} $ and
$\mathcal B = \hat n_i \hat n_j A_{ij}$.
Velocity moments in terms of the derivatives in $J_{\hat n}$ can now be
obtained from the Taylor expansion
\begin{equation}
  \widetilde{\mathcal M}(\vec J, \vec k)
  = \sum_{\ell=0}^\infty \frac{i^\ell}{\ell!}
  \lb J_{\hat n} \rb^\ell \widehat{\Xi}^{(\ell)}_{\hat n}(\vec k),
   \quad{\rm i.e.}\quad
  \widehat{\Xi}^{(\ell)}_{\hat n}(\vec k) =
  (-i)^\ell \frac{\partial^\ell }{\partial J_{\hat n}^\ell}
  \widetilde{\mathcal M}( \vec J, \vec k)\Big|_{\vec J=0}.
\end{equation}
We note that equivalently one could take the derivative relative to the
logarithmic growth rate $f$ given that it always appears as $f J_{\hat n}$
in Eq. \eqref{eq:moment_gen_fnc} above.
This guarantees that the $n^{\rm th}$ velocity moment is proportional to
$f^n$, so the moment expansion can alternatively be considered as a Taylor
expansion of the full RSD spectrum in powers of $f$ \cite{Sug16}.

To obtain an explicit form for the moments it is useful to split them into
odd and even groups.  We write
\eq{
\widetilde \Xi^{(2\ell)}_{\hat n}(k,\nu)
&=f^{2\ell} \sum_{m = 0}^{\ell} \widetilde \Xi^{(2\ell)}_{m}(k)\ \nu^{2m} \non\\
&= f^{2\ell} \sum_{n=0}^{\ell} \frac{(-1)^n}{2^{\ell + n}} \frac{(2\ell)!}{(\ell - n)! (2n)!}
  \int d^3q\ e^{i\vec{k}\cdot\vec q}~ \mathcal A^{2n} \mathcal B^{\ell -n} e^{ - \frac{1}{2} k_i k_j A_{ij} }, \non\\
\widetilde \Xi^{(2\ell+1)}_{\hat n}(k,\nu)
 &=i f^{2\ell+1} \sum_{m = 0}^{\ell} \widetilde \Xi^{(2\ell+1)}_{m}(k)\ \nu^{2m+1} \non\\
 & = i f^{2\ell+1}
  \sum_{n=0}^{ \ell } \frac{(-1)^n}{2^{\ell + n+1}} \frac{(2\ell+1)!}{(\ell - n)! (2n+1)!}
 \int d^3q\ e^{i\vec{k}\cdot\vec q}~\mathcal A^{2n+1} \mathcal B^{\ell -n} e^{ - \frac{1}{2} k_i k_j A_{ij} },
 \label{eq:vel_moments}
}
where in the first lines we have separated the scale and angle dependence,
implicitly defining the reduced velocity moments $\widetilde \Xi^{(\ell)}_{m} (k)$ via
the $m^{\rm th}$ derivative of the moments themselves:
$(1/m!)(\partial_\nu)^m \widetilde{\Xi}^{(2\ell)}_{\hat n}(\vec k) \big|_{\nu=0}$.
The reduced moments are thus a function only of the amplitude of the wave
vector, $k$.
It is convenient to perform this separation since each moment contains only a
finite number of powers of $\nu$, which can be seen explicitly in
Refs.~\cite{SelMcD11, Vla12} or Appendix \ref{sec:and_decomp_ofvelocities}.

\begin{figure}
\begin{center}
\resizebox{\columnwidth}{!}{\includegraphics{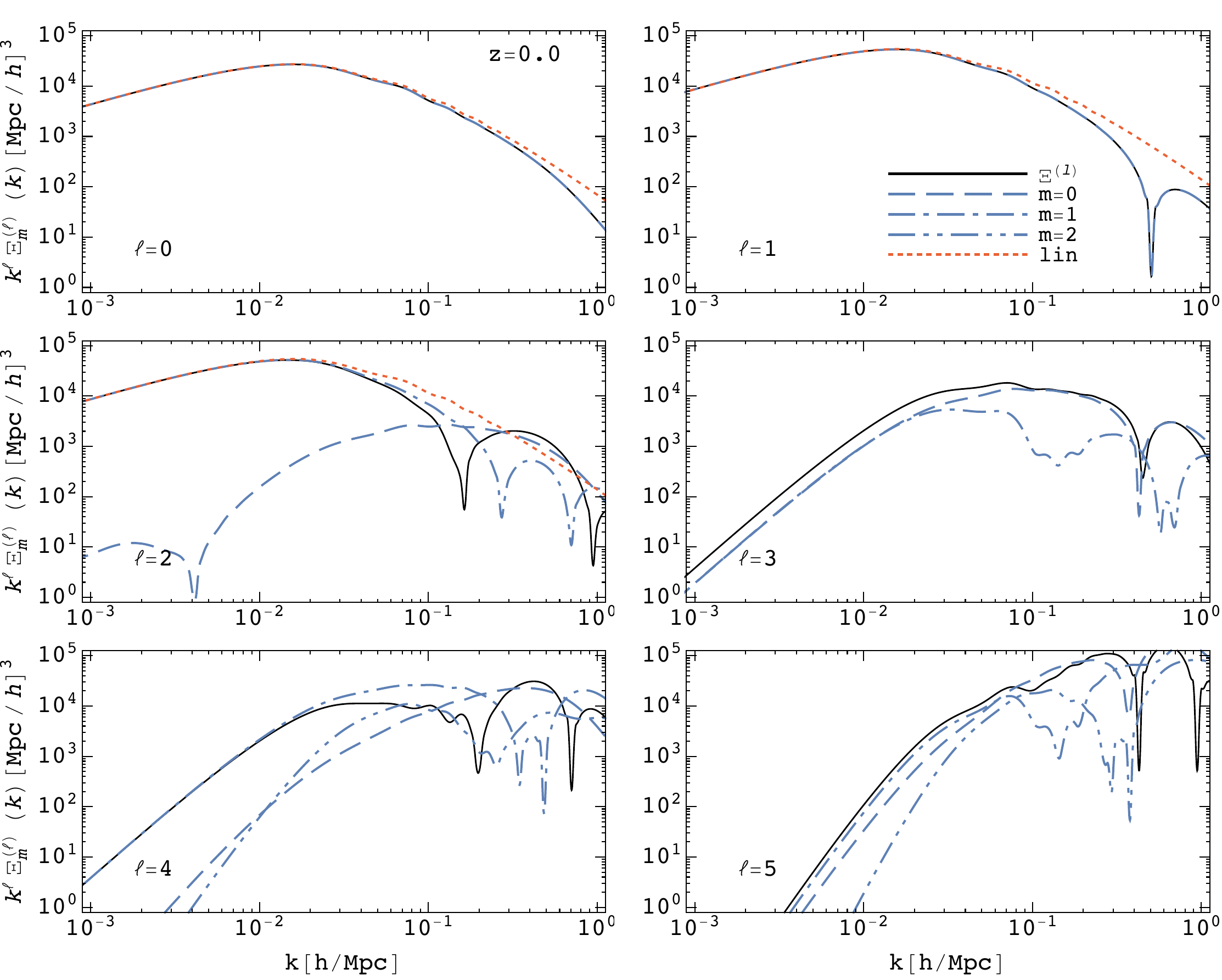}}
\end{center}
\vspace*{-5mm}
\caption{Velocity moments, $k^{\ell}\widetilde{\Xi}^{(\ell)}_m$, in
Fourier space as a function of $k$, for $\Lambda$CDM at $z=0$.
Moments are computed in Zeldovich approximation and separated according to
the angular dependence (as in Eq.~\ref{eq:reduced_vec_moments}):
$m=0$ (long dashed lines), 
$m=1$ (dash-dotted lines), $m=2$ (dash-dot-dotted lines).
The black solid lines show the sum for $\nu=1$, i.e.~each $m$-mode weighted
maximally.  The linear theory results are also shown for $\ell=0$, $1$ and
$2$ (red dotted lines) and are each proportional to $P_L$.}
\label{fig:Mom_DirectLagrangian}
\end{figure}

We delegate the explicit derivation of the reduced velocity moments,
$\widetilde{\Xi}^{(\ell)}_{m}(k)$, to Appendix \ref{sec:red_vel_moments},
where it is shown that
\eq{
\widetilde \Xi^{(2\ell)}_{m} (k)
&= 4\pi \sum_{s=0}^{\infty} \int q^2dq ~ e^{-\frac{1}{2} k^2 (X+Y)} \xi^{(2\ell)}_{m,s} (k,X,Y) \lb \frac{kY}{q} \rb^s j_s(qk),  \non\\
\widetilde \Xi^{(2\ell+1)}_{m} (k)
&= 4\pi \sum_{s=0}^{\infty} \int q^2dq ~ e^{-\frac{1}{2} k^2 (X+Y)} \xi^{(2\ell+1)}_{m,s} (k,X,Y) \lb \frac{kY}{q} \rb^s j_s(qk),
\label{eq:reduced_vec_moments}
}
with the integrand functions given by
\eq{
 \xi^{(2\ell)}_{m,s} (X,Y) &= \sum_{n=0}^{2\ell} \frac{F^{(2\ell)}_{m,n} (k,X,Y)} {(k^2 Y/2)^n} U(-n, s -n +1, k^2Y/2), \non\\
 \xi^{(2\ell+1)}_{m,s} (X,Y) &=  \sum_{n=0}^{2\ell+1} \frac{F^{(2\ell+1)}_{m,n} (k,X,Y)} {(k^2 Y/2)^n} U(-n, s -n +1, k^2Y/2),
 \label{eq:xi_integrands}
}
and the $F^{(\ell)}_{m,n}$ given explicitly by Eq.~\eqref{eq:F_fncs},
with $U(a,b,z)$ is Tricomi's confluent hypergeometric function as before.

Fig.~\ref{fig:Mom_DirectLagrangian} shows the first few velocity moments for
$\Lambda$CDM at $z=0$.  The red lines for $\ell=0$, 1 and 2 show the linear
theory predictions which are all proportional to $P_L$ as plotted.  We see
that the Zeldovich solutions reduce to linear theory at low $k$ as expected
(see below).
Non-linear evolution changes the behavior at large $k$ and generates higher
$\ell$ moments, which are highly suppressed at low $k$ in the way they are
plotted here but become comparable to $\ell\le 2$ for $k$ at or above the
non-linear scale.  By contrast the $\nu$ dependence is less straightforward.
While $m\le\ell$, the higher $m$ modes for a given $\ell$ are not suppressed
at low $k$ in the way the higher $\ell$ modes are.
For example for $\ell=2$ and $\ell=4$
(two lower left panels of Fig.~\ref{fig:Mom_DirectLagrangian})
the $m=1$ mode dominates over $m=0$ and $m=2$ at low $k$.
We shall compare this expansion to the others in \S\ref{sec:Fourier_comparison}.

In Eq.~\eqref{eq:vel_moments} we have expressed the velocity moments as
an expansion in powers of $\nu$.  It will be useful later to rearrange these
expressions to give the multipole expansion of the moments.
Using the relations
\eeq{
  \nu^{2\ell} =  \sum_{i=0}^\ell (-1)^i c^{{\rm e,}(\ell)}_i  \mathcal P_{2i}(\nu), \qquad {\rm and} \qquad
  \nu^{2\ell+1}=  \sum_{i=0}^\ell (-1)^i c^{{\rm o,}(\ell)}_i  \mathcal P_{2i+1}(\nu), 
}
where
\eeq{
c^{{\rm e,}(\ell)}_n =
 \frac{(-1)^n (4n+1)(2\ell)!}{2^{\ell-n}(\ell-n)!(2\ell+2n+1)!!},
 \qquad {\rm and} \qquad 
c^{{\rm o,}(\ell)}_n  =
 \frac{(-1)^n (4n+3)(2\ell+1)!}{2^{\ell-n}(\ell-n)!(2\ell+2n+3)!!}.
\label{eq:c_coefs}
}
we have straightforwardly the velocity moments in terms of Legendre polynomials
\eq{
\widetilde \Xi^{(2\ell)}_{\hat n}(k,\nu)
&=f^{2\ell} \sum_{i=0}^\ell (-1)^i \lb \sum_{m = 0}^{\ell} c^{{\rm e,}(m)}_i  \widetilde \Xi^{(2\ell)}_{m}(k) \rb \mathcal P_{2i}(\nu), \non\\
\widetilde \Xi^{(2\ell+1)}_{\hat n}(k,\nu)
&=i f^{2\ell+1} \sum_{i=0}^\ell (-1)^i \lb \sum_{m = 0}^{\ell} c^{{\rm o,}(m)}_i \widetilde \Xi^{(2\ell+1)}_{m}(k) \rb  \mathcal P_{2i+1}(\nu).
 \label{eq:vel_moments_legendre}
}

Finally we note that the moment expansion is particularly convenient for
comparing to Eulerian methods.  In particular it is easy to obtain the
Kaiser result as indicated in Eq.~\eqref{eq:Kaiser_general}, so let us
derive the leading RSD contributions in this framework.  We have
\eq{
P_s(\vec k) &=
\widetilde \Xi^{(0)}_{\hat n} (\vec k)
+ i \nu k \widetilde \Xi^{(1)}_{\hat n} (\vec k)
-\frac{1}{2} \nu^2 k^2 \widetilde \Xi^{(2)}_{\hat n} (\vec k)
+\ldots \\
& = \int d^3 r ~ e^{i \vec k \cdot \vec r}
\la \big( 1 + \df(\vec x) \big)  \big( 1 + \df(\vec x') \big) \ra
+ i \nu k \int d^3 r ~ e^{i \vec k \cdot \vec r}
\la \big( 1 + \df(\vec x) \big)  \big( 1 + \df(\vec x') \big) \Delta u_{\hat n} \ra \non\\
&\hspace{5cm}-\frac{1}{2} \nu^2 k^2  \int d^3 r ~ e^{i \vec k \cdot \vec r}
\la \big( 1 + \df(\vec x) \big)  \big( 1 + \df(\vec x') \big) \Delta u_{\hat n}^2 \ra
+\ldots \non\\
&= \int d^3 r ~ e^{i \vec k \cdot \vec r} \la \df(\vec x) \df(\vec x') \ra
+i \nu k \int d^3 r ~ e^{i \vec k \cdot \vec r}
\Big( \la \df(\vec x') u_{\hat n}(\vec x) \ra - \la \df(\vec x) u_{\hat n}(\vec x') \ra \Big)\non\\
&\hspace{4.5cm} + \nu^2 k^2  \int d^3 r ~ e^{i \vec k \cdot \vec r} \la u_{\hat n}(\vec x) u_{\hat n}(\vec x') \ra
+\ldots, \non
}
where we have dropped terms proportional to Dirac $\delta$-functions.
Assuming that the vector components of velocity can be neglected and
only the scalar part contributes, i.e.~we can write
$\vec{v}(\vec k) =(i \vec k /k^2)\theta(\vec k)$,
and switching back to the peculiar velocity $\vec v = \mathcal H \vec u $
we get
\eq{
P^s(\vec k) &= P_{\df\df}(k) + i \nu k \mathcal H^{-1}
\Big( P_{u_{\hat n} \df} (\vec k) - P_{\df u_{\hat n}} (\vec k) \Big)
+  \nu^2 k^2 \mathcal H^{-2}  P_{u_{\hat n} u_{\hat n}} (\vec k) \non\\
&= P_{\df\df}(k) - 2\nu^2 \mathcal H^{-1} P_{\df \theta} (k)
+  \nu^4 \mathcal H^{-2}  P_{\theta \theta} (k)  \quad .
}
In linear theory $\theta = -f\mathcal{H}\df$ so we have
\eeq{
  P^s(\vec k) = \big(1 + f \nu^2 \big)^2 P_L(k),
}
i.e.~the well known Kaiser formula \cite{Kai87}.
The same limit can of course also be obtained via the direct Lagrangian
approach, as argued in the earlier section, or from the streaming model
by expanding the exponential and keeping only the leading, linear terms.

\subsection{Streaming models}

The streaming models arise from the cumulant expansion of $\mathcal{M}$
or $\widetilde{\mathcal{M}}$.  In either space it is straightforward to
show that $P_s(\vec{k})$ depends only upon even powers of
$\nu=\hat{k}\cdot\hat{n}$, and that for any power of $\nu$ only a finite
number of cumulants contribute.  We shall be interested in how the
expansion approaches the full Zeldovich result.

As mentioned above, there are two developments of the streaming model.
One applies the cumulant theorem to the generating function in configuration
space and the other to the generating function in Fourier space.  Since
the cumulants are constructed from the moments, and the moments are Fourier
transform pairs, they in principle contain the same information if carried
to infinite order.  However, in practice, the non-local nature of the Fourier transform 
and the need to truncate the expansion at finite order
makes their behavior very different, as we will show.   First though we
develop the streaming models more fully within the Zeldovich approximation.

\subsubsection{Fourier space}

The Gaussian streaming model in configuration space is well known (see
earlier discussion and references) and will be developed in
\S\ref{sec:GSM_config_space}.
The alternative formalism applies the cumulant expansion in Fourier space
and is new to this paper.  The extension of both the configuration-space
and Fourier-space results beyond $2^{\rm nd}$ order is also new to this
paper.

We can obtain cumulants from the $\widetilde{\Xi}^{(m)}_{\hat n}$ through
Eq.~\eqref{eq:k-cums}.  A general form of this transformation is given by 
\eeq{
\widetilde{\mathcal C}^{(\ell)}_{\hat n} (k, \nu)
  = \sum_{i=1}^\ell \frac{(-1)^{i-1} (i-1)!}{\left[ 1+\Delta^2(k) \right]^i} 
  B_{\ell,i} \lb \widehat{\Xi}^{(1)}_{\hat n}(k, \nu), \ldots,
                 \widehat{\Xi}^{(\ell - i + 1)}_{\hat n}(k, \nu) \rb,
\label{eq:bell_expansion}
}
where $B_{\ell,i} \lb x_1, \ldots, x_{\ell - i + 1} \rb$ are partial
Bell polynomials.
The power spectrum is
\begin{equation}
 \frac{k^3}{2\pi^2}P^{s}(k, \nu) 
  =  \left(1+\Delta^2\right)
  \exp \Bigg[ i (\nu k) \widetilde{\mathcal C}^{(1)}_{\hat n} (k, \nu) 
  - \frac{(\nu k)^2 }{2}  \widetilde{\mathcal C}^{(2)}_{\hat n} (k, \nu)
  + \ldots  \Bigg] - 1.
  \label{eq:Fourier_streaming_power_spectrum}
\end{equation}
By analogy to the configuration space case, we can call the Gaussian streaming model (GSM) the truncation at
the second cumulant $\widetilde{\mathcal C}^{(2)}_{\hat n}$ and dropping the higher ones.
As noted previously, this form provides a huge simplifcation over
the `usual' streaming model result in that the connection between
real and redshift space is algebraic.  The structure of the redshift
space terms is also particularly clear, and this form is reminiscent
of the older `dispersion' models which multiplied the linear theory
result by a phenomenological damping
\cite{Pea92,Par94,PD94,BPH96,Ham98,HatCol99,Whi01,Sco04,TNS10}.
We shall compare this expansion to the others in \S\ref{sec:Fourier_comparison}.

To bring out the correspondance with the dispersion models more clearly and
to highlight the structure of the finger of god terms, let us consider
$\mathcal{C}^{(2)}$ which derives from $\widetilde{\Xi}^{(2)}$.
In PT $\widetilde{\Xi}^{(2)}$ contains a term going as $P_L\int P_L$,
which is UV-sensitive.
This gives a contribution to $\widetilde{\mathcal{C}}^{(2)}$ that looks
like a constant.  Thus a piece of $\mathcal{K}$ is
$\exp[-k_\parallel^2\sigma^2]$ for some $\sigma^2$.
On small scales $\Delta^2\gg 1$ and we have
$P^s(k,\nu)\approx P(k)\exp[-k_\parallel^2\sigma^2]$, which is one
of the common forms for the old dispersion models \cite{Pea92,Par94}.
It is interesting to note that the dispersion model approximation may
explicitly break translational invariance (depending upon how $\sigma$
is computed) though this is preserved in the full cumulant form.

\subsubsection{Configuration space}
\label{sec:GSM_config_space}

\begin{figure}
\begin{center}
\resizebox{\columnwidth}{!}{\includegraphics{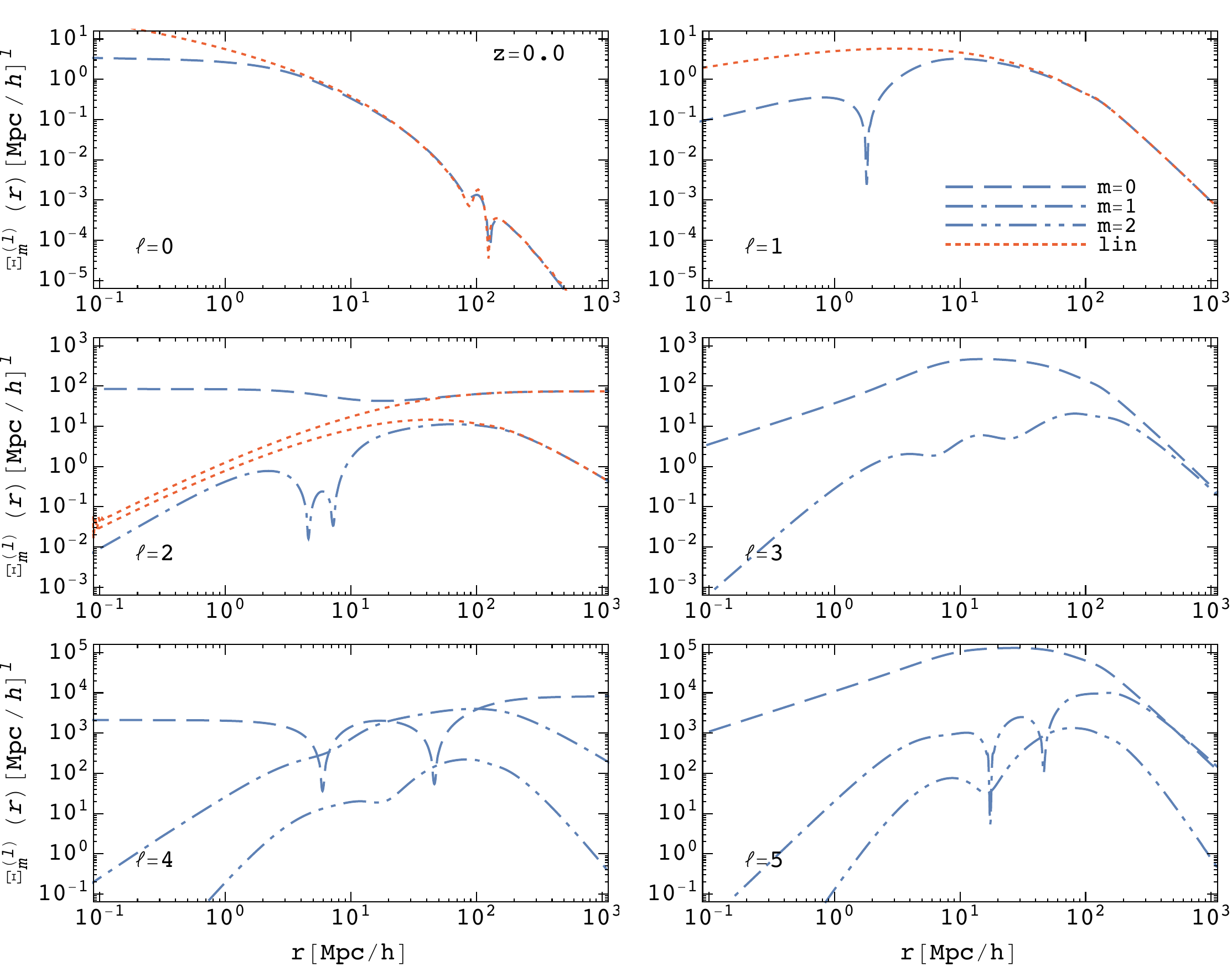}}
\end{center}
\vspace*{-5mm}
\caption{Velocity moments, $\Xi^{(\ell)}_m$ in configuration space as a function of $r$, for $\Lambda$CDM at $z=0$.
Moments are computed in Zeldovich approximation and separated according to
the angular dependence:
$m=0$ (long dashed lines), 
$m=1$ (dash-dotted lines), $m=2$ (dash-dot-dotted lines).
The linear theory results are also shown for $\ell=0$, $1$ and $2$ (red dotted lines) and are each proportional to $P_L$.}
\label{fig:Mom_config_space}
\end{figure}

In order to obtain the streaming model in configuration space given in
Sec.~\ref{sec:SM_config_space} we first need to obtain the configuration
space ingredients, i.e.~cumulants ${\mathcal C}^{(\ell)}_{\hat n}$
and moments $\Xi^{(\ell)}_{\hat n}$ in configuration space. 
We start by Fourier transforming the moments, using the angular dependence
given in Eq.~\eqref{eq:vel_moments}.
We can write
\eq{
\Xi^{(2\ell)}_{\hat n}\lb r,\hat n\cdot \hat r \rb & =  f^{2\ell} \sum_{m = 0}^{\ell} \int \frac{d^3 k}{(2\pi)^3} ~ \widetilde \Xi^{(2\ell)}_{m}(k) \lb \hat k\cdot\hat n \rb^{2m}  e^{-i \vec k \cdot \vec r}, \non\\
\Xi^{(2\ell+1)}_{\hat n}\lb r,\hat n\cdot \hat r \rb & = i  f^{2\ell+1} \sum_{m = 0}^{\ell} \int \frac{d^3 k}{(2\pi)^3} ~ \widetilde \Xi^{(2\ell+1)}_{m}(k) \lb \hat k\cdot\hat n \rb^{2m+1}  e^{-i \vec k \cdot \vec r}. 
}
Using Legendre tensors defined by 
\eeq{
\left\{ \hat r_{i_1}\ldots \hat r_{i_L} \right\}^{L}_{\ell}
= (2\ell +1) \int \frac{d\Omega_k}{4\pi}~ \hat k_{i_1}\ldots \hat k_{i_{L}} \mathcal P_{\ell} \lb \hat k \cdot \hat r \rb, 
}
we have
\eeq{
\int \frac{d\Omega_k}{4\pi}~ \hat k_{i_1}\ldots \hat k_{i_{L}} e^{-i \vec k \cdot \vec r} 
= \sum_{\ell=0}^\infty i^\ell j_\ell (k r) \left\{ \hat r_{i_1}\ldots \hat r_{i_L} \right\}^{L}_{\ell}.
}
This leads to the useful angular integrals 
\eq{
\int \frac{d \Omega_k}{4\pi} ~ \lb \hat k \cdot \hat n \rb^{2L}  e^{-i \vec k \cdot \vec r} 
&=  \sum_{\ell=0}^{L} c^{{\rm e,}(L)}_\ell \mathcal P_{2\ell} (\nu_r)  j_{2\ell} (k r), \non\\
\int \frac{d \Omega_k}{4\pi} ~ \lb \hat k \cdot \hat n \rb^{2L+1}  e^{-i \vec k \cdot \vec r}
&=- i \sum_{\ell=0}^{L} c^{{\rm o,}(L)}_\ell \mathcal P_{2\ell+1} (\nu_r)  j_{2\ell+1} (k r),
\label{eq:angle_integrals}
}
where the coefficients $c^{{\rm e,}(L)}_\ell$ and $c^{{\rm o,}(L)}_\ell$ are
given in Eq.~\eqref{eq:c_coefs}.
Finally we have for the configuration space velocity moments 
\eq{
\Xi^{(2\ell)}_{\hat n}\lb r, \nu_r \rb
& =  f^{2\ell} \sum_{q=0}^{\ell}  \mathcal P_{2q} (\nu_r) \int \frac{k^2 d k}{2\pi^2} ~  \lb \sum_{m = 0}^{\ell} c^{{\rm e,}(m)}_q \widetilde \Xi^{(2\ell)}_{m}(k) \rb j_{2q} (k r), \non\\
\Xi^{(2\ell+1)}_{\hat n}\lb r, \nu_r \rb 
& = f^{2\ell+1}  \sum_{q=0}^{\ell}   \mathcal P_{2q+1} (\nu_r) 
 \int \frac{k^2 d k}{2\pi^2} ~ \lb \sum_{m = 0}^{\ell} c^{{\rm o,}(m)}_q \widetilde \Xi^{(2\ell+1)}_{m}(k) \rb j_{2q+1} (k r).
\label{eq:config_space_vel_moms}
}
In analogy to what we show in Fig.~\ref{fig:Mom_DirectLagrangian}, angle multipoles of these velocity moments, $\Xi^{(\ell)}_{q}$,\footnote{Note 
that we have also added the zero-lag contributions to the real space multiple moments. In Fourier space these are proportional to the zero $k$-mode.} 
are shown in Fig.~\ref{fig:Mom_config_space}.
Velocity moments, $\Xi^{(\ell)}_{\hat n}$, can of course be represented also in terms of power series in
configuration space angles $\nu_r = \hat{n}\cdot\hat{r}$ by expanding
the Legendre polynomials $\mathcal{P}_{\ell}$.
The configuration space cumulants, ${\mathcal C}^{(\ell)}_{\hat n}$,
are then given in terms of the moments, $\Xi^{(2\ell)}_{\hat n}$, by
expressions analogous to Eq.~\eqref{eq:bell_expansion}. 

The redshift space power spectrum in terms of the configuration space
streaming model is given by Eq.~\eqref{eq:PS_stream_mom}.
The redshift space distortion effects are contained in the kernel
$\mathcal{K}$, and given that $\xi(r)$ is isotropic the angular dependent
kernel can be decomposed as
\eq{
\ln\mathcal{K}(\vec{k},\vec{r}) = \sum_{\ell=1}^\infty \frac{i^\ell}{\ell!} (k \nu )^\ell \mathcal C^{(\ell)}_{\hat n} (r, \mu)
=  \sum_{n=0}^\infty X_n  \frac{\mu^n}{n!} = X_0 + \ln \Bigg[ \sum_{n=0}^\infty B_n\lb X_1, \ldots, X_n \rb \frac{\mu^n}{n!} \Bigg] 
}
where $B_n(X_1,\ldots,X_n)$ is the complete, exponential Bell polynomial and
we have introduced angle series coefficients 
\eq{
X_{2m}(k,r,\nu)&= (2m)! \sum_{\ell = 1}^\infty \frac{(-1)^\ell}{(2\ell)!} (\nu f k)^{2\ell} \mathcal C^{(2\ell)}_{m} (r), \non\\
X_{2m+1}(k,r,\nu)&= i (2m+1)! \sum_{\ell = 0}^\infty \frac{(-1)^\ell}{(2\ell+1)!} (\nu f k)^{2\ell+1} \mathcal C^{(2\ell+1)}_{m} (r) .
\label{eq:X_coefs}
}
In analogy to the velocity moment decomposition in
Eq.~\eqref{eq:vel_moments} we can decompose the cumulants as
\eq{
\mathcal C^{(2\ell)}_{\hat n} (r, \mu) &= f^{2\ell} \sum_{m=0}^\ell \mathcal C^{(2\ell)}_{m} (r) \mu^{2m}, \non\\
\mathcal C^{(2\ell+1)}_{\hat n} (r, \mu) &= f^{2\ell+1} \sum_{m=0}^\ell \mathcal C^{(2\ell+1)}_{m} (r) \mu^{2m+1}. \non
}
For the angular integration in the power spectrum expression,
Eq.~\eqref{eq:PS_stream_mom}, we can now write
\eq{
\int d\Omega_r \ e^{i \vec k \cdot \vec r} \mathcal{K}(\vec{k},\vec{r})
&= e^{X_0}\sum_{L=0}^\infty \frac{1}{L!} B_L\lb X_1, \ldots, X_L \rb \int d\Omega_r \ \mu^L e^{i \vec k \cdot \vec r} \non\\
&= 4\pi e^{X_0} \sum_{\ell=0}^{\infty} \mathcal P_{\ell} (\nu) \Bigg[ \sum_{L=0}^\infty \frac{c^{(L)}_\ell}{L!} B_L \lb X_1, \ldots, X_L \rb \Bigg]  j_{\ell} (k r) 
\label{eq:SM_angle_int}
}
where the coefficient, $c^{(L)}_\ell$, is given by combining the odd and
even coefficients given in Eq.~\eqref{eq:c_coefs}.
Explicitly we can write the values as
\eq{
c^{(L)}_\ell = \frac{i^\ell(2\ell+1)L!}{2^{(L-\ell)/2}(\tfrac{1}{2}(L-\ell))!(L+\ell+1)!!} 
\begin{cases} 
1, ~~{\rm if} ~~ L+\ell ~~ {\rm even} \\
0, ~~{\rm otherwise}
\end{cases}.
}
Finally for the configuration space streaming model power spectrum we have
\eq{
\label{eq:cofig_space_stream_PS}
(2\pi)^3 \df^K(\vec k) + P_s(\vec k) & =   \int d^3r\ e^{i \vec k \cdot \vec r} \big[ 1 + \xi(r) \big]  \mathcal{K}(\vec{k},\vec{r}) \\
& =  4\pi \sum_{\ell=0}^{\infty} \mathcal P_{\ell} (\nu)  \int r^2 dr\ \big[ 1 + \xi(r) \big] e^{X_0}  \Bigg[ \sum_{L=\ell}^\infty \frac{c^{(L)}_\ell}{L!} B_L \lb X_1, \ldots, X_L \rb \Bigg]  j_{\ell} (k r). \non 
}
An alternative strategy for obtaining the power spectrum, taking into
account just the Gaussian parts (i.e.~truncation at the second cumulant),
would be to directly transform the redshift space correlation function.
This approach has been discussed in context of the linear theory results
for density peaks in Ref.~\cite{Desj10}.
Extending this approach beyond the second cumulant depends on efficient
evaluation of the streaming model correlation function, which we discuss
in Sec.~\ref{sec:streaming_config_space}.

It is instructive to take the linear theory limit of the expression above.
We start by writing the linear theory, configuration space velocity
moments given in Eq.~\eqref{eq:config_space_vel_moms}.
Only first three velocity moments are non-vanishing.
Moreover, given that $\widetilde \Xi^{(2)}_{0}$ has no contribution in
linear theory and that $\Xi^{(0)}_{0} = P_L$,
$\Xi^{(1)}_{0} = - 2 P_L/k$, $\Xi^{(2)}_{1} = - 2 P_L/k^2$, it follows 
\eq{
\Xi^{(0)}_{\hat n}\lb r, \nu \rb  & =  \int \frac{k^2 d k}{2\pi^2} ~ P_L(k) j_{0} (k r), \\
\Xi^{(1)}_{\hat n}\lb r, \nu \rb  & = - 2 f  \nu \int \frac{k d k}{2\pi^2} ~ P_L(k) j_1(k r), \non\\
\Xi^{(2)}_{\hat n}\lb r, \nu \rb  & =  - f^2   \frac{2}{3}  \int \frac{d k}{2\pi^2} ~  P_L(k) j_{0} (k r) - f^2 \frac{2}{3}  \lb 1 - 3\nu^2 \rb \int \frac{d k}{2\pi^2} ~  P_L(k) j_{2} (k r). \non
}
The configuration space velocity cumulants coincide with the moments in
linear theory
($\mathcal C^{(0)}_{\hat n} = \Xi^{(0)}_{\hat n}$,
 $\mathcal C^{(1)}_{\hat n}\approx \Xi^{(1)}_{\hat n}$ and
 $\mathcal C^{(2)}_{\hat n}\approx \Xi^{(2)}_{\hat n}$)
so using Eq.~\eqref{eq:X_coefs} we have
\eeq{
  X_{0}= - \frac{1}{2} (\nu f k)^{2} \Xi^{(2)}_{0} (r) \quad , \quad
  X_{1}= i \nu f k \Xi^{(1)}_{0} (r) \quad {\rm and} \quad
  X_{2}= - (\nu f k)^2  \Xi^{(2)}_{1} (r) \quad .
}
Collecting all this into Eq.~\eqref{eq:cofig_space_stream_PS} and using
the explicit forms $B_0=1$, $B_1(X_1)=X_1$ and $B_2(X_1,X_2) = X^2_1+X_2$,
the linear theory power spectrum is given by
\eeq{
\frac{P_s(\vec k)}{4\pi} =  
\int r^2 dr\ \left[ \xi(r) + X_0 + \tfrac{1}{6} X_2 \right]j_{0}(kr) 
  + i \mathcal P_1 (\nu)  \int r^2 dr\  X_1 j_{1}(kr) 
  - \frac{1}{3} \mathcal P_2 (\nu)  \int r^2 dr\  X_2 j_{2}(kr), \non
}
which upon using the integral representation of the Dirac delta function
\eeq{
 \int r^2 dr\  j_{n}(k'r)j_{n}(kr) = \frac{\pi}{2k^2} \df^K\lb k' - k \rb,
}
immediately gives the Kaiser result,
$P_s(\vec k)=  \lb 1 + f \nu^2 \rb ^2 P_L(k)$.

Finally, let us note that the computation procedure for the configuration
space streaming model described in this section is not the only possibility.
One alternative is to use methods already presented when computing the direct
Lagrangian approach in Sec.~\ref{sec:direct_lag}.
In particular the integrals given near Eq.~\eqref{eq:int_mu_n} can be applied
to solve the angular integral in Eq.~\eqref{eq:SM_angle_int}.
We have tried this and checked that the obtained results are consistent.
We find that this leads to a somewhat more challenging numerical problem
with slower convergence and thus we did not pursue this method further. 

\subsection{Comparison of different Fourier space methods}
\label{sec:Fourier_comparison}

\begin{figure}
\begin{center}
\resizebox{\columnwidth}{!}{\includegraphics{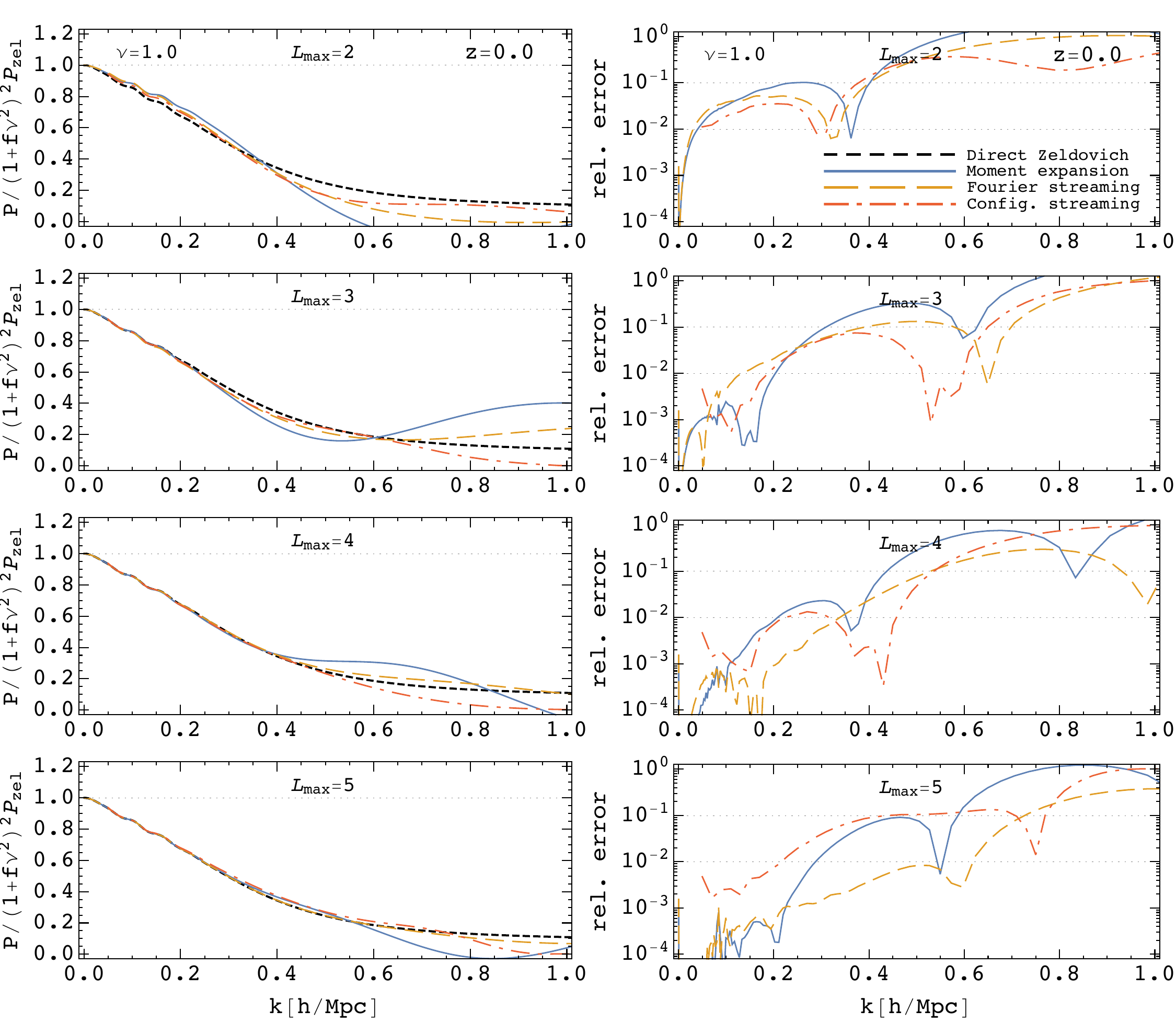}}
\end{center}
\vspace*{-5mm}
\caption{Comparison of different models in Fourier space, i.e.~for the
line-of-sight power spectrum.  The left panels show the ratio of each model
to the `Kaiser-Zeldovich' model, to reduce the dynamic range for plotting
purposes.  The dashed black line indicates the full Zeldovich calculation,
solid blue the moment expansion, dashed orange the Fourier-space streaming
model and dot-dashed red the configuration-space streaming model.
The right panels show the relative error, compared to the full Zeldovich
calculation.  The rows show how the convergence is improved by including
more terms in the expansion, with the first row being the common `Gaussian'
approximation for the streaming models. 
See text for further discussion.}
\label{fig:Fourier_comparison_nu10}
\end{figure}

The line-of-sight power spectra for the different methods introduced in
the previous sections are compared in Fig.~\ref{fig:Fourier_comparison_nu10}.
The left hand panels show the ratio of each expression to the
`Kaiser-Zeldovich' model, $P=(1+f\nu^2)P_{\rm Zel}$, to reduce the dynamic
range for plotting purposes.  The right hand panels show the relative error
in each expansion compared to the exact result (our direct Zeldovich
expansion, shown as the dashed black line in the left panels).
The different rows show how the convergence is improved by including more
terms in the expansion with the first row being the common `Gaussian'
approximation for the streaming models.  With our new formalism we are able
to extend both the configuration-space and Fourier-space models beyond the
$L_{\rm max}=2$/Gaussian approximation\footnote{In general, with $L_{\rm max}$ we label the
highest cumulant or moment that is included in the truncation in order to study the performance of each expansion.} to assess the convergence of the
cumulant expansion(s).

All models approach the correct answer, and linear theory, as $k\to 0$.
On `linear scales', $k<0.1\,h\,{\rm Mpc}^{-1}$, the models all perform at
the percent level.  However they deviate rapidly when moving to smaller
scales.  In general it appears the cumulant expansions out-perform the
moment expansions at higher $k$, but the absolute performance of all of
the models is relatively poor on these scales.
All of the models show improvement when going beyond $L_{\rm max}=2$.
While at low $k$ the moment expansion does as well as the cumulant expansion,
the latter performs better at intermediate and high $k$ and converges
slightly more quickly with increasing $L_{\rm max}$.

\begin{figure}
\begin{center}
\resizebox{\columnwidth}{!}{\includegraphics{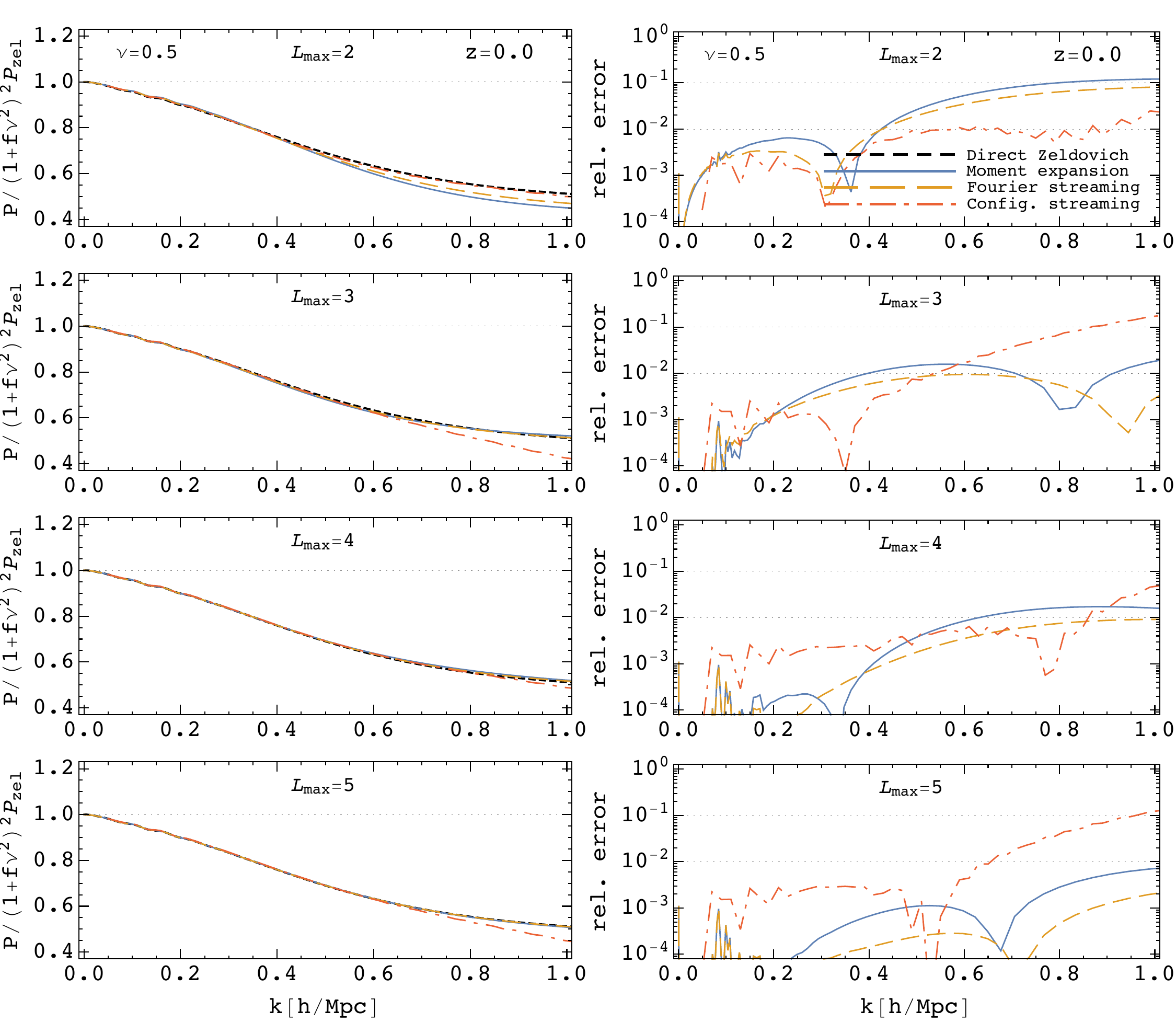}}
\end{center}
\vspace*{-5mm}
\caption{As for Fig.~\ref{fig:Fourier_comparison_nu10}, but for $\nu=0.5$,
i.e.~a mode with $k_z=0.5\,k$.  All of the models perform considerably better
as $\nu$ is decreased, eventually approaching the exact result as $\nu\to 0$.
See text for further discussion.}
\label{fig:Fourier_comparison_nu05}
\end{figure}

The line-of-sight power spectrum represents the worst-case scenario for
the models, which all reproduce the transverse power spectrum exactly by
construction.  Fig.~\ref{fig:Fourier_comparison_nu05} is like
Fig.~\ref{fig:Fourier_comparison_nu10}, except for $\nu=0.5$.  Note the
relative error is much smaller in this case for all of the models.
The moment expansion and the streaming models show sub-percent agreement
with the full Zeldovich calculation well into the non-linear regime.
The convergence of the expansions with increasing $L_{\rm max}$ is very
rapid, with the streaming model showing fraction of a percent performance
for all plotted scales by $L_{\rm max}=5$.

\begin{figure}
\begin{center}
\resizebox{\columnwidth}{!}{\includegraphics{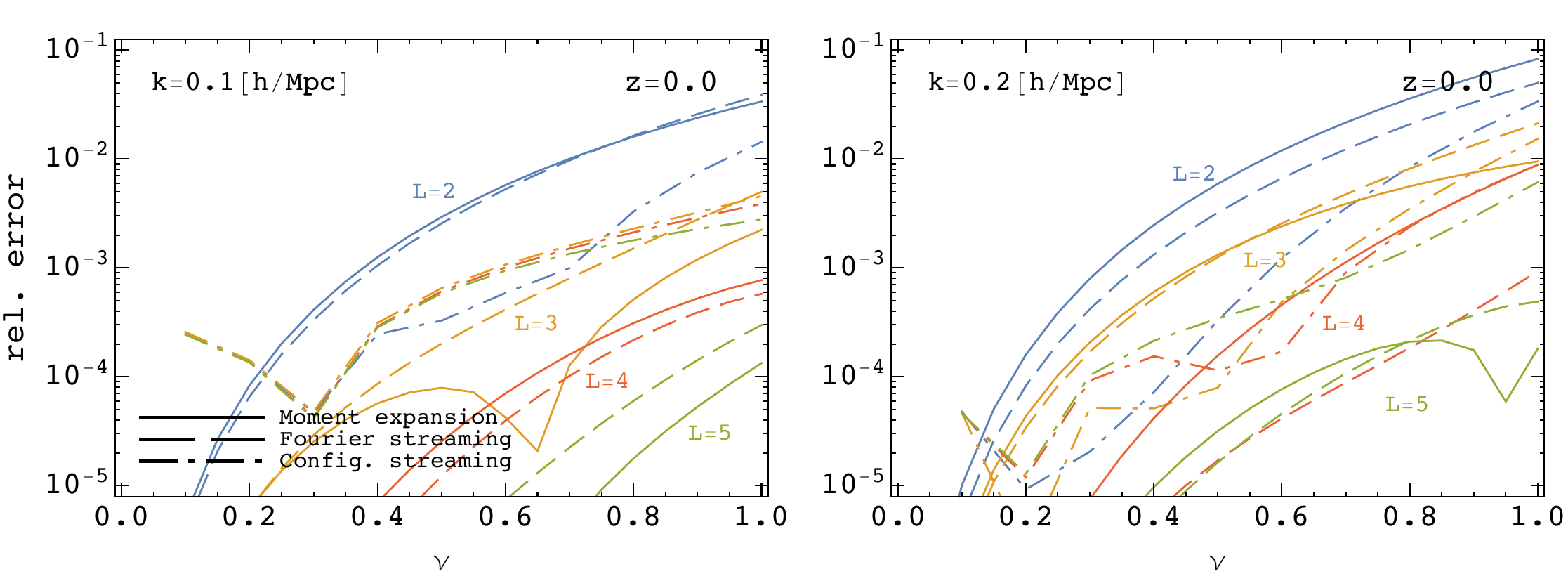}}
\end{center}
\vspace*{-5mm}
\caption{The error in the redshift-space power spectrum on quasi-linear scales
(left: $k=0.1\,h\,{\rm Mpc}^{-1}$ and right: $k=0.2\,h\,{\rm Mpc}^{-1}$) as a
function of $\nu=k_z/k$ for our different models (solid: moments,
dashed: Fourier-space streaming and dot-dashed: configuration-space streaming).
We see that the performance of all of the models is a strong function of $\nu$.
All models have zero error as $\nu\to 0$ by construction, and every model
performs significantly worse for $\nu\approx 1$ than for intermediate values
of $\nu$.  The configuration-space streaming model does well for small
$L$, but the Fourier-space model improves most rapidly with increasing $L$.
See text for further discussion and implications.}
\label{fig:Fourier_comparison_vs_nu}
\end{figure}

For all of the models we find that the errors are a strong function of $\nu$.
Fig.~\ref{fig:Fourier_comparison_vs_nu} shows the same models as
Figs.~\ref{fig:Fourier_comparison_nu10} and \ref{fig:Fourier_comparison_nu05},
but now at $k=0.1$ and $k=0.2\,h\,{\rm Mpc}^{-1}$
(both quasi-linear scales at $z=0$) as a function of $\nu$.
The steep dependence of the error on $\nu$ is not too surprising, but could
have consequences for comparison with observation.
Observational probes which are restricted to high $\nu$, such as $21\,$cm
interferometry in the presence of a foreground wedge \cite{Kovetz17},
present a particular challenge for perturbation theory approaches.
Similarly, methods for dealing with observational systematics which require
modeling of high multipole moments (e.g.~Ref.~\cite{Han17}) place stringent
demands upon the theory.
Conversely observational methods which downweight the line-of-sight modes
(e.g.~the $\hat{\xi}$ of Ref.~\cite{Rei14}) could reduce the systematic
error in comparison with theory while only modestly increasing the statistical
error.  An alternative, although similar, approach would be to include a
`theoretical error' \cite{Baldauf16} which is a steep function of $\nu$ when
performing fits.
We shall defer consideration of these approaches to future work.

\begin{figure}
\begin{center}
\resizebox{\columnwidth}{!}{\includegraphics{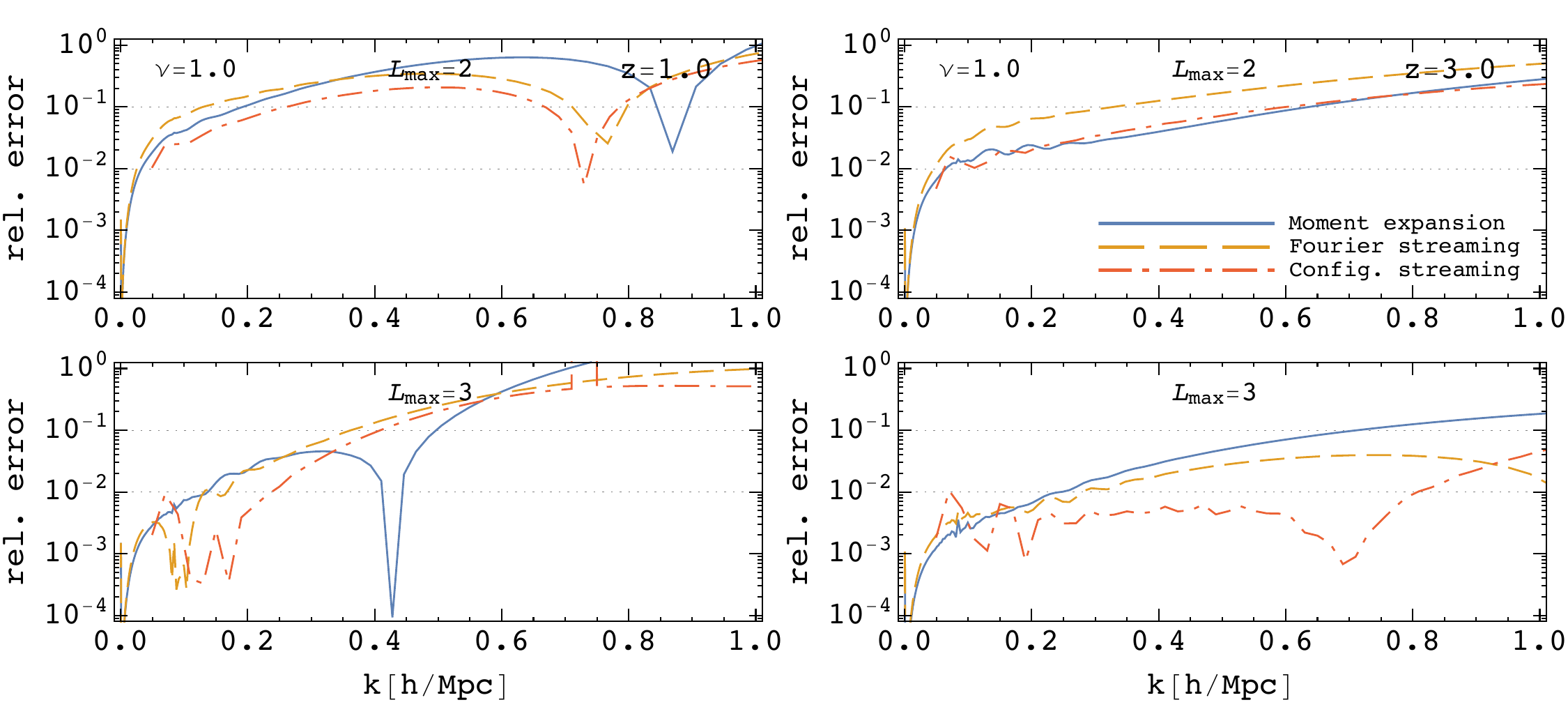}}
\end{center}
\vspace*{-5mm}
\caption{The error in our models at higher redshift.  The lines are as in
Fig.~\ref{fig:Fourier_comparison_nu10}: solid blue for the moment expansion,
dashed orange for the Fourier-space streaming model and dot-dashed red for
the configuration-space streaming model.  Plotted is the relative error at
$\nu=1$ (i.e.~the line of sight), compared to the full Zeldovich calculation,
vs.~$k$.  The rows show $L_{\rm max}=2$ and $3$ to indicate how the convergence
is improved by including more terms in the expansion (with the first row being
the common `Gaussian' approximation for the streaming models).
The left panels show $z=1$ and the right panels $z=3$.
See text for further discussion.}
\label{fig:Fourier_comparison_vs_z}
\end{figure}

\begin{figure}
\begin{center}
\resizebox{\columnwidth}{!}{\includegraphics{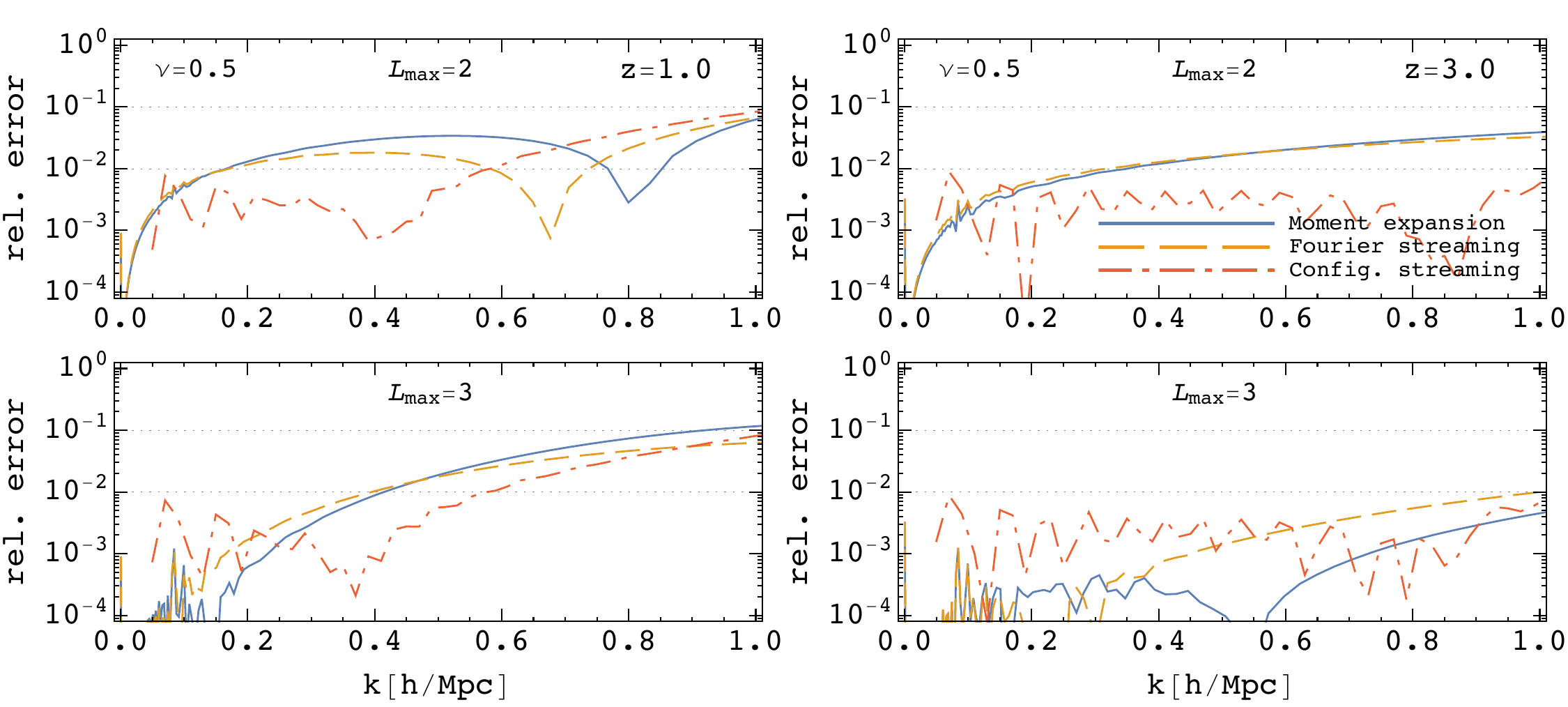}}
\end{center}
\vspace*{-5mm}
\caption{Same as figure \ref{fig:Fourier_comparison_vs_z} for $\nu=k_z/k=0.5$.
All of the models perform better for lower $\nu$ at all redshifts.
See text for further discussion.}
\label{fig:Fourier_comparison_vs_z05}
\end{figure}

Finally, each of the models converges more quickly to the full result at
high redshift than at low redshift.  The relevant expansion parameter in
this case is $fD$ rather than simply $D$ (the linear growth rate), and this
actually peaks near $z\simeq 0.5$ for currently favored cosmological models.
Figs.~\ref{fig:Fourier_comparison_vs_z} and \ref{fig:Fourier_comparison_vs_z05}
show the relative error for each of our models at $\nu=1$ and $\nu=0.5$ for
$z=1$ and $z=3$.  Note by $z=3$ the best models are performing much better
than a percent at $\nu=0.5$ over the entire range of scales shown, with the
configuration-space streaming model achieving this even for $L_{\rm max}=2$.
Even at $z=3$, however, the models have super-percent errors for $\nu=1$.

\section{Redshift-space distortions in configuration space}
\label{sec:config_space}

In this section we will consider the performance of each of our models in
configuration space, i.e.~for the redshift-space correlation function. 
We look at the convergence of the moment expansion approach as well
as the two streaming models, compared to the full Zeldovich result.
In particular we are able to go beyond leading terms in Fourier streaming
model to look at the convergence of the cumulant expansion.
In the case of the configuration space streaming model we shall stick
to the Gaussian case, commenting later on what methods are required to
efficiently move beyond this.
As well as in the previous section, most of the tool developed to compute
the cumulant and moment expansion are independent of the Zeldovich dynamics
and are fully applicable in the full nonlinear case as well as in case of
biased tracers (see \S\ref{sec:bias}).

\subsection{Direct integration in configuration space}

First we focus on obtaining directly the Zeldovich result for the RSD correlation function,
given that it will serve as our benchmark result that we contrast to the other expansions.
Obtaining the direct expression for Zeldovich RSD correlation function
is relatively simple, since the two dimensional integration can be easily
performed numerically.  Given that the integral for $P(k)$ is Gaussian we have
\eq{
  \xi_s(\vec r)  &= \int \frac{d^3k}{(2\pi)^3}\ e^{- i\vk \cdot \vec{r}}
  \int d^3q ~ e^{i\vk \cdot \vec{q}} e^{-\frac{1}{2} k_i k_j A^s_{ij}}
  = \int \frac{d^3q}{(2\pi)^{3/2} \sqrt{{\rm det} A^s } }
  \ e^{\frac{1}{2}(\vec q-\vec r)_i(\vec q-\vec r)_j [A^s]^{-1}_{ij}},
}
where again $A^s_{ij} = R_{i\ell} R_{jm} A_{\ell m}$.
The integrand in this case is not oscillatory and can be directly integrated
in $\vec q$ variable, e.g.~Ref.~\cite{Whi14}.
It is interesting to compare the structure of the integrand to its Fourier
counterpart, $k_i k_j R_{i\ell} R_{jm} A_{\ell m}$, given in
Eq.~\eqref{eq:kkRRA}. 
First we note that if $\hat k\cdot\hat n \to 0$ one obtains the real space
limit $k_i k_j R_{i\ell} R_{jm} A_{\ell m} =k^2 \left(X + \mu^2 Y \right)$,
also if $f \to 0$.  This does not happen in configuration space and we do not
obtain the real space result for any value of $\hat r\cdot\hat n$.
The limits $\hat r\cdot\hat n \to 0$ and $f\to 0$ are not the same. 
This is already clear from the form of the integral measure,
$({\rm det} A^s) = (1+f)^2 X^2 (X+Y)$, 
where no matter the value of $\hat r\cdot\hat n$, the $(1+f)^2$ factor is
present.  A similar thing happens for the term in exponent,
$(\vec q - \vec r)[A^s]^{-1}(\vec q - \vec r)$.
For this reason the error on the redshift-space correlation function does
not have to go to zero as $\nu\to 0$.

Next we turn to looking at the performance of the three models;  
moment expansion, Fourier and configuration space streaming models 
in configuration space, and study the convergence focusing on Zeldovich dynamics
as a guideline for more general cases. 

\subsection{Moment expansion in configuration space}

In \S\ref{sec:mom_exp} we explored the moment expansion in Fourier space.
To obtain results for the redshift-space correlation function corresponding
to this expansion we can simply Fourier transform it.
Given that the redshift-space correlation function is a Fourier transform
of the redshift-space power spectrum, and the latter is given a sum of
moments, the correlation function will also be given as a sum of Fourier
transforms of the individual terms.
Starting from Eq.~\eqref{eq:PS_momentsum} we have
\eq{
 \xi_s(\vec r) &= \sum_{\ell=0}^\infty \frac{i^\ell}{\ell!} \int \frac{d^3 k}{(2\pi)^3} \lb k \nu \rb^\ell \widetilde \Xi^{(\ell)}_{\hat n}(\vec k) e^{-i \vec k . \vec r} \quad .
 }
Using the expansion in the powers of $\nu$, given in Eq.~\eqref{eq:vel_moments}, we have 
\eq{
\xi_s(\vec r) 
&=   \sum_{n=0}^{\infty} \mathcal P_{2n} (\nu_r) \int \frac{k^2 d k}{2\pi^2} \Bigg[ 
 \sum_{\ell=0}^\infty \frac{(-1)^\ell}{(2\ell)!}  (f k)^{2\ell} \sum_{m = 0}^{\ell} \bigg( c^{{\rm e,}(\ell+m)}_n  \widetilde \Xi^{(2\ell)}_{m}(k) \non\\
&\hspace{6.5cm}  - \frac{fk}{2\ell+1} c^{{\rm e,}(\ell+m+1)}_n \widetilde \Xi^{(2\ell+1)}_{m}(k) \bigg) \Bigg]  j_{2n} (k r).
} 
We shall consider the convergence of this expression below.  Before that we
consider the linear (Kaiser) theory limit of this result.
In linear theory only $\widetilde \Xi^{(0)}$, $\widetilde \Xi^{(1)}$ and
$\widetilde \Xi^{(2)}$ contribute and we have
\eq{
\xi_s(\vec r) &=
 \sum_{n=0}^{\infty} \mathcal P_{2n} (\nu_r) \Bigg[
  c^{{\rm e,}(0)}_n \int \frac{d k}{2\pi^2} k^{2} \widetilde \Xi^{(0)}_{0}(k) j_{2n} (k r) 
- f \int \frac{dk}{2\pi^2} k^3 c^{{\rm e,}(1)}_n \widetilde \Xi^{(1)}_{0}(k) j_{2n} (k r) \non\\
&\hspace{4cm} - \frac{1}{2}  f^2 \int \frac{d k}{2\pi^2} k^4 \lb c^{{\rm e,}(1)}_n \widetilde \Xi^{(2)}_0(k) + c^{{\rm e,}(2)}_n \widetilde \Xi^{(2)}_1(k) \rb j_{2n} (k r) \Bigg]
\label{eq:xi_multipoles}
}
Given the $c^{{\rm e,}\ell}_n$ coefficients in Eq.~\eqref{eq:c_coefs},
and the fact that in linear theory $\Xi^{(2)}_0=0$, we have for the
correlation function multipoles 
\eq{
\xi^{(2\ell)}_s(r) &= \frac{4\ell +1 }{2} \int_{-1}^1 d\nu_r~ \xi_s(\vec r) \mathcal P_{2\ell} (\nu_r) \non\\
&= \int \frac{d k}{2\pi^2} k^{2} \lb 
c^{(0)}_\ell \widetilde \Xi^{(0)}_{0}(k) - f c^{(1)}_\ell  k \widetilde \Xi^{(1)}_{0}(k) -  \frac{1}{2} f^2 c^{(2)}_\ell k^2 \widetilde \Xi^{(2)}_1(k) \rb j_{2\ell} (k r).
\label{eq:xi_multipoles_kaiser}
}
Using again the linear theory expressions $\widetilde \Xi^{(0)}_{0} = P_L$,
$\widetilde \Xi^{(1)}_{0} = - 2 P_L/k$ and $\widetilde\Xi^{(2)}_{1}=-2P_L/k^2$
it directly follows
\eq{
\xi^{(0)}_s(r) &= \lb1 + \frac{2}{3}f  + \frac{1}{5} f^2 \rb \int \frac{k^2 d k}{2\pi^2}~P_L(k) j_{0} (k r), \non\\
\xi^{(2)}_s(r) &= - \lb \frac{4}{3}f  + \frac{4}{7} f^2 \rb \int \frac{k^2 d k}{2\pi^2}~P_L(k) j_{2} (k r), \non\\
\xi^{(4)}_s(r) &= \frac{8}{35} f^2 \int \frac{k^2 d k}{2\pi^2}~P_L(k) j_{4} (k r).
\label{eq:Kaiser_multipoles_config}
}
These results are, of course, in agreement with the well-known result that
the Fourier and configuration space multipoles are simply linked by spherical
Bessel transforms \cite{Ham92}
\eeq{
\xi^{(\ell)}_s(r) = i^\ell \int \frac{k^2 d k}{2\pi^2} ~P^{(\ell)}_s(k) j_{\ell} (k r) \quad , \quad
P^{(\ell)}_s(k) = \frac{2\ell +1 }{2} \int_{-1}^1 d\nu~ P_s(\vec k) \mathcal P_{\ell} (\nu).
}

\subsection{Fourier streaming models in configuration space}

Next we move to the new Fourier version of the streaming model
(\S\ref{subsec:FS_model}). 
One of the simplifying features of this representation is that it can be
transformed directly into configuration space (in contrast to the
configuration-space streaming model where the translations are more complex).
We present these results in this subsection. 
By Fourier transforming Eq.~\eqref{eq:Fourier_streaming_power_spectrum}
we have
\eq{
\xi_s(\vec r) &= \int \frac{d^3k}{(2\pi)^3} P^{s}(k, \nu) e^{ - i \vec k . \vec r} \non\\
&= \int \frac{d^3k}{(2\pi)^3}~ e^{ - i \vec k . \vec r} \frac{2\pi^2}{k^3} \lb \left[1+\Delta^2(k)\right]
   \exp \left\{ \sum_{\ell=1}^\infty \frac{i^\ell}{\ell!} (\nu k)^\ell \widetilde{\mathcal C}^{(\ell)}_{\hat n} (k, \nu) \right\} - 1 \rb.
}
In order to proceed we need to compute the angular integral and thus we first
collect the $\nu$ dependence in the exponent
\eq{
  \sum_{\ell=1}^\infty \frac{i^\ell}{\ell!} (k \nu )^\ell
  \mathcal C^{(\ell)}_{\hat n} (k, \nu)
  &= \sum_{n=1}^\infty \nu^{2n} \sum_{\ell = 1}^\infty
  \frac{(-1)^\ell}{(2\ell)!} (fk)^{2\ell}
  \lb \widetilde{ \mathcal C}^{(2\ell)}_{n-\ell} (k)
  +  \frac{2\ell}{f k} \widetilde{ \mathcal C}^{(2\ell-1)}_{n-\ell} (k) \rb
  \nonumber \\
  &= \sum_{m=1}^\infty Y_{m}(k) \frac{\nu^{m}}{m!},
}
where we have defined angle power coefficients, $Y_m$, in analogy to
Eq.~\eqref{eq:X_coefs}:
\eeq{
  Y_{m}(k) = \lb 1+(-1)^m \rb \frac{m!}{2} \sum_{\ell = 1}^\infty \frac{(-1)^\ell}{(2\ell)!} (fk)^{2\ell} \lb \widetilde{ \mathcal C}^{(2\ell)}_{m/2-\ell} (k) +  \frac{2\ell}{f k} \widetilde{ \mathcal C}^{(2\ell-1)}_{m/2-\ell} (k) \rb.
\label{eq:Y_func}
}
This allows us to do the angular part of the $\vec k$ integral.
Using Eq.~\eqref{eq:angle_integrals}
\eq{
\int \frac{d\Omega_k}{4\pi} \ e^{-i \vec k \cdot \vec r} \exp\Bigg[ \sum_{\ell=1}^\infty \frac{i^\ell}{\ell!} (k \nu )^\ell \widetilde{\mathcal C}^{(\ell)}_{\hat n} (k, \nu) \Bigg]
&= \sum_{L=0}^\infty \frac{1}{L!} B_L\lb Y_1, \ldots, Y_L \rb  \sum_{\ell=0}^{\infty} (-1)^\ell c^{(L)}_\ell \mathcal P_{\ell} (\nu_r)  j_{\ell} (k r)\non\\
&= \sum_{\ell=0}^{\infty} (-1)^\ell \mathcal P_{\ell} (\nu_r) \Bigg[ \sum_{L=0}^\infty \frac{c^{(L)}_\ell}{L!} B_L \lb Y_1, \ldots, Y_L \rb \Bigg]  j_{\ell} (k r), \non
}
where $B_L$ are again the Bell polynomials of order $L$.
Note that the coefficients $c^{(L)}_\ell$ are identically zero if $\ell>L$,
which limits the number of terms, i.e.~the number of multipole moments,
in the first sum above. 
This is of direct practical use since the truncation of the sum has to be
enforced only in the $L$ variable.  
For the numerical implementations we consider below we find that truncation
at $L\leq10$ gives well converged results on scales above $\sim1\,h^{-1}$Mpc.

The correlation function is thus given in terms of the Legendre polynomials
that describe the angular dependence and scale dependent terms that are
obtained as spherical Bessel transforms of the Fourier space quantities
\eq{
\xi_s(\vec r) = \int \frac{dk}{k}~&\Delta^2(k)  j_{0} (k r) \\
& + \sum_{\ell=1}^{\infty} (-1)^\ell \mathcal P_{\ell} (\nu_r) \int \frac{dk}{k}~ \left(1+\Delta^2 (k) \right) \Bigg[ \sum_{L=1}^\infty \frac{c^{(L)}_\ell}{L!} B_L \lb Y_1, \ldots, Y_L \rb \Bigg]  j_{\ell} (k r). \non
}
Given the multipoles defined by Eq.~\eqref{eq:xi_multipoles},
and that only even $\ell$ survive, we finally have
\eq{
\xi^{(2\ell)}_s(r) = \int \frac{dk}{k}~ \Bigg( \df^K_{\ell,0} &\Delta^2(k)
+ \left(1+\Delta^2(k)\right) \sum_{L=1}^\infty \frac{c^{{\rm e,}(L)}_{\ell}}{(2L)!} B_{2L} \lb 0, Y_2, 0, \ldots, Y_{2L} \rb \Bigg) j_{2\ell} (k r),
}
where again the coefficients $c^{{\rm e,}(L)}_{\ell}$ are given by
Eq.~\eqref{eq:c_coefs}, $B_L$ are ordinary the Bell polynomials and the scale
dependent functions, $Y_{2n}$, are given in Eq.~\eqref{eq:Y_func} above.

As earlier it is instructive to see how the linear theory result emerges from
the solution above.  First we note that only two of the $Y_{2n}$ terms survive,
explicitly $Y_{2}(k)= -2fk\widetilde{ \mathcal C}^{(1)}_{0}(k)=4f\Delta_L^2(k)$
and $Y_{4}(k)=-12(fk)^2\widetilde{ \mathcal C}^{(2)}_{1}(k)=24f^2\Delta_L^2(k)$.
Keeping only the linearised contributions in the Bell polynomials we
immediately regain the linear theory formula in the form
\eq{
\xi^{(2\ell)}_s(r) &= \lb \df^K_{\ell,0} + 2 f c^{{\rm e,}(1)}_{\ell} + f^2  c^{{\rm e,}(2)}_{\ell} \rb  \int \frac{dk}{k}~\Delta_L^2  j_{2\ell} (k r).
}
Using the values for the $c^{{\rm e,}(L)}_{\ell}$ coefficients, we promptly
recover the linear configuration space multipoles given in
Eq.~\eqref{eq:Kaiser_multipoles_config}.

\subsection{Streaming models in configuration space}
\label{sec:streaming_config_space}

As a last expansion of the RSD contributions in configuration space we
consider the configuration space streaming model introduced in
\S\ref{sec:SM_config_space}.
This method is the most challenging to evaluate beyond the second
cumulant (Gaussian case), given that the methods needed differ from ones
we have been developing so far.
Brute force expansion in beyond Gaussian terms is of course possible but
is labor intensive and does not guarantee fast convergence.
An alternative is to use the Edgeworth expansion
(proposed in e.g.~Ref.~\cite{Uhl15} and studied in Ref.~\cite{Bianchi16})
but we do not pursue this direction further here. 

Fourier transforming the result from Eq.~\eqref{eq:PS_stream_mom} we obtain
\eeq{
 1 + \xi_s(\vec r) =   \int d^3r'\ \big[ 1 + \xi(r') \big]  \int \frac{d^3 k}{(2\pi)^3} e^{ - i \vec k . \lb \vec r - \vec r' \rb} \mathcal{K}(\vec{k}, \vec r'),
 \label{eq:xi_SM}
}
where the configuration streaming kernel is given by Eq.~\eqref{eq:SM_Kfnc}.
Using the notation  $\hat n\cdot\vec k = k_\pp$ and
$\hat n\cdot\vec r = r_\pp$ and noting that the kernel, $\mathcal{K}$,
depends only on $k_\pp$ we can perform the integration in
$\vec k_\perp = \vec k - k_\pp \hat n$ to get
\eeq{
 \int \frac{d^3 k}{(2\pi)^3} e^{ - i \vec k . \Delta \vec r} \mathcal{K}(\vec{k}, \vec r - \Delta \vec r)
=  \df^D \lb \Delta \vec r_\perp \rb  \int \frac{d k_\pp}{2\pi} e^{ - i k_\pp \Delta r_\pp} \mathcal{K}(k_\pp, \vec r - \Delta r_\pp \hat n),
}
where the argument of the Dirac delta function is the perpendicular component $ \vec r_\perp = \vec r - r_\pp \hat n$.
The full correlation function can thus be written as
\eeq{
  1 + \xi_s(\vec r)  
  =   \int d \Delta r_\pp \ \big[ 1 + \xi\lb | \vec r - \Delta r_\pp \hat n | \rb \big] 
   \int \frac{d k_\pp}{2\pi} e^{ - i k_\pp \Delta r_\pp} \mathcal{K}(k_\pp, \vec r - \Delta r_\pp \hat n),
}
where we can use $|\vec r'| = | \vec r - \Delta r_\pp \hat n | = \sqrt{ r^2 + \Delta r_\pp^2 - 2 r_\pp \Delta r_\pp}$ 
and $\hat n\cdot \vec r' = \hat n\cdot (\vec r - \Delta r_\pp \hat n) = r_\pp -  \Delta r_\pp  =  r \nu_r - \Delta r_\pp$.
Finally, the explicit result is given as
\eq{
  1 + \xi_s(r,\nu_r)  
  &=   \int \frac{dx\,dy}{2\pi} e^{ - i x y} \ \Big[ 1 + \xi\lb  \sqrt{ r^2 + x^2 - 2 \nu_r r x} \rb \Big]  \\
  &\hspace{2cm} \times  \exp \Bigg[ \sum_{\ell=1}^\infty \frac{i^\ell}{\ell!} y^\ell \mathcal C^{(\ell)}_{\hat n} \lb \sqrt{ r^2 + x^2 - 2 \nu_r r x}, r \nu_r - x \rb \Bigg]. \non
}
Unfortunately, due to the oscillatory nature of the integrand this expression
is not particularly useful in this form. 
We will not explore the general result further here, but will focus on 
the simpler case when cumulant expansion is truncated at second cumulant.
In the case of $\ell\le 2$ the integration in $y$ can be performed analytically
giving
\eeq{
 \int \frac{dy}{2\pi} e^{ - i y \lb x - \mathcal C^{(1)}_{\hat n} \rb - \frac{1}{2} y^2 \mathcal C^{(2)}_{\hat n} } 
 = \frac{1}{\sqrt{2\pi \mathcal C^{(2)}_{\hat n}} } \exp \left\{ - \lb x - \mathcal C^{(1)}_{\hat n} \rb^2 / \lb 2 \mathcal C^{(2)}_{\hat n} \rb \right\},
}
and thus the correlation function is given by the standard result
\eeq{
  1 + \xi_s(r,\nu_r)  
  =   \int_{-\infty}^\infty dx \  \frac{ \Big[ 1 + \xi\lb x,r,\nu_r \rb \Big]}{\sqrt{2\pi \mathcal C^{(2)}_{\hat n} \lb x,r,\nu_r \rb} } \exp \left\{ - \frac{\left[ x - \mathcal C^{(1)}_{\hat n} \lb x,r,\nu_r \rb \right]^2}{ 2 \mathcal C^{(2)}_{\hat n} \lb x,r,\nu_r \rb } \right\} .
\label{eq:gauss_SM_cf}
}

Given that we have derived the linear theory results in all the earlier
cases it is natural to comment how the same result follows here.
To this end it is not very convenient to use Eq.~\eqref{eq:gauss_SM_cf}
directly, but instead we start from Eq.~\eqref{eq:xi_SM} and expand the
exponential containing the cumulants.
Linearising the cumulants (as done in the prior subsection for the Fourier
streaming results) we get 
\eq{
 1 + \xi_s(\vec r) 
 &=   \int \frac{d^3 k}{(2\pi)^3} e^{ - i \vec k . \vec r} \int d^3r'\ e^{i \vec k \cdot \vec r'} \big[ 1 + \xi(r') \big]  
 \lb 1 + i k_\pp \mathcal C^{(1)}_{\hat n} (\vec{r}') - \frac{1}{2} k_\pp^2 \mathcal C^{(2)}_{\hat n} (\vec{r}') \rb \non\\
  &=   \int \frac{d^3 k}{(2\pi)^3} e^{ - i \vec k . \vec r} 
   \lb  1 +  \widetilde \Xi^{(0)}_{\hat n} (\vec k) + i k_\pp \widetilde \Xi^{(1)}_{\hat n} (\vec k)   - \frac{1}{2} k_\pp^2 \widetilde \Xi^{(2)}_{\hat n} (\vec k)\rb.
}
This is the familiar, linearized, moment expansion result.  By using the
multipoles we obtain the result in Eq.~\eqref{eq:xi_multipoles_kaiser}.

\subsection{Comparison of models in configuration space}
\label{sec:streaming_config_space_comparison}

\begin{figure}
\begin{center}
\resizebox{\columnwidth}{!}{\includegraphics{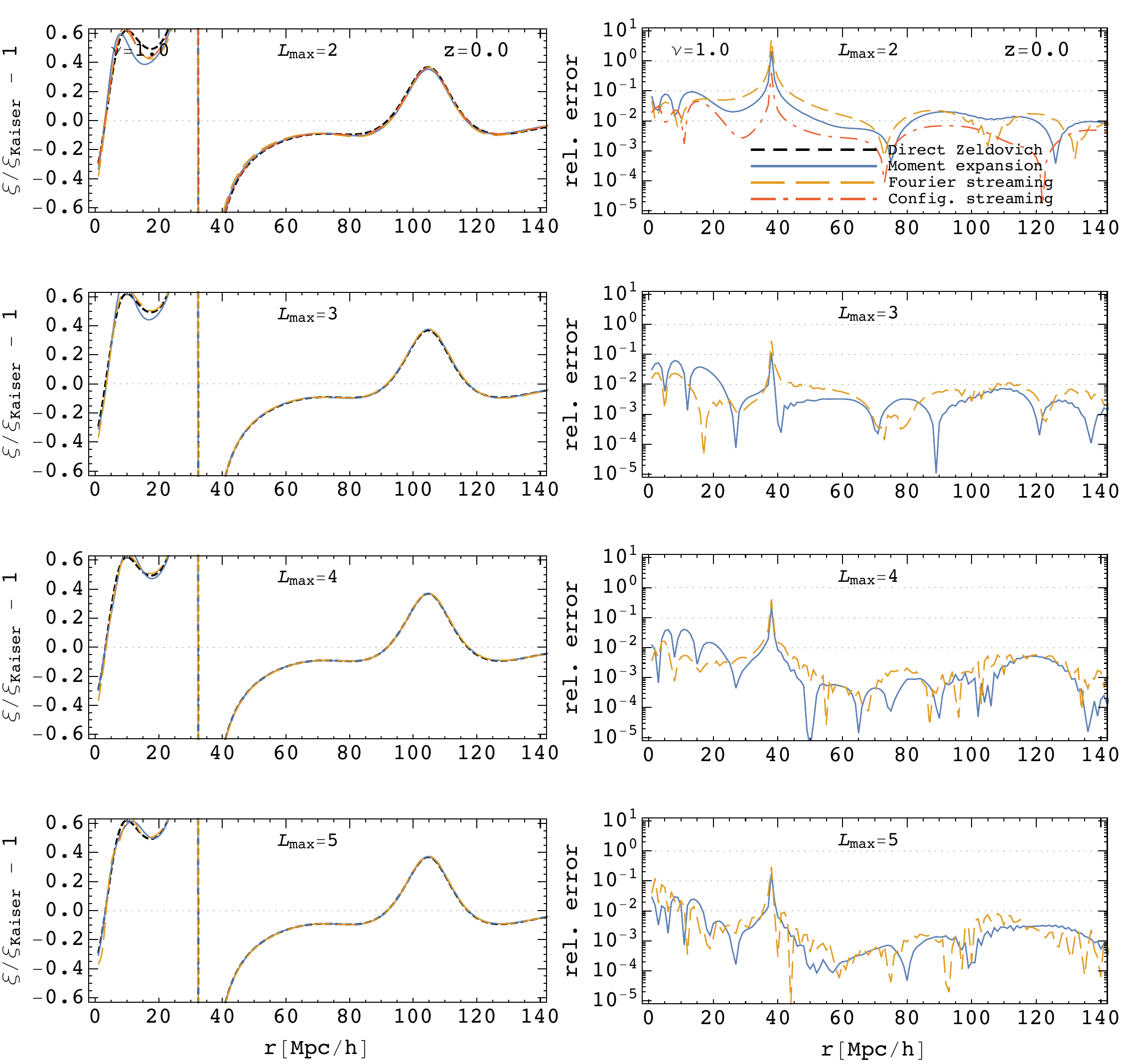}}
\end{center}
\vspace*{-5mm}
\caption{Comparison of the different models in configuration space, i.e.~for
the line-of-sight correlation function.  This is the configuration-space analog
of Fig.~\ref{fig:Fourier_comparison_nu10}.  As in that figure, the left panels
show the ratio of each model to the `Kaiser-Zeldovich' model (minus 1),
to reduce the dynamic range for plotting purposes.
The dashed black line indicates the full Zeldovich calculation,
solid blue the moment expansion, dashed orange the Fourier-space streaming
model and dot-dashed red the configuration-space streaming model.
The right panels show the relative error, compared to the full Zeldovich
calculation.  The rows show how the convergence is improved by including
more terms in the expansion, with the first row being the common `Gaussian'
approximation for the streaming models.  Note that we have gone beyond
the Gaussian streaming model in both Fourier and configuration space except
for the configuration-space streaming model.
See text for further discussion.}
\label{fig:Config_comparison_nu10}
\end{figure}

\begin{figure}
\begin{center}
\resizebox{\columnwidth}{!}{\includegraphics{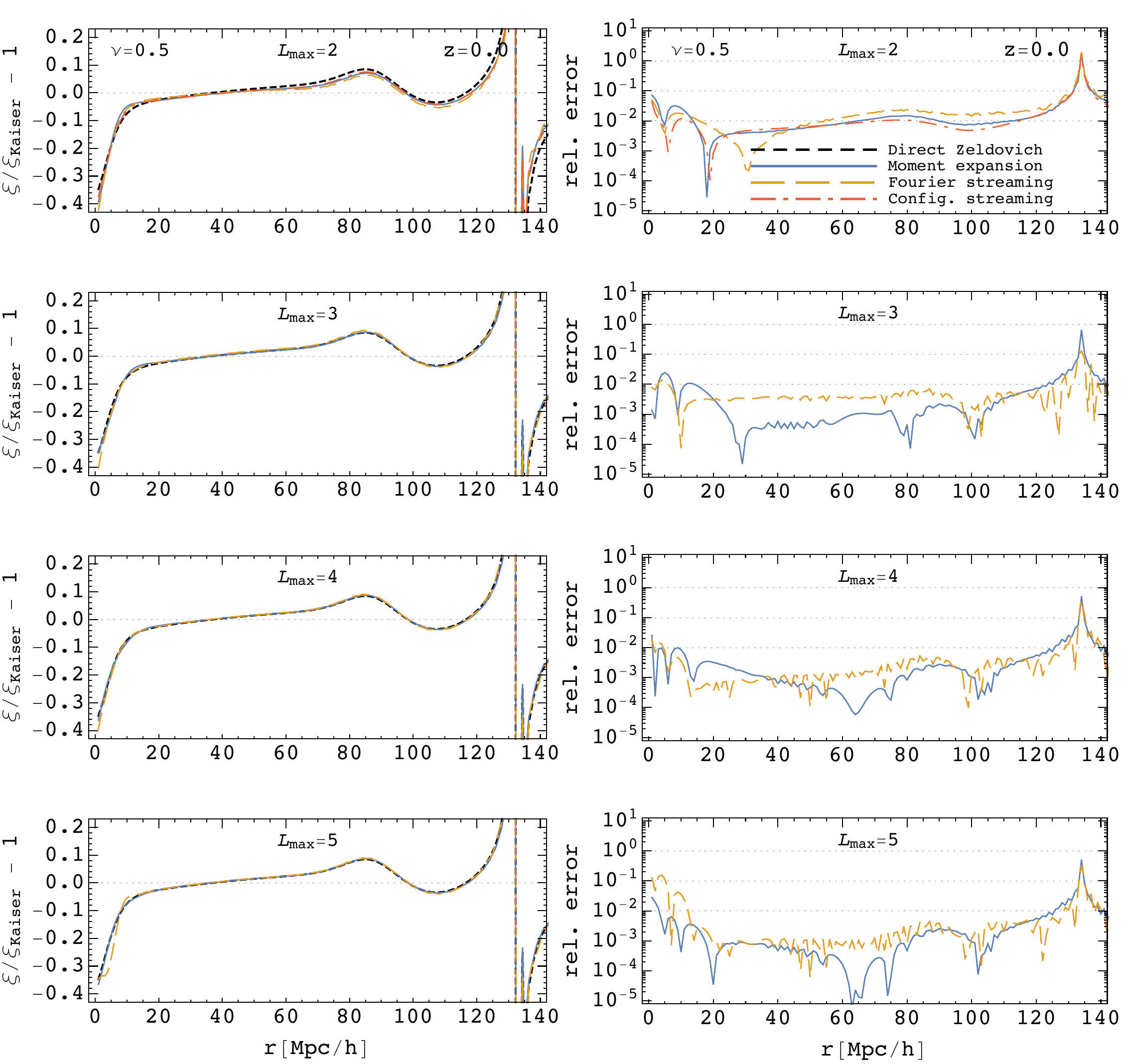}}
\end{center}
\vspace*{-5mm}
\caption{The same as Fig.~\ref{fig:Config_comparison_nu10} but for $\nu=0.5$.}
\label{fig:Config_comparison_nu05}
\end{figure}

\begin{figure}
\begin{center}
\resizebox{\columnwidth}{!}{\includegraphics{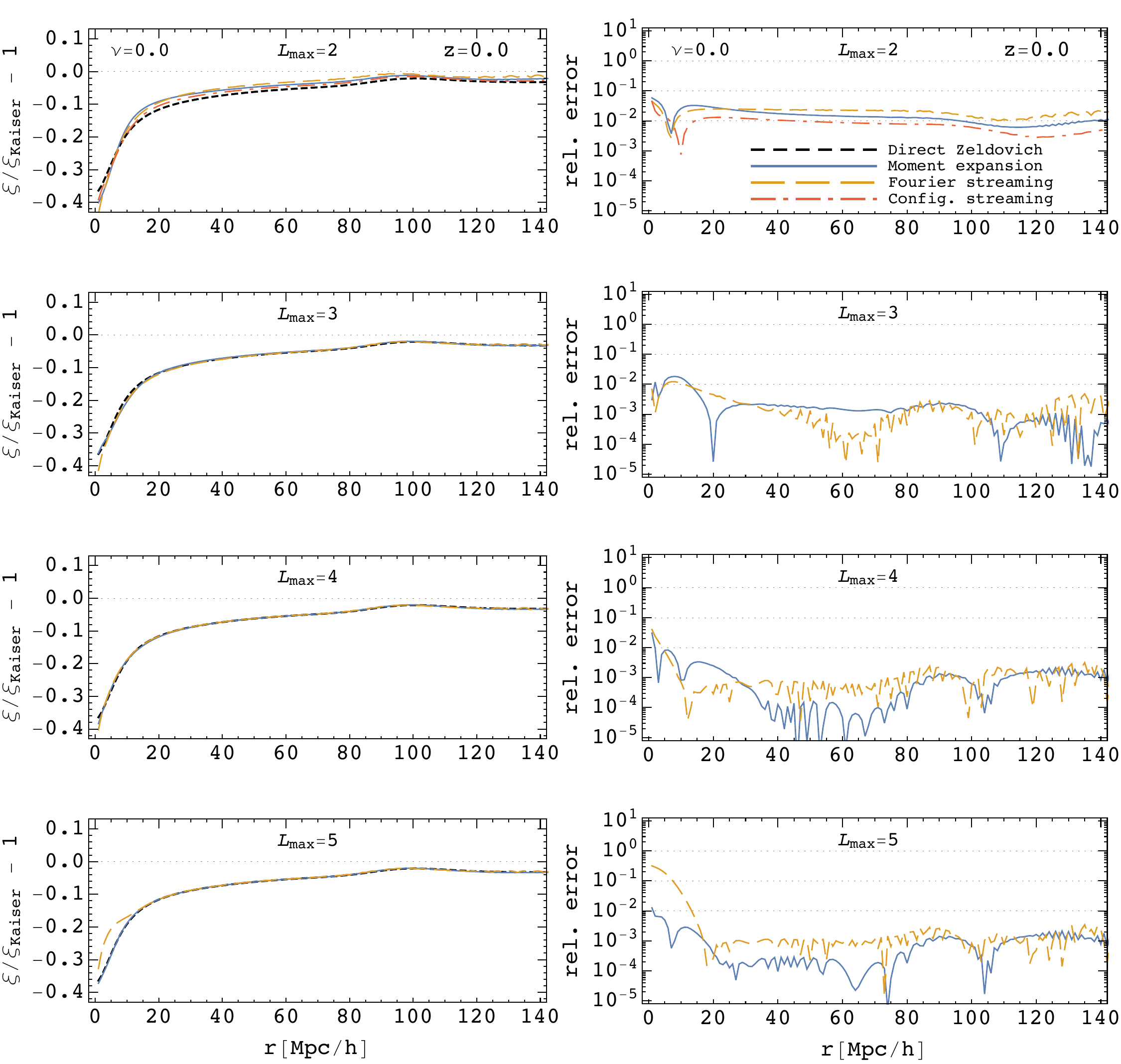}}
\end{center}
\vspace*{-5mm}
\caption{The same as Fig.~\ref{fig:Config_comparison_nu10} but for $\nu=0.0$.}
\label{fig:Config_comparison_nu00}
\end{figure}

Finally we compare the different expansions, above, to the full Zeldovich
calculation.  Fig.~\ref{fig:Config_comparison_nu10} compares the different
models to the full Zeldovich calculation for the line-of-sight correlation
function, $\xi_s(r,\nu=1)$.
In Fig.~\ref{fig:Config_comparison_nu05} and \ref{fig:Config_comparison_nu00}
equivalent plots are shown for the case $\xi_s(r,\nu=0.5)$ and
$\xi_s(r,\nu=0.0)$ respectively.
The figures show the absolute and relative convergence of the three approaches
discussed in the earlier sections to the directly computed Zeldovich
correlation function.

As we found for the Fourier-space statistics, the models perform less well
as $\nu\to 1$, with the error growing as a steep function of $\nu$.
Unlike in the Fourier-space case we no longer expect zero error as $\nu\to 0$.
Mirroring the discussion in \S\ref{sec:Fourier_comparison} this could have
implications for the way in which models are compared to data or the kinds
of applications for which perturbative models are appropriate, but the details
will differ from the Fourier-space case.

As mentioned earlier for the configuration space streaming model we performed
only the $L=2$ Gaussian case calculation, and we find that already at
$L_{\rm max}=2$ it performs at the level of $\sim1\%$ precision on scales
larger than $20\,h^{-1}$Mpc, outperforming both the moment expansion and the
Fourier cumulant models (for $L_{\rm max}=2$) for all angles. 
The moment expansion performs very well, reaching subpercent accuracy for
scales larger than $20\,h^{-1}$Mpc for $L_{\rm  max}\leq3$, and reaching
sub-permille accuracy for $L_{\rm max}=5$.
On these scales we see that Fourier streaming model is performing as well as
the moment expansion for all angles and $L_{\rm max}$ values.
The slight benefit of the Fourier streaming model can be noticed on scales
smaller than $20\,h^{-1}$Mpc, where for $L_{\rm max}\geq3$ it typically
provides slightly better performance, i.e. faster convergence to the full result.

\section{Including bias expansion and non-linear dynamics}
\label{sec:bias}

So far we have studied the effects of redshift-space mapping and explored the
convergence properties of three different approaches within the Zeldovich
approximation.  As has been stressed before, our expansion and the mapping of
cumulants and moments to the redshift-space power spectra and correlation
functions are valid also in the fully nonlinear and biased case.
In this section we provide the description for the first three moments 
in the Lagrangian Effective Field Theory (LEFT) context, taking into account
biasing and nonlinear effects up to 1-loop. 
We study the density two point function, pairwise velocity and velocity
dispersion.
These quantities have been also studied in detail in the same framework
in Ref.~\cite{VlaCasWhi16},and we repeat some of the derivations provided
there but also provide the complete expression in Fourier space
(Ref.~\cite{VlaCasWhi16} focused primarily on configuration space).
The results presented here can be readily used with the expansion presented
in the earlier sections, either in the Fourier or configuration space versions.

Since the Fourier space streaming model takes as ingredients velocity cumulants, 
it is useful to note that contributions up to the second cumulant (velocity dispersion) 
are the only non-vanishing contributions at one-loop PT. This also implies that working 
at the same PT order, all contributions to higher velocity moments come from disconnected 
graphs obtainable from these cumulants. This is, of course, related to the order counting in 
our PT prescription. Alternatively, one could organise the PT using as the leading order 
results from Sec. \ref{sec:direct_lag}. There we show that linear displacements contribute 
to all of the velocity cumulants. Nonetheless, we will not pursue this line of investigation here and will 
leave it for future work.

We start the discussion by quickly reviewing the bias expansion, but 
for more detailed discussion we refer readers to
Refs.~\cite{McDRoy09,Sen14,Mirb14,Ang15,Des16}.
Postulating that the number of biased objects is preserved by nonlinear
evolution we have a continuity equation
\eeq{
(1+\df_a(\vec x,\tau)) d^3 x = (1+\df_a(\vec q, \tau_{\rm in})) d^3 q,
\label{eq:continuity_halos}
}
Let us consider the biasing map $\df_a(\vec q, \tau_{\rm in})$, which we
assume is a continuous and smooth function that can be expanded in powers
of a characteristic (inverse distance) scale. In our case the scale will be
the Lagrangian radius of the biased object (e.g.~a protohalo) with
associated wavenumber $k_L$.
If we choose the initial time $\tau_{\rm in}$ early enough, all the dark
matter fields in the problem can be considered as linear.
An extensive list of bias parameters in Lagrangian space can be found in
Ref.~\cite{Des16}.  The first few terms are
\eq{
\df_a(\vec q) &= b_\df ~ :\df_L: (\vec q) \non\\
& ~ + b_{\df^2} :\df^2_L: (\vec q) + b_{s^2} :s_L^2:(\vec q)  \non\\ 
& ~ + b_{\df^3} :\df^3_L: (\vec q) + b_{\df s^2} :\df_L:(\vec q) :s^2_L: (\vec q) + b_{s^3} :s^3_L: (\vec q) \non\\
& ~ + b_{\partial^2\df} \frac{\partial_q^2}{k_L^2} :\df_L: (\vec q) + {\rm ``stochastic"} + \ldots
}
where $: O :$ represents the renormalised bias term (these will be defined explicitly further below), and $k_L$
is a characteristic physical scale related to Lagrangian halo radius. We have also defined the shear operator
\eq{
\hat s_{ij} (\vec q) = \frac{\partial_i\partial_j}{\partial^2} - \frac{1}{3} \df^K_{ij}, ~~~ 
{\rm and~the~corresponding~Fourier~operator~is} ~~ \hat s_{ij} (\vec p) = \frac{p_i p_j}{p^2} - \frac{1}{3} \df^K_{ij}. \non
}
Note that, by construction, independent third order bias terms like the $\psi$,\footnote{At third order 
Ref.~\cite{McDRoy09} introduces an independent bias term $\psi$ that can explicitly be written as 
\eeq{
\psi(\vec x) = \theta(\vec x) - \df(\vec x) - \frac{2}{7}s^2(\vec x) + \frac{4}{21}\df^2(\vec x), \non
}
where $ \theta$ is the velocity divergence defined as $\theta = \vec \nabla \cdot \vec u$. 
Note that all the fields above, including the shear field, should be considered including the nonlinear corrections.}
vanish in the initial conditions if we restrict ourselves to linear initial
dynamics.  This means that terms of this form, in this bias picture where biasing is set in the initial conditions, 
arise in Eulerian space only due to nonlinear evolution, and thus should not be considered as free biasing
coefficients (this is analogous to the so-called `co-evolution picture').
We use the notation
\eq{
s_{L, ij} (\vec q) &= \hat s_{ij} (\vec q)  \df_L (\vec q) = \left( \frac{\partial_i\partial_j}{\partial^2} - \frac{1}{3} \df^K_{ij} \right) \df_L (\vec q)
}
and also define renormalised operators, where the trivial zero-lag parts
are subtracted from the higher operators, so that:
\eq{
:\df_L: &= \df_L, \non\\
:\df_L^2: &= \df_L^2 - \la \df_L^2 \ra = \df_L^2 - \sigma_L^2, \non\\
:s_L^2: &= s_L^2 - \la s_L^2 \ra = s_L^2 - \tfrac{2}{3}\sigma_L^2, \non\\
:\df_L^3: &= \df_L^3 - 3 \la \df_L^2 \ra \df_L -  \la \df_L^3 \ra =  \df_L^3 - 3 \sigma_L^2 \df_L,\non\\
:s_L^3: &= s_L^3 - \la s_L^3 \ra =  s_L^3,
\label{eq:bias_operators}
}
An alternative would be to use the biasing prescription given in
e.g.~Ref.~\cite{Mat08b} (see also Ref.~\cite{Aviles18} for recent discussion),
where the biases are defined in full resummed form, rather then perturbatively.
In terms of generating functions we can rewrite the given density field 
\eq{
\df_a(\vec q) 
& = \hat \df_a(\lambda,  \gamma, \eta) e^{i \lambda \df_L (\vec q) +  i \gamma s^2_L (\vec q) + i \eta s^3_L (\vec q)} \Big|_{\lambda = \gamma = \eta = 0}, \non
}
where $\hat \df_a(\lambda,  \gamma, \eta)$ is the biasing operator acting
on the bias generating function
$\exp[ i \lambda \df_L +  i \gamma s^2_L  + i \eta s^3_L ]$.
Explicitly we can write 
\eq{
1 +  \hat \df_a(\lambda,  \gamma, \eta) = 1 & - i \lb b_\df - 3 \sigma_L^2 b_{\df^3} \rb \partial_{\lambda} 
- b_{\df^2} \lb \partial^2_{\lambda} + \sigma_L^2 \rb + i b_{\df^3} \partial^3_{\lambda} \non\\
& - i b_{s^2} \partial_{\gamma}  - b_{\df s^2} \partial_{\lambda} \partial_{\gamma} - i b_{s^3} \partial_{\eta} \non\\ 
& - i b_{\partial^2\df} \frac{\partial_q^2}{k_L^2} \partial_{\lambda} + {\rm ``stochastic"} + \ldots .
}
The generating function for any cross-moments of pairwise velocity for two
generic biased tracers in terms of the displacement field then generalizes to
\eq{
1 + \mathcal M_{ab} \big( \vec J, \vec r \big)
& = \int \frac{d^3 k}{(2\pi)^3} d^3 q ~ e^{- i \vec k \cdot ( \vec r- \vec q )} ~
\la  \big[1+ \df_{a}(\vec q)\big] \big[ 1+\df_{b}(\vec q) \big] \exp\Big[ i \vec k \cdot  \Delta (\vec q) + i J^j \Delta'_j (\vec q)\Big] \ra \non\\
& =  \big[1+ \hat \df_a(\lambda_1,  \gamma_1, \eta_1)\big] \big[ 1+ \hat \df_b(\lambda_2,  \gamma_2, \eta_2) \big] 
\int \frac{d^3 k}{(2\pi)^3} d^3 q ~ e^{- i \vec k \cdot ( \vec r- \vec q )} ~ \la e^{iX} \ra \bigg |_{  \lambda = \gamma = \lambda = 0  } \non,
}
where
\eq{
X = \lambda_1 \df_L (\vec q_1) + \lambda_2 \df_L (\vec q_2)  &+ \gamma_1 s_L^2 (\vec q_1) + \gamma_2 s_L^2 (\vec q_2) \non\\
&+ \eta_1 s_L^3 (\vec q_1) + \eta_2 s_L^3 (\vec q_2)  + \vec k \cdot  \Delta (\vec q) + J^j \Delta'_j (\vec q).
}
This provides us with the means to compute the moments, and thus the cumulants,
for biased tracers up to the terms we used in the bias expansion. 

Let us focus for a moment on the stochastic term.
In Lagrangian coordinates we are trying to describe the overdensity of
discrete objects in terms of continuous dark matter fields.
In order to be able to achieve that an auxilliary stochastic field, $\epsilon$,
has to be added to the set of our bias operators.
Intuitively, if we are describing very sparse objects this field has to play
the role of noise.  Thus we can write  
\eq{
\df_a(\vec q) = \bar \df_a(\vec q) + \epsilon_a(\vec q),
}
where $\bar \df_a(\vec q)$ is the average over the stochastic distribution,
i.e.~assuming we have a PDF, $p_a[\epsilon]$, associated with the random
variable $\epsilon$ such that
$\int [\mathcal D \epsilon] \epsilon p_a[\epsilon] = 0$, we have 
\eq{
\bar \df_a(\vec q, \tau) = \int [\mathcal D \epsilon] \df_a(\vec q, \tau) p_a[\epsilon](\vec q, \tau).
}
For $\bar \df_a$ we can assume the usual bias expansion in terms of the
operators in Eq.~\eqref{eq:bias_operators}.
For the field in the Eulerian coordinates we then have 
\eq{
(2\pi)^3\df^D(\vec k) + \df_{a}(\vec k) 
& = \int d^3 q ~ e^{ i \vec k \cdot \vec q} ~ \left[ 1 + \df_a (\vec q) \right]
e^{i \vec k \cdot \Psi(\vec q)} \\
& = \int d^3 q ~ e^{ i \vec k \cdot \vec q} ~ \left[ 1 + \bar \df_a (\vec q) \right]
e^{i \vec k \cdot \Psi(\vec q)} 
+ \int d^3 q ~ e^{ i \vec k \cdot \vec q} ~ \epsilon_a (\vec q )
e^{i \vec k \cdot \Psi(\vec q)}. \non
}
Given that, by construction, the stochastic field does not correlate with
the $\bar \df_a$, this gives us the auto power spectrum
\eq{
(2\pi)^3\df^D(\vec k)  + P_a(k) &=  \int d^3 q ~ e^{ i \vec k \cdot \vec q} ~
\la \left[ 1+ \bar \df_a (\vec q_1) \right] \left[1 + \bar \df_a(\vec q_2) \right] e^{ i \vec k \cdot  \Delta (\vec q, \tau) } \ra \non\\
&\hspace{4cm} +  \int d^3 q ~ e^{ i \vec k \cdot \vec q} ~
\la \epsilon_a (\vec q_1) \epsilon_a (\vec q_2) e^{ i \vec k \cdot  \Delta (\vec q, \tau) } \ra
}
Focusing on the last term, and assuming also that $\Delta$ and $\epsilon_a$
are uncorrelated to be consistent with the definition of $\epsilon_a$, we
can introduce
$\la\epsilon_a(\vec q_1)\epsilon_a(\vec q_2)\ra' = \xi_{\epsilon_a}(\vec q)$.
Assuming Poisson statistics for $\epsilon_a$ we have a constant Fourier space
variance, i.e.~$\xi_{\epsilon_a}(\vec q)\sim \df^D(\vec q)$.
Thus
\eq{
\int d^3 q ~ e^{ i \vec k \cdot \vec q} ~
\xi_{\epsilon_a} (\vec q, \tau) \la e^{ i \vec k \cdot  \Delta (\vec q, \tau) } \ra
\sim {\rm const}_a ~ \la e^{ i \vec k \cdot  \Delta (\vec q, \tau) } \ra_{\vec q \to 0} \sim {\rm const}_a.
}
It is interesting to note that the constant term, under the assumptions above,
is not sensitive to RSD and should have the same value for all angle bins
(or equivalently, contribute only to $\ell=0$).
The noise should depend only on the tracer type (thus the $a$ label on the
constant). 
If we allow a correlation with the stochastic contributions to the dynamical
fields, $\Delta$, it should be clear from above that we will get additional
scale dependence and that will, of course, also affect redshift space.
Moreover, similar analysis can be performed if we have scale-dependent
stochasticity 
$\xi'_{\epsilon_a}(\vec q)\sim\partial^2/k_\epsilon^2\df^D(\vec q)$.  From
there we can also conclude that such terms, even though they carry new bias
coefficients, should not be affected by the redshift space mapping
(under the assumptions above).  One caveat to this statement is the
possibility of anisotropic selection effects (e.g.\ Ref.~\cite{Hirata09}),
which would introduce additional line-of-sight-dependent bias operators
and stochastic components (e.g.\ Ref.~\cite{Desjacques2018}).  These terms
arise from survey non-idealities rather than dynamical processes and their
impact would need to be addressed on a case-by-case basis.

\subsection{Two point function in real space}
\label{sec:two_points}

In this subsection we focuse on providing the theory for the two point
halo correlation function and power spectrum in real space.
Formally this is the zeroth moment of the real space generating function
\eeq{
\xi_{ab} = \Xi_{0}(\vec r) =  1+ \mathcal M_{ab} \big( \vec J=0, \vec r \big).
}
Using the Lagrangian framework we have set up and applying the cumulant
expansion at 1-loop we see that we get contributions only from the first
two cumulants 
\eeq{
\log \la e^{iX} \ra = \sum_{n=2}^\infty \frac{i^n}{n!} \la X^n \ra_c = - \frac{1}{2} \la X^2 \ra_c - \frac{i}{6} \la X^3 \ra_c + \ldots.
}
Given that terms involving third order shear field $s_{L}^3$ do not contribute
at 1-loop and we have 
\eq{
- \frac{1}{2} \la X^2 \ra_c = & - \frac{1}{2} k_ik_j A_{ij} - \lambda_1\lambda_2 \xi_L - (\lambda_1 + \lambda_2 )k_i U^{10}_i \non\\
 & - \gamma_1\gamma_2 \zeta_L - (\gamma_1+\gamma_2) k_i V^{10}_i 
 - \frac{1}{2} (\lambda_1^2 + \lambda_2^2) \sigma_{\df_L}^2 \non\\
 - \frac{i}{6} \la X^3 \ra_c =& - \frac{i}{6} k_ik_jk_l W_{ijk}
- i \lambda_1 \lambda_2 k_i U^{11}_i
- \frac{i}{2} (\lambda_1 + \lambda_2 ) k_i k_j A^{10}_{ij}
- \frac{i}{2} \lb \lambda_1^2 + \lambda_2^2 \rb k_i U^{20}_{i} \non\\
& - \frac{i}{2}(\gamma_1 + \gamma_2 ) k_i k_j A^{20}_{ij} 
- i (\gamma_1 \lambda_1 + \gamma_2 \lambda_2) k_i V^{11}_i
- i (\gamma_1 \lambda_2 + \gamma_2 \lambda_1) k_i V^{12}_i \non\\
& - i (\gamma_1 \lambda_1 \lambda_2 + \gamma_2 \lambda_1 \lambda_2 ) \chi^{11}
 - \frac{i}{2} (\gamma_1 \lambda_2^2 + \gamma_2 \lambda_1^2 ) \chi^{12} 
 - \frac{i}{2} (\gamma_1 \lambda_1^2 + \gamma_2 \lambda_2^2 ) \sigma^4_{\df_L^2 s^2_L},
 }
where we have introduced
$\sigma^4_{\df_L^2 s^2_L} = \la \df_{L}^2 s_{L}^2 \ra_c$,
and we have \cite{VlaCasWhi16}
\eq{
A_{ij}      &= \la \Delta_i \Delta_j \ra_c, ~~
\xi_L       = \la \df_{L1} \df_{L2} \ra_c, ~~
U^{10}_i = \la \df_L \Delta_i \ra_c, ~~
\zeta_L = \la s^2_{L1} s^2_{L2} \ra_c, ~~
V^{10}_i = \la s^2_L \Delta_i \ra_c, \non\\
W_{ijk}   &= \la \Delta_i \Delta_j \Delta_l \ra_c, ~~
U^{11}_i = \la \df_{L1}\df_{L2} \Delta_i \ra_c, ~~
A^{10}_{ij} = \la \df_L \Delta_i \Delta_j \ra_c, ~~
U^{20}_{i} = \la \df_L^2 \Delta_i \ra_c,  \non\\
&\hspace{2.6cm}
A^{20}_{ij} = \la s^2_L \Delta_i \Delta_j \ra_c,~
V^{11}_{i} =\la s_{L1}^2 \df_{L1} \Delta_i \ra_c, ~
V^{12}_{i} =\la s_{L1}^2 \df_{L2} \Delta_i \ra_c, \non\\
&\hspace{2.6cm}
\chi^{11} = \la s_{L1}^2 \df_{L1} \df_{L2} \ra_c, ~
\chi^{12} = \la s_{L1}^2 \df_{L2}^2 \ra_c.
\label{eq:correlators}
}
This produces a number of cross and auto correlators of the bias operators and
displacement field.  The purely displacement terms, $A_{ij}$ and $W_{ijk}$, are 
the same as in the unbiased case (i.e.~the dark matter case) and we refer the
reader to Ref.~\cite{VlaWhiAvi15} for a detailed discussion of this case.
Terms represented by $\xi_L$, $\zeta_L$ and $\chi^{12}$ ($\chi^{11}$ does not
contribute) are pure bias correlation terms, and describe the proto-field of
a biased tracer.  The rest of the terms are correlations of bias terms and
dynamical displacement terms.

Acting with the biasing operators on the cumulants and expanding all but
linear displacements up to 1-loop order we have 
\begin{align}
\big[1&+ \hat \df_{a}(\lambda_1,  \gamma_1, \eta_1)\big]
 \big[ 1+\hat \df_{b}(\lambda_2,  \gamma_2, \eta_2) \big] \la e^{iX} \ra = 
 e^{- \frac{1}{2} k_ik_j A_{L, ij}} \Bigg\{
  1 - \frac{1}{2} k_ik_j A^{\rm loop}_{ij} - \frac{i}{6} k_ik_jk_l W^{\rm loop}_{ijk} \non\\
&~~~
- b_{\df,\{a,b\}} \Big( k_i k_j A^{10}_{ij} - 2 i k_i U^{10}_i \Big) 
+ b_{\df,a}b_{\df,b} \Big( \xi_L + i k_i U^{11}_i - k_i k_j U^{10}_i U^{10}_j  \Big) \non\\
&~~~ 
+ b_{\df^2,\{a,b\}}\Big( i k_i U^{20}_i - k_i k_j U^{10}_i U^{10}_j  \Big)
+ b_{\df^2,\{a} b_{\df,b\}} \Big( 2 i k_i U^{10}_i  \xi_L \Big) 
+ b_{\df^2,a} b_{\df^2,b} \bigg( \frac{1}{2} \xi_L^2 \bigg) \non\\
&~~~ 
- b_{s^2,\{a,b\}} \Big(  k_i k_j A^{20}_{ij} - 2 i k_i V^{10}_i  \Big)
+ b_{\df,\{a} b_{s^2,b\}}  \Big( 2 i k_i V^{12}_i \Big)
+ b_{s^2,\{a} b_{\df^2,b\}} \chi^{12} 
+ b_{s^2,a}b_{s^2,b}  \zeta_L \non\\
&~~~
- \frac{1}{2}\alpha_0 k^2
+ i b_{\partial^2 \df,\{a,b\}} \bigg( 2 k_i \frac{\partial^2}{k_L^2} U^{(1)}_i \bigg)
+ b_{\df,\{a}  b_{\partial^2 \df,b\}} \Big( 2 \frac{\partial^2}{k_L^2} \xi_L \Big)  + \ldots
 \Bigg\}  +  {\rm ``stochastic"},
\end{align}
where we introduced the notation $b_{\{a,b\}} = \frac{1}{2}(b_a+b_b)$.
For evaluation purposes we can use $k_L=1h\,{\rm Mpc}^{-1}$ even though
it should be noted that a better estimate could be obtained based on the
sizes and masses of halos.  The 1-loop halo power spectrum can thus be
expressed as 
\eq{
(2\pi)^3 \df^D(k) + P_{ab}(k) = \int d^3 q~ e^{ i \vec k \cdot \vec q}~ \big[1+ \hat \df_a \big]  \big[1+ \hat \df_b\big] \la e^{iX} \ra + {\rm ``stochastic"} ,
} 
which can be written as a sum of terms
\eq{
P_{ab}(k) =&  \lb 1 - \tfrac{1}{2} \alpha_{0} k^2 \rb P_{\rm Zel} +  \non\\
&+ b_{\df,\{a,b\}} P_{\df} 
+ b_{\df,a}b_{\df,b} P_{\df\df} + b_{\df^2,\{a,b\}} P_{\df^2}
+ b_{\df,\{a} b_{\df^2,b\}} P_{\df \df^2}
+ b_{\df^2,a} b_{\df^2,b} P_{\df^2\df^2} \non\\
& 
+ b_{s^2,\{a,b\}} P_{s^2}
+ b_{\df,\{a} b_{s^2,b\}} P_{\df s^2}
+ b_{\df^2,\{a}  b_{s^2,b\}} P_{\df^2 s^2}
+ b_{s^2,a}b_{s^2,b} P_{s^2s^2} \non\\
& 
+ b_{\partial^2 \df,\{a,b\}} P_{\partial^2 \df}
+ b_{\df,\{a}  b_{\partial^2 \df,b\}} P_{\df \partial^2 \df}
+ P_{\epsilon_a \epsilon_b}.
\label{eq:PS_real}
}
Each of the terms above can be expressed as an integral over $\mu$ and
written as a sum of spherical Bessel functions using Eq.~\eqref{eq:int_mu_n}.
The terms are given in Appendix \ref{app:explicitPS_CF}.
The counter term, $\alpha_{0}$, is capturing the leading 1-loop UV
dependence due to nonlinear dynamics.
For more in depth discussion on this dependence in Lagrangian dynamics and counterterms we
refer a reader to Ref.~\cite{VlaWhiAvi15}.
The two derivative terms, $P_{\partial^2 \df}$ and $P_{\df \partial^2 \df}$,
differ from the $k^2 P_{\rm Zel}$ term only up to resummed long displacements
and can in principle be gathered into one term.
Expanding also those long displacements we can recover the Eulerian derivative
terms \cite{Ang15,Des16}.
Very recently an analogous biasing model has been studied at the field level,
where is showed good performance \cite{Schmittfull18}.

Analogously we can write the expression for the halo correlation function
\eeq{
1 + \xi_h(r) = \int \frac{d^3kd^3q}{(2\pi)^3}~ e^{-i \vec k \cdot (\vec r-\vec q)}~\la e^{iX} \ra.
}
which is also given as a sum of individual contributions
\eq{ 
 \xi_{ab}(r) =& \xi_{\text{Zel}} + \xi_{\text{loop}}
       + \alpha_{0}~ \xi^{(\alpha)}_{\text{c.t.}} \non\\
& + b_{\df,\{a,b\}} \xi_{\df}
       + b_{\df,a}b_{\df,b}  \xi_{\df\df}
       + b_{\df^2,\{a,b\}} \xi_{\df^2}
       + b_{\df,\{a} b_{\df^2,b\}} \xi_{\df\df^2}
       + b_{\df^2,a} b_{\df^2,b} \xi_{\df^2\df^2} \non\\
&+ b_{s^2,\{a,b\}} \xi_{s^2}
       + b_{\df,\{a} b_{s^2,b\}} \xi_{\df s^2}
       + b_{\df^2,\{a} b_{s^2,b\}} \xi_{\df^2 s^2} 
       + b_{s^2,a}b_{s^2,b} \xi_{s^2 s^2} \non\\
&+ b_{\partial^2 \df,\{a,b\}} \xi_{\partial^2 \df}
+ b_{\df,\{a}  b_{\partial^2 \df,b\}} \xi_{\df \partial^2 \df}
+ \xi_{\epsilon_a \epsilon_b} .
\label{eq:xi_real}
}
where again the explicit form of the individual contributions is given in Appendix~\ref{app:explicitPS_CF}.

\subsection{The mean pairwise velocity}

Next in the moment hierarchy is the mean pairwise velocity term,
i.e.~the first velocity moment.  We are interested in obtaining the 1-loop
contributions (see also Ref.~\cite{WanReiWhi14})
\eq{
\Xi_{i} (\vec r) = (1+\xi(r)) v_{12,i}(\vec r)& = 
\int \frac{d^3 k}{(2\pi)^3} d^3 q ~ e^{- i \vec k \cdot ( \vec r- \vec q )} 
(-i) \frac{\partial}{\partial J_i}\la e^{iX} \ra \Big|_{\vec J=0} \\
&= \sum_{n=1}^\infty \frac{i^n}{n!}
\int \frac{d^3 k}{(2\pi)^3} d^3 q ~ e^{- i \vec k \cdot ( \vec r- \vec q )} 
\la Y^n \Delta'_i \ra_c \exp \Bigg[ \sum_{l=2}^\infty \frac{i^l}{l!} \la Y^l \ra_c \Bigg]. \non
}
If we look at the pairwise velocity correlator we have 
\eeq{
\widetilde \Xi_{i} (\vec k) =  \int d^3 r~ e^{i \vec k \cdot \vec r}
\left[1+\xi(r)\right] v_{12,i}(\vec r).
}
The natural basis to project this vector field onto is the projection onto
the $\hat k$ vector, thus we have
$\widetilde \Xi_{i}(\vec k) = \widetilde \Xi^{(0)}_1(k)  \hat k_i $,
and similarly $v_{12,i}(\vec r) = v_{12}(r) \hat r_i$.
It follows that
\eq{
\left[ 1+\xi(r)\right] v_{12}(r) &= - i \int \frac{k^2 d k}{2\pi^2}~ \widetilde \Xi_1(k)  j_{1}(kr), \non\\
\widetilde \Xi_1(k) &= 4 \pi i \int r^2 d r~ \big[1+\xi(r) \big] v_{12}(r)  j_{1}(kr).
}
Using the above relation we have for the pairwise velocity power spectrum,
i.e.~the Fourier representation of the first velocity moment 
\eeq{
k \widetilde \Xi^{(0)}_1(k)  =  \big[1+ \hat \df_{a}\big] \big[ 1+\hat \df_{b} \big] \sum_{n=1}^\infty \frac{i^n}{n!}
\int d^3 q ~ e^{ i \vec k \cdot \vec q} ~
k^i \la Y^n \Delta'_i \ra_c \exp \Bigg[ \sum_{l=2}^\infty \frac{i^l}{l!} \la Y^l \ra_c \Bigg],
}
where we also added the biasing operators.
If we consider the bias terms up to $\df_L$ we have 
\eeq{
i Y = i \lambda_1 \df_{L1} + i \lambda_2 \df_{L2} + i \vec k \cdot \Delta(\vec q).
\label{eq:Y_bias}
}
We have dropped all of the shear contributions in the higher velocity terms,
the motivation being the realisation that these contributions are rather
small already in the pure density statistics on the scales of interest
\cite{VlaCasWhi16} and they are even less relevant in the higher velocity
statistics.  The higher-order biasing terms can of course be added in
analogous ways to the earlier section if desired.
For a consistent treatment of these terms in configuration space we refer
the reader to Ref.~\cite{VlaCasWhi16}.

Given that we are concerned with the 1-loop calculation here, relevant
contributions that enter to the dynamics and bias generating function,
$\exp \big[ \sum_{\ell=2}^\infty (i^\ell)/(\ell!) \la Y^\ell \ra_c \big]$,
are determined by two terms
\eq{
i k_i \la Y \Delta'_i \ra_c & = i f(\tau) ( \lambda_1 + \lambda_2) k_i \lb U^{\rm lin}_{10,i} + 3 U^{\rm loop}_{10,i} \rb 
+ i f(\tau) k_i k_j \lb A^{\rm lin}_{ij} + 2 A^{\rm loop}_{ij} \rb, \non\\
-\frac{1}{2} k_i \la Y^2 \Delta'_i \ra_c 
& =  - f(\tau) \lb \lambda_1^2 + \lambda_2^2 \rb k_i U^{20}_i - 2 f(\tau) \lambda_1 \lambda_2 k_i U^{11}_i \non\\
&\hspace{3cm}  -  \frac{3}{2}f(\tau) \lb \lambda_1 + \lambda_2 \rb k_i k_j A^{10}_{ij} 
- \frac{2}{3} f(\tau) k_ik_jk_l W_{ijl}. 
} 
Where we had to explicitly split the linear and loop contributions of the
$U_{10,i}$ and $A_{ij}$ terms given that the loop contributions enter with
different prefactors than earlier.
All of the terms are the same as those in Eq.~\eqref{eq:correlators} and
explicit linear and 1-loop representations can be found in
Appendix \ref{app:explicitPS_CF}.

We have for the power spectrum
\eq{
   k \widetilde \Xi^{(0)}_1(k) &= f(\tau) \int d^3 q ~ e^{ i \vec k \cdot \vec q} ~ e^{- \frac{1}{2} k_ik_j A_{L, ij}} 
   \bigg( i k_i k_j \lb A^{\rm lin}_{ij} + 2 A^{\rm loop}_{ij} \rb  - \frac{2}{3} k_ik_jk_l W_{ijl}  \\
&~~~ + b_{\df,\{ab\}} \Big( 
  2 k_i \lb U^{\rm lin}_{10,i} + 3 U^{\rm loop}_{10,i} \rb
 + i 3 k_i k_j A^{10}_{ij} - 2 k_i k_j k_lA^{\rm lin}_{ij} U^{\rm lin}_{10,l}
 \Big) \non\\
&~~~ +  b_{\df,\{a}b_{\df,b\}} \Big( 
  2 k_i U^{11}_i +  i k_i k_j A^{\rm lin}_{ij} \xi_L + 2 i k_i k_j U^{\rm lin}_{10,i}U^{\rm lin}_{10,j} 
  \Big) \non\\
&~~~ +  b_{\df^2,\{ab\}} \Big( 
   2 k_i U^{20}_i +  i 2 k_i k_j U^{\rm lin}_{10,i}U^{\rm lin}_{10,j} 
  \Big) + b_{\df^2,\{a} b_{\df,b\}} 2 k_i U^{\rm lin}_{10} \xi_L + \ldots
     \bigg). \non
}
Term by term this result can be separated into the scale dependent spectra and 
biasing terms so that 
\eq{
k \widetilde \Xi^{(0)}_1(k) = f(\tau) 
\Big( P^{01}_{\rm loop} + \alpha_{1} k^2 P_{\rm Zel} 
&+ b_{\df,\{a,b\}} P^{01}_{\df} 
+ b_{\df,a}b_{\df,b} P^{01}_{\df\df} \\
&+ b_{\df^2,\{ab\}} P^{01}_{\df^2} 
+ b_{\df,\{a} b_{\df^2,b\}} P^{01}_{\df\df^2} +\ldots \Big), \non
}
where the $\alpha_{1}$ is again the leading counterterm coefficient.
The continuity equation gives a simple relation for the purely dynamical
spectra, coming from $A$ and $W$ terms,
\eq{
  P^{01}_{\rm loop} (k) = P^{01}_{\rm Zel}(k) + 2 P_{\rm loop}(k),
}
where for the first term we have
$P^{01}_{\rm Zel}(k) = (1/2)(\partial/\partial D_+)P_{\rm Zel}(k)$.
The explicit form of each of the bias spectra can be found in 
Appendix \ref{app:explicitPS_CF}.
For the real-space pairwise velocity we then obtain
\eq{
\Xi_{01}(r) = \hat r_i \Xi_{i}(r)
  = f(\tau) \Big( \Xi^{01}_{\rm loop} + \alpha_{1} \Xi^{01}_{\rm c.t.}
  &+ b_{\df,\{ab\}} \Xi^{01}_{\df}  
  + b_{\df,a}b_{\df,b} \Xi^{01}_{\df\df} \\
  &+ b_{\df^2,\{ab\}} \Xi^{01}_{\df^2} +
  b_{\df,\{a} b_{\df^2,b\}} \Xi^{01}_{\df\df^2} +\ldots \Big), \non
}
where we can identify
$\Xi^{01}_{\rm loop}(r)=\Xi^{01}_{\rm Zel}+2\Xi_{\rm 1-loop}$.
Again all the explicit formulae for individual bias correlation functions
can be found in Appendix~\ref{app:explicitPS_CF}.

\subsection{The pairwise velocity dispersion}
\label{eq:pair_disp}

The final thing in this section is the 1-loop velocity dispersion
(i.e.~the second velocity moment), for biased tracers.
We use the same biasing assumptions as in Eq.~\eqref{eq:Y_bias}.
First we start taking the derivatives of the generating function 
\eq{
\Xi_{ij} (r) &= (1+\xi(r)) \big[ \sigma_{12,ij}(r) + v_{12,i}(r) v_{12,j}(r) \big] = 
-\frac{\partial}{\partial J_{i} \partial J_{j}} \Big[ 1+ \mathcal M \big( \vec J, \vec r \big) \Big]_{J=0} \non\\
&= \int \frac{d^3 k}{(2\pi)^3} d^3 q ~ e^{- i \vec k \cdot ( \vec r- \vec q )} ~ 
 \la \Delta'_i\Delta'_j \ra_c \exp \Bigg[ \sum_{n=2}^\infty \frac{i^n}{n!} \la Y^n \ra_c \Bigg] \non\\
&\hspace{1cm} + \sum_{n=1}^\infty \frac{i^n}{n!} 
\int \frac{d^3 k}{(2\pi)^3} d^3 q ~ e^{- i \vec k \cdot ( \vec r- \vec q )} ~ 
\Bigg( 
\sum_{m=1}^n \binom{n}{m} \la Y^{m} \Delta'_i \ra_c \la Y^{n-m} \Delta'_j \ra_c \non\\
& \hspace{6cm} +  \la Y^n \Delta'_i\Delta'_j \ra_c \Bigg)
\exp \Bigg[ \sum_{n=2}^\infty \frac{i^n}{n!} \la Y^n \ra_c \Bigg].
\label{eq:Xi2_expansion}
}
As discussed in Appendix~\ref{sec:and_decomp_ofvelocities} we can 
decompose the pairwise velocity dispersion in three components
\eeq{
\Xi_{ij} (r) = \la \big[1+\df(x) \big] \big[1+\df(x') \big] \Delta u_i(x)\Delta u_j(x')  \ra
= 2 \lb \xi^{ij}_{02}(0) + \xi^{ij}_{02}(r) - \xi^{ij}_{11}(r) \rb,
}
where first term is the zero-lag (point) contribution, second is the
kinetic energy tensor term (correlated with the density) and the last
is the momentum field correlation
\eq{
\xi^{ij}_{02}(0) &= \la \big[1+\df(x) \big] u_i(x) u_j(x) \ra = \sigma^2_{02} \df^K_{ij}, \non\\
\xi^{ij}_{02}(r) &= \la \df(x) \big[1+\df(x') \big]  u_i(x') u_j(x') \ra, \non\\
\xi^{ij}_{11}(r) &= \la \big[1+\df(x) \big] u_i(x) \big[1+\df(x') \big] u_j(x') \ra.
}
The Fourier-space representation of the velocity dispersion term is then
\eq{
(2\pi)^3 \sigma^2_{02} \df^K_{ij} \df^D_{\vec k} + \widetilde \Xi_{ij}(k) =  \int d^3 r~ e^{i \vec k \cdot \vec r} ~ \Xi_{ij} (r),
}
where if we use the Lagrangian multipole decomposition we can write 
$\widetilde \Xi_{ij}(k)  = \df^K_{ij} \widetilde \Xi^{(0)}_{2}(k) + \frac{3}{2}\lb \hat k_i \hat k_j - \frac{1}{3} \df^K_{ij} \rb  \widetilde \Xi^{(2)}_{2}(k) $,
and similarly  $\Xi_{ij}(r)  = \df^K_{ij} \Xi^{(0)}_{2}(r) - \frac{3}{2} \lb \hat r_i \hat r_j - \frac{1}{3} \df^K_{ij} \rb \Xi^{(2)}_{2}(r) $.
Note that this decomposition is somewhat different than the one we used in
\S\ref{sec:mom_exp}, where $\Xi^{(m)}_2$ were scale dependent spectra
multiplying different powers of $\nu$.  To avoid multiplying notation we
will keep the same labels here but keep this change in definition in mind. 
These scalar components then transform as
\eq{
 \widetilde \Xi^{(0)}_{2}(k) &=\frac{1}{3}\df^K_{ij}  \widetilde \Xi_{ij}(k) = - (2\pi)^3 \sigma^2_{02} \df^D_{\vec k} + 4 \pi \int r^2 dr~ \Xi^{(0)}_2(r) j_0(kr) , \non\\
 \widetilde \Xi^{(2)}_{2}(k) &=\lb \hat k_i \hat k_j - \frac{1}{3} \df^K_{ij} \rb  \widetilde \Xi_{ij}(k) = 4 \pi \int r^2 dr ~ \Xi^{(2)}_2(r) j_2(kr).
\label{eq:tensor_decpmposition}
}
It is useful at this point to give the connecting relations to frequently
used alternative decompositions (see e.g.~Ref.~\cite{WanReiWhi14}):
\eeq{
\sigma^2_{12,nm} = \sigma^2_\pp \hat r_n \hat r_m + \sigma^2_\perp \lb \df^K_{nm} - \hat r_n \hat r_m \rb,
}
so that
$\sigma^2_\pp = \sigma^2_{12,nm}\hat r_n \hat r_m$ and
$\sigma^2_\perp = \lb \sigma^2_{nm}\df^K_{nm} - \sigma^2_\pp \rb /2$.
Connecting this to the notation given in Eq.~\eqref{eq:tensor_decpmposition}
and the discussion above we have 
\eeq{
\Xi^{(0)}_{2}(r) / \lb 1+\xi(r) \rb = \frac{1}{3} \lb 2\sigma_\perp^2 + \sigma_\pp^2 \rb, \qquad 
\Xi^{(2)}_{2}(r) / \lb 1+\xi(r) \rb = \frac{2}{3} \lb \sigma_\perp^2 - \sigma_\pp^2 \rb,
}
or inversely
\eeq{
\lb 1+\xi(r) \rb \sigma_\pp^2 = \Xi^{(0)}_{2}(r) - \Xi^{(2)}_{2}(r), \qquad
\lb 1+\xi(r) \rb \sigma_\perp^2 = \Xi^{(0)}_{2}(r) + \frac{1}{2} \Xi^{(2)}_{2}(r).
}

We consider the perturbative, 1-loop contributions to
Eq.~\eqref{eq:Xi2_expansion}.
Considering the first term ($n=0$) we have:
\eq{
&\lb \la \Delta'_i\Delta'_j \ra_c^{\rm lin} + \la \Delta'_i\Delta'_j \ra_c^{\rm 1-loop} \rb 
\bigg(
 1 - \lambda_1\lambda_2 \xi_L
 - (\lambda_1 + \lambda_2 )k_l U^{\rm lin}_{10,l}  \ldots \bigg) \\
& \hspace{1.5cm} = f^2 \bigg(
 A_{ij}^{\rm lin} + 4 \lb A_{ij,22}^{\rm 1-loop}+\tfrac{3}{4} A_{ij,13}^{\rm 1-loop} \rb
 - \lambda_1\lambda_2 \xi_L A_{ij}^{\rm lin}
 - (\lambda_1 + \lambda_2 )k_l U^{\rm lin}_{10,l}  A_{ij}^{\rm lin}\ldots \bigg) \non
}
For the second term ($n=1$) the contributions up to 1-loop are:
\eeq{
i \la Y \Delta'_i\Delta'_j \ra_c
=f^2 \bigg( 2 i (\lambda_1 +\lambda_2 ) A_{ij}^{10} 
+   i k_l \lb 2 W_{ijl} - W^{(112)}_{ijl}  \rb \bigg). 
}
For the third term ($n=2$) the contribution up to 1-loop are:
\eq{
& - \frac{1}{2} \lb 2 \la Y \Delta'_i \ra_c \la Y \Delta'_j \ra_c +
\la Y^2 \Delta'_i\Delta'_j \ra_c  \rb \non\\
&= - f(\tau)^2 \bigg(  ( \lambda_1 + \lambda_2)^2 U^{\rm lin}_{10,i}U^{\rm lin}_{10,j}
 + (\lambda_1 + \lambda_2) k_n  \lb  A^{\rm lin}_{in} U^{\rm lin}_{10,j} + A^{\rm lin}_{jn}U^{\rm lin}_{10,i} \rb
 +  k_n k_m A^{\rm lin}_{in} A^{\rm lin}_{jm} + \ldots \bigg). \non
}
The Fourier space representation of the velocity dispersion is then
\eq{
\widetilde \Xi_{ij}(k) &= f^2
\int d^3 q ~ e^{ i \vec k \cdot \vec q} e^{- \frac{1}{2} k_ik_j A_{{\rm lin}, ij}}  \bigg( 
A_{ij}^{\rm lin} + 4 \lb A_{ij,22}^{\rm 1-loop}+\tfrac{3}{4} A_{ij,13}^{\rm 1-loop} \rb + i k_l \lb 2 W_{ijl} - W^{(112)}_{ijl}  \rb \non\\
&\hspace{1cm}  -  k_n k_m A^{\rm lin}_{in} A^{\rm lin}_{jm}
+  b_{\df,\{ab\}} \lb i k_l U^{\rm lin}_{10,l}  A_{ij}^{\rm lin} +  2 A_{ij}^{10} + 2i k_n  A^{\rm lin}_{in} U^{\rm lin}_{10,j} \rb \non\\
& \hspace{1cm}  +  b_{\df,\{ab\}} \lb \xi_L A_{ij}^{\rm lin}  + 2 U^{\rm lin}_{10,i}U^{\rm lin}_{10,j} \rb + b_{\df^2,\{ab\}}  U^{\rm lin}_{10,i}U^{\rm lin}_{10,j} 
 + \ldots \bigg).
}
Individually, in terms of components, this gives
\eq{
\widetilde \Xi^{(0)}_{2}(k) &= f^2 \Big[ \widetilde \Xi^{(0)}_{2,{\rm loop}} +\alpha^{(0)}_{2} \widetilde \Xi^{(0)}_{2, \rm c.t.} + b_{\df,\{ab\}} \widetilde \Xi^{(0)}_{2, \df} 
                                              + b_{\df,a} b_{\df,b} \widetilde \Xi^{(0)}_{2, \df\df} + b_{\df^2,\{ab\}} \widetilde \Xi^{(0)}_{2, \df^2} \Big], \non\\
\widetilde \Xi^{(2)}_{2}(k) &=  f^2 \Big[ \widetilde \Xi^{(2)}_{2,{\rm loop}}+\alpha^{(2)}_{2} \widetilde \Xi^{(2)}_{2, \rm c.t.}  + b_{\df,\{ab\}} \widetilde \Xi^{(2)}_{2, \df}
                                              + b_{\df,a} b_{\df,b} \widetilde \Xi^{(2)}_{2, \df\df} + b_{\df^2,\{ab\}} \widetilde \Xi^{(2)}_{2, \df^2} \Big],
}
where $\alpha^{(0)}_{2}$ and $\alpha^{(2)}_{2}$ are the leading counterterm coefficients.
In configuration space we analogously have
\eq{
\Xi^{(0)}_{2}(r) &= f^2 \Big[ \Xi^{(0)}_{2,{\rm loop}} +\alpha^{(0)}_{2} \Xi^{(0)}_{2, \rm c.t.} + b_{\df,\{ab\}} \Xi^{(0)}_{2, \df}
                                              + b_{\df,a} b_{\df,b} \Xi^{(0)}_{2, \df\df} + b_{\df^2,\{ab\}} \Xi^{(0)}_{2, \df^2} \Big], \non\\
\Xi^{(2)}_{2}(r) &=  f^2 \Big[ \Xi^{(2)}_{2,{\rm loop}}+\alpha^{(2)}_{2} \Xi^{(2)}_{2, \rm c.t.}  + b_{\df,\{ab\}} \Xi^{(2)}_{2, \df}
                                              + b_{\df,a} b_{\df,b} \Xi^{(2)}_{2, \df\df} + b_{\df^2,\{ab\}} \Xi^{(2)}_{2, \df^2} \Big].
}
For both Fourier and configuration space these individual terms are
given explicitly in Appendix \ref{app:explicitPS_CF}.

\section{Application to the bispectrum}
\label{sec:bispectrum}

Our focus so far has been on 2-point statistics, which form a complete
description of zero-mean Gaussian fields.  However there is great interest
in non-Gaussian statistics, either primordial or those which evolve due to
non-linear structure formation.  In this section we show how the formalism
developed above can be applied to higher-order functions.

Historically it has been hard to handle redshift-space distortions for
higher order functions, due to the plethora of vectors involved
\cite{Ber02}.  However, since the signal-to-noise ratio and information in
the 3D N-point functions is larger than in their projected counterparts,
but the 3D versions can only be measured in redshift space, there is ample
motivation to investigate this problem.
The new Fourier cumulant expansion is particularly interesting in this
regard, as it provides a relatively straightforward way to implement the
redshift-space mapping and even a helpful bookkeeping device for organizing
the terms.

The redshift-space mapping given in Eq.~\eqref{eqn:delk_def} also allows us
to directly predict the higher N-point functions.  For example, the 3-point
function in Fourier space (the bispectrum) can be derived from
\eq{
  \widetilde{\mathcal{M}}^{abc}(\vec J_1, \vec J_2;\vec k_1, \vec k_2) &=
  \frac{k_1^3k_2^3}{4\pi^4}\int d^3r_{12}\ e^{i\vec{k}_1\cdot\vec{r}_1+i\vec{k}_2\cdot\vec{r}_2} \non\\
&\qquad~~~ \times  \left\langle(1+\delta_a(\vec{x}))(1+\delta_b(\vec{x}'))(1+\delta_c(\vec{x}''))
  e^{i \vec{J}_1 \cdot\Delta u_{ac} + i \vec{J}_2 \cdot\Delta u_{bc}}\right\rangle,
}
where $d^3r_{12} = d^3r_{1}d^3r_{2}$, and where the three-point moment
generating function is
\eq{
1 + \mathcal{M}^{abc}(\vec J_1, \vec J_2;\vec{r}_1,\vec{r}_2) =  \left\langle(1+\delta_a(\vec{x}))(1+\delta_b(\vec{x}'))(1+\delta_c(\vec{x}''))
  e^{i \vec{J}_1 \cdot\Delta u_{ac} + i \vec{J}_2 \cdot\Delta u_{bc}}\right\rangle .
}
From the structure above it is relatively straightforward to see how to
generalise it to the higher N-point functions.  This could be of interest
if one would like to investigate the non-Gaussian part of the redshift-space
power spectrum covariance matrix.

\subsection{Moment expansion}

We expand the exponential term in $\mathcal{M}$ in order to obtain the three point
(density weighted) moments of the velocity field:
\eq{
  \Xi_{i_1\ldots i_n,j_1\ldots j_m}(\vec{r}_1,\vec{r}_2) &=
  \left\langle \big(1 + \df_a(\vec{x})\big)\big(1+\df_b(\vec{x}')\big)(1+\delta_c(\vec{x}'')) \right. \non\\
&\hspace{4cm} \times \left. \Delta u_{ac,i_1}\ldots\Delta u_{ac,i_n}\Delta u_{bc,j_1}\ldots\Delta u_{bc,j_n}\right\rangle .
}
We can write the bispectrum as the sum of the Fourier transforms of
these moments,
\eq{
 B^{abc}_s(\vec{k}_1, \vec{k}_2)
  &= \sum_{n,m=0}^\infty \frac{i^{n+m}}{n!m!} k_{1,i_1} \ldots k_{1,i_n} k_{2,j_1} \ldots k_{2,j_m}
  \widetilde \Xi_{i_1\ldots i_n,j_1\ldots j_m}(\vec{k}_1,\vec{k}_2) \\
  &= \sum_{n,m=0}^\infty \frac{i^{n+m}}{n!m!} k_{1,i_1} \ldots k_{1,i_n} k_{2,j_1} \ldots k_{2,j_m}
  \int d^3r_{12}\ \Xi_{i_1\ldots i_n,j_1\ldots j_m}(\vec{r}_1,\vec{r}_2) e^{i\vec{k}_1\cdot\vec{r}_1+i\vec{k}_2\cdot\vec{r}_2}. \non
}
This is analogous to our moment expansion for the power spectrum given in Eq.~\eqref{eq:PS_momentsum}.
This approach would be equivalent to any version of direct Eulerian
perturbation theory approaches to bispectrum in redshift space, as
is also the case for the power spectrum.

\subsection{Fourier space cumulant expansion}

In Fourier space we can perform a similar cumulant expansion as was done
for the power spectrum, i.e.~we have
$\widetilde{\mathcal{Z}}^{abc}(\vec{J}_1,\vec{J}_2; \vec{k}_1,\vec{k}_2) =
   \ln \big[1 + \widetilde{\mathcal{M}}^{ab}(\vec{J}_1,\vec{J}_2; \vec{k}_1,\vec{k}_2) \big]$
and expanding in $\vec{J}_1$ and $\vec{J}_2$ we get the three point cumulants. First few of them are
\begin{gather}
\widetilde{\mathcal C}^{(0)}( \vec k_1,  \vec k_2 ) = \ln \big[ 1+\Delta^2_{abc}( \vec k_1,  \vec k_2 ) \big], \\
\widetilde{\mathcal C}^{(1)}_{i}( \vec k_1,  \vec k_2 ) = \widehat{\Xi}_i( \vec k_1,  \vec k_2 )/\big[ 1+\Delta^2_{abc} \big], ~~~
\widetilde{\mathcal C}^{(1)}_{j}( \vec k_1,  \vec k_2 ) = \widehat{\Xi}_j( \vec k_1,  \vec k_2 )/\big[ 1+\Delta^2_{abc} \big], \non\\
\widetilde{\mathcal C}^{(2)}_{i_1i_2}( \vec k_1,  \vec k_2 ) = \widehat{\Xi}_{i_1i_2}(\vec k)/\big[ 1+\Delta^2_{ab} \big] - \widetilde{C}^{(1)}_{i_1}\widetilde{C}^{(1)}_{i_2} , ~~
\widetilde{\mathcal C}^{(2)}_{j_1j_2}( \vec k_1,  \vec k_2 ) = \widehat{\Xi}_{j_1j_2}(\vec k)/\big[ 1+\Delta^2_{ab} \big] - \widetilde{C}^{(1)}_{j_1}\widetilde{C}^{(1)}_{j_2} , \non\\
\widetilde{\mathcal C}^{(2)}_{i_1j_1}( \vec k_1,  \vec k_2 ) = \widehat{\Xi}_{i_1j_1}(\vec k)/\big[ 1+\Delta^2_{ab} \big] - \widetilde{C}^{(1)}_{i_1}\widetilde{C}^{(1)}_{j_1} , \non
\end{gather}
where we use the notation $\Delta^2=k_1^3k_2^3 B(\vec k_1,\vec k_2)/(4\pi^4)$
and $\widehat{\Xi}=k_1^3k_2^3 \widetilde{\Xi}(\vec k_1,\vec k_2)/(4\pi^4)$
in analogy to the power spectrum case.
In analogy to what we had before, we can introduce the three point kernel
\begin{equation}
  \ln \widetilde{\mathcal{K}}^{abc}(\vec{k}_1,\vec{k}_2) =
  \sum_{n+m=1}^\infty \frac{i^{n+m}}{n!m!} k_{1,i_1} \ldots k_{1,i_n} k_{2,j_1} \ldots k_{2,j_m}
  \widetilde{\mathcal C}^{(n+m)}_{i_1\ldots i_n,j_1\ldots j_n} (\vec{k}).
\end{equation}
We note that the translation kernel in this case depends on $\vec k$ only
and not on $\vec r$ as was the case above.
This is significant since in order to compute the power spectrum no additional
Fourier transform is needed.
It should be clear that the kernel $\widetilde{\mathcal{K}}^{abc}$
is not simply the Fourier transform of $\mathcal{K}^{abc}$, and
neither do $\widetilde{\mathcal{Z}}^{abc}$ and
$\mathcal{Z}^{abc}$ form a Fourier transform pair.

The redshift-space bispectrum can thus be written
\begin{align}
 \label{eq:Fourer_bis}
 1+ \Delta^2_{s, abc} & (\vec{k}_1,\vec{k}_2) =
  \exp\left[ \widetilde{\mathcal{Z}}^{abc}(\vec{k}_1,\vec{k}_2; \vec{k}_1,\vec{k}_2) \right]  \\
  &= \big[1+\Delta^2_{abc}\big]\widetilde{\mathcal{K}}^{abc}(\vec{k}_1,\vec{k}_2) \non \\
  &=  \big[1+\Delta^2_{abc}\big]
  \exp \Bigg[ \sum_{n+m=1}^\infty \frac{i^{n+m}}{n!m!} k_{1,i_1} \ldots k_{1,i_n} k_{2,j_1} \ldots k_{2,j_m}
  \widetilde{\mathcal C}^{(n+m)}_{i_1\ldots i_n,j_1\ldots j_n} (\vec{k}_1,\vec{k}_2) \Bigg]~. \non
\end{align}
The structure of this expression is again similar to the power spectrum case
(see Eq.~\ref{eq:Fourier_streaming_power_spectrum}).
We see that in this Fourier space representation all the RSD effects are
contained in a kernel, $\widetilde{\mathcal{K}}$, that has a simple form
as a sum of cumulants.
Given this structure it might be appealing to study the properties of this
expansion in the form of the observables, where we take a log of the ratio
of bispectra, i.e.~$\log \lb \big[1+ \Delta^2_{s, abc}\big]/\big[1+\Delta^2_{abc}\big] \rb$.

\section{Conclusions}
\label{sec:conclusions}

We have investigated the use of several expansions of the
real-to-redshift-space mapping, with a focus on the power spectrum and
correlation function.
We reviewed the velocity moment expansion approach and the
configuration space cumulant expansion.
We also presented a novel, Fourier-based streaming model, characterized by
a simple, algebraic, form and rapid convergence.
We showed how to systematically extend the evaluation of all of these
approaches in both Fourier and configuration space, in a manner that is
independent of the way the respective ingredients are computed.
This gives an efficient algorithm for computing the redshift-space correlation
function and power spectrum, which can be made arbitrarily accurate for a
given dynamics. The ingredients can be supplied either by perturbation theory (taking care
of the consistent expansion in a given parameter) or from some other means,
e.g.\ fits to N-body simulations or emulators.
Some of the relations we derived give support to earlier, phenomenological
models for redshift-space power spectra while showing how to extend the
approximations in a controlled manner.
In the follow-up work we intend to perform the comparative study 
of this framework using the N-body simulations and PT results presented
in Sec. \ref{sec:bias}

Apart from this, we presented the first, complete computation of the
redshift-space power spectrum within the Zeldovich approximation, that
we then used as a toy model with which to test the convergence properties
of moment expansion and the two streaming models.
As part of this calculation we also evaluated the arbitrary-order velocity
moments (and subsequently cumulants) within the Zeldovich approximation.
This allowed us to perform detailed convergence and performance studies of
these models at various redshifts and configurations.

All of the expansion schemes work well at large scales and agree to high
precision.  Depending on the order of truncation of the expansion, agreement
gradually deteriorates as we go to smaller scales.
We found that in the power spectrum, the Fourier based streaming model
performed best when $L_{\rm max}=4$ or $5$, and for lower $L_{\rm max}$
the performance was similar to other models.
For the correlation function, and at low order ($L_{\rm max}=2$), the
configuration-space streaming model performed best over all scales,
while for higher orders ($L_{\rm max}>2$) the Fourier streaming model
and moment expansion improved quickly and became comparable.
The agreement between all of the schemes and the full expression was, as
expected, worst for the modes along the line of sight and best for the
modes transverse to the line of sight.
The size of the error scaled as a relatively high power of the (cosine of the)
angle to the line of sight, $\nu$.  This suggests that comparisons between
data and an expansion truncated at any finite order could be enhanced by the
use of statistics which downweight the line-of-sight modes compared to
traditional multipole expansions.  Alternatively, it suggests that lower
order perturbative models are not applicable to certain situations where high
accuracy for $\nu\approx 1$ is required.

While the formalism is much more general, our numerical comparisons employed
an approximate dynamics (the Zeldovich approximation) and are considering only unbiased
tracers.  The reader should exercise caution in assessing the absolute
numerical convergence of any of these schemes in more realistic scenarios
involving full nonlinear, tracer dynamics.  However, we expect the trends
and relative behaviors to be quite robust.

Being perturbative, all of the expansions perform better at high redshift
where the expansion parameters are small. In this regard, it is worth noting
that the relevant expansion parameter for redshift-space effects is $fD$,
with $f$ the growth rate and $D$ the linear growth factor.  In currently
favored cosmologies this actually peaks near $z\simeq 0.5$, and falls more
slowly than $D$ to earlier times.  The relative improvement of perturbative
schemes with redshift are thus expected to be worse for redshift-space
statistics than real-space statistics.

One of the interesting features of our newly developed Fourier-space streaming
model is that the relation between the redshift-space and real-space power
spectra is analytic (in a manner reminiscent of phenomenological dispersion
models).  In fact directly from Eq.~\eqref{eq:Fourier_streaming_power_spectrum}
we see
\begin{equation}
  \ln\frac{1+\Delta^2_s(k,\nu)}{1+\Delta^2(k)} =
  i(\nu k)\widetilde{C}_{\hat{n}}^{(1)}(k,\nu)
  -\frac{(\nu k)^2}{2}\widetilde{C}_{\hat{n}}^{(2)}(k,\nu) + \cdots ,
\end{equation}
where we note again that $\nu$ is the usual cosine of the angle to the
line of sight and $\widetilde{C}^{(\ell)}$ are velocity two-point cumulants
given in Eq.~\eqref{eq:k-cums}.
In principle, the left-hand side can be measured from data, and the
velocity cumulants (including the finger-of-god terms) can be inferred from
the angle and scale dependence of the result.
This newly developed Fourier-space streaming model also provides a simple
framework for applying RSD effects to higher N-point functions.
In Eq.~\eqref{eq:Fourer_bis} we give a simple cumulant expansion for the
redshift-space bispectrum that follows the same structure as the power
spectrum expression above. 

While our numerical comparisons focus on matter statistics, we review the
more realistic scenario of biased tracers and nonlinear dynamics
(up to 1-loop in Lagrangian perturbation theory).
We give the explicit expressions for the configuration and Fourier space
two point functions, as well as pairwise velocity and velocity dispersion.
These ingredients are equivalent to ones presented in Ref.~\cite{VlaCasWhi16} 
in configuration space for studying the redshift space correlation function. 
Together with our new Fourier version of the streaming model these 
ingredients give an elegant and practical description of the redshift space power spectrum.

Finally we note that redshift-space distortions form just one example of a
``shifted'' field, in which the object is displaced from its true position.
Other examples of shifted fields arise in the context of initial condition
reconstruction or density field reconstruction \cite{ESSS07} for baryon
acoustic oscillations \cite{Wei13,PDG18} and CMB lensing \cite{LewCha06,Han10}.
During reconstruction objects are deliberately displaced during the data
analysis in order to reduce the impact of non-linear evolution on the
measurement of the distance scale.  In CMB lensing the photon's angular
positions are remapped by gravitational deflections along their path to the
observer.  However there are many aspects of these examples which are similar,
and a unified treatment is both possible and desirable.
We intend to return to this in a future publication.

\acknowledgments
We would like to thank E. Castorina, T. Fujita, V. Desjacques, M. Ivanov, A. Raccanelli, F. Schmidt, U. Seljak, S. Sibiryakov 
and M. Simonovi\' c for useful discussions during the preparation of this manuscript.
M.W.~is supported by the U.S. Department of Energy and the NSF.
This research has made use of NASA's Astrophysics Data System.

\appendix

\section{Angle decomposition of velocity moments.}
\label{sec:and_decomp_ofvelocities}

Following the angular decomposition procedure pioneered in \cite{SelMcD11, Vla13} 
in this appendix we decompose the pairwise velocity moments showing the
angular structure only based on the rotation symmetries and independently of the perturbative arguments.
Pairwise velocity moments are Eulerian quantities that are defined as
\eeq{
\Xi_{i_1,\ldots,i_N}(\vec{r}) = \la \big(1 + \df(\vec{x})\big)\big(1+\df(\vec{x}')\big) \Delta u_{i_1}\ldots\Delta u_{i_N} \ra,
}
where $\Delta u_i = u_i(\vec x') - u_i(\vec x)$.
Given that we are interested in projections along the line of sight $\hat n$ we have
\eq{
\Xi^{(N)}_{\hat n}(\vec{r}) &= \la \big(1 + \df(\vec{x})\big)\big(1+\df(\vec{x}')\big) \Delta u_{\hat n}^N \ra \non\\
&= \sum_{n=0}^N (-1)^n \binom{N}{n} \la \vec T^{(N-n)}_{\hat n}(\vec x') \vec T^{(n)}_{\hat n}(\vec x)  \ra,
}
where
\eeq{
 \vec T^{(L)}_{\hat n}(\vec x) = \big(1 + \df(\vec{x})\big) u^{L}_{\hat n}(\vec x) .
 }
Fourier transform of these moments can be decomposed as
\eeq{
 \vec T^{(L)}_{\hat n}(\vec k) = \sum_{l = L, L-2, \ldots} \sum_{m= -l}^l T_l^{L,m}(k) Y_{lm}(\hat k).
}
Two point correlation function is than
\eeq{
\la \vec T^{(L)}_{\hat n}(\vec k) | \vec T^{(L')}_{\hat n}(\vec k')  \ra
=  \sum_{\substack{l = L, L-2, \ldots \\ l' = L', L'-2, \ldots }} \sum_{m= -{\rm min}(l,l')}^{{\rm min}(l,l')} \la T_l^{L,m}(k) | T_{l'}^{L',-m}(k') \ra Y_{l m}(\hat k)  Y_{l' -m}(\hat k'),
}
and the power spectrum is then given as
\eeq{
P_{LL'} (k, \mu) =  \sum_{\substack{l = L, L-2, \ldots \\ l' = L', L'-2, \ldots }} \sum_{m= -{\rm min}(l,l')}^{{\rm min}(l,l')} P^{L,L', m}_{l, l'}(k) \mathcal P_l^m(\mu) \mathcal P_{l'}^{-m}(\mu),
}
where, omitting the Dirac delta functions, we used straightforward definitions $\la \vec T^{(L)}_{\hat n} | \vec T^{(L')}_{\hat n} \ra' = P_{LL'}$ and 
$\la T_l^{L,m} | T_{l'}^{L',-m}\ra' = P^{L,L', m}_{l, l'}$ .
We can note that $P_{LL'}  = P_{L' L}^* $, so, without loss of generality, we can assume $L \leq L'$ and get
\eeq{
P_{LL'} (k, \mu) =  \sum_{\substack{l = L, L-2, \ldots \\ l' = L', L'-2, \ldots }} \sum_{m= -l}^{l} P^{L,L', m}_{l, l'}(k) \mathcal P_l^m(\mu) \mathcal P_{l'}^{-m}(\mu).
}
Here we can use the standard relation for the Wigner rotation matrices
\eeq{
\mathcal D^{(j_1)}_{m_1m'_1}(R) \mathcal D^{(j_2)}_{m_2m'_2}(R) = \sum_{j, m, m'} \la j_1 j_2; m_1 m_2 | j_1 j_2; j m \ra \la j_1 j_2; m'_1 m'_2 | j_1 j_2; j m' \ra \mathcal D^{(j)}_{m m'} (R),
}
when we set $m'_1= 0$ and $m'_2 = 0$ and using $\mathcal D^{(j)}_{m0} = Y^{m*}_{l}$ we have
\eeq{
\mathcal P^{m}_{l}(\mu) \mathcal P^{-m}_{l'}(\mu) = \sqrt{\frac{(l+m)! (l'-m)!}{(l-m)! (l'+m)!}} \sum_{\ell} \la l l'; m -m | l l' ; \ell 0 \ra \la l l' ; 0 0 | l l'; \ell 0 \ra \mathcal P_{\ell}(\mu).
}
Collecting this gives the velocity moment spectra in therm of a sum of just one Legendre polynomial 
\eeq{
P_{LL'} (k, \mu) =  \sum_{\substack{l = L, L-2, \ldots \\ l' = L', L'-2, \ldots }} \sum_{\ell = | l -l' |}^{l+l'}  \la l l' ; 0 0 | l l'; \ell 0 \ra  C^{L,L', \ell}_{l, l'}(k) \mathcal P_{\ell}(\mu),
}
where the scale dependent part is re-expressed as a sum of $P^{L,L', m}_{l, l'}$ spectra as
\eeq{
 C^{L,L', \ell}_{l, l'}(k)  =  \sum_{m= -l}^{l}  \la l l'; m -m | l l' ; \ell 0 \ra \sqrt{\frac{(l+m)! (l'-m)!}{(l-m)! (l'+m)!}} P^{L,L', m}_{l, l'}(k) .
}
Finally for the $N-$th pairwise velocity moment  we then have
\eq{
\Xi^{(N)}_{\hat n}(\vec{k})
&= \sum_{n=0}^N (-1)^n \binom{N}{n} P_{n, N-n} (k, \mu) \non\\
&= (-1)^{N/2} \df^K_{N/2, \floor{N/2} }  \binom{N}{N/2} P_{N/2, N/2} (k, \mu) \non\\
&\qquad +  \sum_{n<N/2} (-1)^n \binom{N}{n} \lb P_{n, N-n} (k, \mu) + (-1)^{N} P^*_{n, N-n} (k, \mu) \rb .
}
These can be separated in the odd and even moments to make the angular dependence more explicit. 
We have for each contribution
\eq{
\Xi^{(2N)}_{\hat n}(\vec{k}) &= (-1)^{N} \binom{2N}{N} P_{N, N} (k, \mu) + 2  \sum_{n<N} (-1)^n \binom{2N}{n} {\rm Re} \Big[ P_{n, 2N-n} (k, \mu) \Big] \non\\
&= \sum_{n = 0}^N P_{2n}(k) \mathcal P_{2n} (\mu), \non\\
\Xi^{(2N+1)}_{\hat n}(\vec{k}) &= 2 i \sum_{n\leq N} (-1)^n \binom{2N+1}{n}  {\rm Im} \Big[ P_{n, 2N-n+1} (k, \mu) \Big] \non\\
&= \sum_{n = 0}^N P_{2n+1}(k) \mathcal P_{2n+1} (\mu).
}
Thus we see that $N-$th moment has contributions of only either odd or even Legendre polynomials,
up to the $N-$th order. Thus in the $N-$th moment all odd or even powers of angle $\mu$ appear 
up to the $\mu^N$. It is thus interesting to note that even though we used the, linear approximation 
for displacement, this Zeldovich approximation, still generated the full RSD angle complexity 
as we see in Sec.~\ref{sec:mom_exp}. This is, of course, not so in the direct Eularian PT 
approaches that, at face value, do not exhibit any resumation of the IR modes \cite{SenZal15,VlaWhiAvi15} (for Eulerian 
based resumation of IR modes e.g. see Ref.~\cite{Blas2016} and in redshift space Ref.~\cite{Ivanov2018}). 

First few examples of these pairwise moments given in terms of the simple velocity moments as given in \cite{SelMcD11, Vla13}
\eq{
\Xi^{(1)}_{\hat n}(\vec{k}) &= P_{01}(k,\mu) - P^*_{01}(k,\mu) \non\\
&= 2 i {\rm Im}[P_{01}(k,\mu)], \non\\
\Xi^{(2)}_{\hat n}(\vec{k}) &= P_{02}(k,\mu) - 2 P_{11}(k,\mu) + P^*_{02}(k,\mu) \non\\
&= 2 {\rm Re}[P_{02}(k,\mu) - P_{11}(k,\mu)], \non\\
\Xi^{(3)}_{\hat n}(\vec{k}) &= P_{03}(k,\mu) - 3 P_{12}(k,\mu) + 3 P^*_{12}(k,\mu) - P^*_{03}(k,\mu) \non\\
&= 2 i {\rm Im}[P_{03}(k,\mu) - 3 P_{12}(k,\mu)], \non\\
\Xi^{(4)}_{\hat n}(\vec{k}) &= P_{04}(k,\mu) - 4 P_{13}(k,\mu) + 6 P_{22}(k, \mu) - 4 P^*_{13}(k,\mu) + P^*_{04}(k,\mu) \non\\
&= 2 {\rm Re} [P_{04}(k,\mu) - 4 P_{13}(k,\mu) + 3 P_{22}(k, \mu) ].
}
This provides the direct link from these moment based approaches to the various streaming approaches (e.g \cite{ReiWhi11,Rei12,WanReiWhi14,VlaCasWhi16}).

\section{Derivation of general velocity moments in Zeldovich approximation}
\label{sec:red_vel_moments}

In the section \ref{sec:mom_exp} we have derived the velocity moments in the Zeldovich approximation.
We have shown that these are given in therms of reduced velocity moments that are implicitly defined by expression \eqref{eq:vel_moments}.
In this section we derive the explicit expressions for these reduced velocity moments. They can be defined via the RSD angle $\nu$ derivatives
of the velocity moments
\eq{
\widetilde \Xi^{(2\ell)}_{m} (k) &= \frac{1}{m!} \lb \frac{d}{d \nu^2} \rb^m  \Bigg[ \sum_{n=0}^{\ell} \frac{(-1)^n}{2^{\ell + n}} \frac{(2\ell)!}{(\ell - n)! (2n)!}
  \int d^3q\ e^{i\vec{k}\cdot\vec q}~ \mathcal A^{2n} \mathcal B^{\ell -n} e^{ - \frac{1}{2} k_i k_j A_{ij} } \Bigg]_{\nu=0} \non\\
  &= \sum_{n=0}^{2\ell}  \int d^3q\ e^{i\vec{k}\cdot\vec q}~ F^{(2\ell)}_{m,n} (X,Y) \mu^{2n} e^{ - \frac{1}{2} k_i k_j A_{ij} }, \non\\
\widetilde \Xi^{(2\ell+1)}_{m} (k) &= \frac{1}{m!} \lb \frac{d}{d \nu^2} \rb^m \frac{1}{\nu}  \Bigg[ \sum_{n=0}^{ \ell } \frac{(-1)^n}{2^{\ell + n+1}} \frac{(2\ell+1)!}{(\ell - n)! (2n+1)!}
 \int d^3q\ e^{i\vec{k}\cdot\vec q}~\mathcal A^{2n+1} \mathcal B^{\ell -n} e^{ - \frac{1}{2} k_i k_j A_{ij} }  \Bigg]_{\nu=0} \non\\
   &= \sum_{n=0}^{2\ell+1}  \int d^3q\ e^{i\vec{k}\cdot\vec q}~ F^{(2\ell+1)}_{m,n} (X,Y) \mu^{2n} e^{ - \frac{1}{2} k_i k_j A_{ij} },
}
where further angular dependence in $\mu^2$ angles is stripped from the integrands and the remaining kernels $F^{(2\ell)}_{m,n}$ are given by
\eq{
F^{(2\ell)}_{m,n} (X,Y) &=  \frac{1}{n!m!} \lb \frac{d}{d \mu^2} \rb^n \lb \frac{d}{d \nu^2} \rb^m  \Bigg[ \sum_{i=0}^{\ell} \frac{(-1)^i}{2^{\ell + i}} \frac{(2\ell)!}{(\ell - i)! (2i)!}
 \int \frac{d \phi}{2\pi}  \mathcal A^{2i} \mathcal B^{\ell -i} \Bigg]_{\nu=\mu=0} \non\\
F^{(2\ell+1)}_{m,n} (X,Y) &= \frac{1}{n!m!} \lb \frac{d}{d \mu^2} \rb^n \lb \frac{d}{d \nu^2} \rb^m \frac{1}{\nu}  \Bigg[ \sum_{i=0}^{ \ell } \frac{(-1)^i}{2^{\ell + i+1}} \frac{(2\ell+1)!}{(\ell - i)! (2i+1)!}
\int \frac{d \phi}{2\pi}~\mathcal A^{2i+1} \mathcal B^{\ell -i}  \Bigg]_{\nu=\mu=0}
\label{eq:F_integral}
}
Using the integral given by Eq. \eqref{eq:int_mu_n} we can integrate over $\mu$ so we get the equations \eqref{eq:reduced_vec_moments}
for the reduced velocity moments
\eq{
\widetilde \Xi^{(2\ell)}_{m} (k)
&= 4\pi \sum_{s=0}^{\infty} \int q^2dq ~ e^{-\frac{1}{2} k^2 (X+Y)} \xi^{(2\ell)}_{m,s} (k, X,Y) \lb \frac{kY}{q} \rb^s j_s(qk),  \non\\
\widetilde \Xi^{(2\ell+1)}_{m} (k)
&= 4\pi \sum_{s=0}^{\infty} \int q^2dq ~ e^{-\frac{1}{2} k^2 (X+Y)} \xi^{(2\ell+1)}_{m,s} (k, X,Y) \lb \frac{kY}{q} \rb^s j_s(qk),
}
with integrands $\xi^{\ell}_{m,s}$ defined as in Eq \eqref{eq:xi_integrands}.

The goal is do derive explicit form for the $F^{(\ell)}_{m,n}$ kernels. The strategy is to consider the integral in Eq. \eqref{eq:F_integral}
and rearrange them in therms of powers of $\mu$ and $\nu$. This will allow us to take easily the derivative in angles.
But first we will remind ourselves what the is the notation; we use $\mathcal A = k_{\{ i} \hat n_{j\}} A_{ij} = 2 \nu k X + 2 \mu \gamma k Y$
and $\mathcal B = \hat n_i \hat n_j A_{ij} = X + \gamma^2 Y$ and $\gamma = \hat n. \hat q = \mu \nu + \eta \cos \phi$ and  $\eta = \sqrt{1-\mu^2}\sqrt{1-\nu^2}$.
We have
\eq{
 \int \frac{d \phi}{2\pi}~ \mathcal A^{2n} \mathcal B^{\ell -n} &= (2k)^{2n} \sum_{k=0}^{2\ell} {}_1G^{\ell, n}_k(\mu, \nu) (\mu \nu)^k\ { }_2F_1\Big( \frac{1-k}{2}, -\frac{k}{2}, 1, \frac{\eta^2}{\mu^2\nu^2} \Big) \non\\
 &=  (2k)^{2n} \sum_{q=0}^{2l}\sum_{p=0}^{l}  {}_1C^{\ell,n}_{2p,2q}\nu^{2p}  \mu^{2q},  \non\\
 \int \frac{d \phi}{2\pi}~ \mathcal A^{2n+1} \mathcal B^{\ell -n} &= (2k)^{2n+1} \sum_{k=0}^{2\ell+1} {}_2G^{\ell, n}_k(\mu, \nu) (\mu \nu)^k\ { }_2F_1\Big( \frac{1-k}{2}, -\frac{k}{2} , 1, \frac{\eta^2}{\mu^2\nu^2} \Big) \non\\
 &=  (2k)^{2n+1} \sum_{q=0}^{2l+1}\sum_{p=0}^{l}  {}_2C^{\ell,n}_{2p,2q}\nu^{2p}  \mu^{2q}
}
where ${ }_2F_1$ is the ordinary hypergeometric function and functions ${}_1G^{\ell, n}$ and ${}_2G^{\ell, n}$ are given by sums
\eq{
{}_1G^{\ell, n}_{k} (\mu, \nu) &= \sum_{m=0}^{k/2} \binom{2n}{k-2m} \binom{\ell - n}{m} X^{\ell + n-(k-m)} Y^{k-m} \nu^{2(n + m)-k} \mu^{k -2 m}, \non\\
{}_2G^{\ell, n}_{k} (\mu, \nu) &= \sum_{m=0}^{k/2} \binom{2n+1}{k-2m} \binom{\ell - n}{m} X^{\ell + n-(k-m)+1} Y^{k-m} \nu^{2(n + m)-k} \mu^{k -2 m}. \non
}
The goal is to extract the coefficients ${}_1C^{\ell,n}$ and ${}_2C^{\ell,n}$ from the formulae above.
To do this we need to first represent the hypergeometric function ${ }_2F_1$ as a series
\eeq{
(\nu \mu)^k { }_2F_1 \Big( (1-k)/2, -k/2, 1, (\eta/\mu\nu)^2 \Big) = \sum_{\substack{i=0\\ j=0}}^{\infty} (-1)^{i+j} c^k_{ij} \nu^{k-2i} \mu^{k-2j},
}
where the coefficient is given by
\eq{
c^k_{ij}&= \sum_{n=0}^{k} 4^{-n} \binom{2n}{n}\binom{k}{2n}\binom{n}{i}\binom{n}{j}
=  \frac{{}_3F_2 (1, (1 - k)/2, -k/2 ; 1-i, 1-j; 1)}{\Gamma( 1-i )\Gamma(1+i)\Gamma( 1-j )\Gamma(1+j)}.
}
Collecting this information we get for the ${}_1C^{\ell,n}$ and ${}_2C^{\ell,n}$ coefficients
\eq{
 {}_1C^{\ell,n}_{p,q} &= \sum_{\substack{r=0\\ s=0}}^{l} (-1)^{r+s} X^{l+n-q-s} Y^{q+s} \binom{l-n}{p+r-n} \binom{2 n}{n-p+q-r+s} c^{p + q + r + s - n}_{r,s}, \non\\
 {}_2C^{\ell,n}_{p,q} &= \sum_{\substack{r=0\\ s=0}}^{l} (-1)^{r+s} X^{l+n-q-s+1} Y^{q+s} \binom{l-n}{p+r-n} \binom{2 n+1}{n-p+q-r+s} c^{p + q + r + s - n}_{r,s}.
 }
Finally the kernels that we wanted to derive have a form
\eq{
F^{(2\ell)}_{p,q} (k,X,Y) &=  \sum_{n=0}^{\ell} \frac{(-1)^n}{2^{\ell - n}} \frac{(2\ell)!}{(\ell - n)! (2n)!} k^{2n} {}_1C^{\ell,n}_{p,q}(X,Y) \non\\
F^{(2\ell+1)}_{p,q} (k,X,Y) &=  \sum_{n=0}^{ \ell } \frac{(-1)^n}{2^{\ell - n}} \frac{(2\ell+1)!}{(\ell - n)! (2n+1)!} k^{2n+1} {}_2C^{\ell,n}_{p,q}(X,Y).
\label{eq:F_fncs}
}
Note that all the sums above are finite, unlike in the previous cases when the full RSD power spectra was computed in Sec \ref{sec:direct_lag}.

\section{Velocity moments up to $L=2$ for biased tracers}
\label{app:explicitPS_CF}

In this appendix section we give the supplementary formulas for the Sec.~\ref{sec:bias}.
All the bias and nonlinear dynamics terms that appear in the integrands of that sections are listed below. 
We first start with the non-linear term of the displacement field. 
The cumulant components can be decomposed as follows \cite{CLPT,VlaSelBal15}:
\eq{
  A_{ij}(q) &= \frac{2}{3}\delta_{ij}\left(\Xi_{0}(0)-\Xi_{0}(q)\right)
  + 2\left(\hat{q}_i\hat{q}_j-\frac{1}{3}\delta_{ij}\right)\Xi_{2}(q), \non\\
  W_{ij\ell}(q) &= \frac{2}{5}\hat{q}_{\{i}\delta_{j\ell\}}\Xi_{1}
  +\frac{3}{5}\left(5\hat{q}_i\hat{q}_j\hat{q}_\ell-
   \hat{q}_{\{i}\delta_{j\ell\}}\right)\Xi_{3},
\label{eqn:AWterms}
}
where we have
\eq{
  \Xi_{0}(q) &= \Xi^{\text{lin}}_{0}(q) +\Xi^{\text{loop}}_{0}(q) 
                    = \int\frac{dk}{2\pi^2} \left[  P_0(k) + \frac{9}{98}Q_1(k) + \frac{10}{21}R_1(k) \right]j_0(kq)  \non\\
  \Xi_{1}(q) &= \Xi^{\text{loop}}_{1}(q) 
                    = \int\frac{dk}{2\pi^2}\left(-\frac{3}{7k}\right) \left[ Q_1(k) - 3Q_2(k) + 2R_1(k) - 6R_2(k) \right]j_1(kq) \non\\
  \Xi_{2}(q) &= \Xi^{\text{lin}}_{2}(q) + \Xi^{\text{loop}}_{2}(q)
                    = \int\frac{dk}{2\pi^2} \left[  P_0(k) + \frac{9}{98}Q_1(k) + \frac{10}{21}R_1(k) \right]j_2(kq) \non\\
  \Xi_{3}(q) &= \Xi^{\text{loop}}_{3}(q) 
                    = \int\frac{dk}{2\pi^2}\left(-\frac{3}{7k}\right) \left[  Q_1(k) + 2Q_2(k) + 2R_1(k) + 4R_2(k) \right]j_3(kq),
\label{eq:Xi_1234}
}
In the rest of the section we also use the notation:
\eq{
X(q) &= \frac{2}{3} \left( \Xi_{0}(0) - \Xi_{0}(q) - \Xi_{2}(q) \right), \qquad Y(q) = 2 ~\Xi_{2}(q), \non\\
V(q) &= \frac{1}{5} \left( 2 \Xi_{1}(q) - 3 \Xi_{3}(q) \right), \qquad\qquad ~~T(q) = 3 ~\Xi_{3}(q).
\label{eq:XYVT}
}
Let us now give the explicit expression for the terms in the Eq. \eqref{eq:correlators}. 
We can split the term relative to the total spin they can carry. First we can consider 
the zero spin term: 
\eq{
\xi_L(q) = \la \df_{L1} \df_{L2} \ra_c &= \int  \frac{p^2 dp}{2 \pi^2} ~ P_L(p) j_{0}(pq), \non\\
\zeta_L(q) = \la s^2_{L1} s^2_{L2} \ra_c 
&= \int \frac{p^2 dp}{2 \pi^2} ~ Q_{s^2s^2}(p) j_{0}(pq) - \lb \tfrac{2}{3} \rb^2  [ \sigma_L^2 ]^2, \non\\
\chi^{11}(q) = \la s_{L1}^2 \df_{L1} \df_{L2} \ra_c &= 0, \non\\
\chi^{12}(q) = \la s_{L1}^2 \df_{L2}^2 \ra_c 
&= \int \frac{p^2 dp}{2 \pi^2} ~ Q_{s^2 \df^2}(p) j_{0}(pq) - \tfrac{2}{3} [ \sigma_L^2 ]^2,
}
Spin one terms are also given by the single scalar oriented in a given direction $\hat q$
\eq{
U^{10}  = \hat q_i U^{10}_i &= \hat q_i \la \df_L \Delta_i \ra_c = \hat q_i \la \df_1 \psi_{2,i} \ra_c
=  - \int \frac{k dk}{2 \pi^2} ~ \Big( P_{L}(k) +  R_{\df \Delta}(k) \Big) j_1( k q), \non\\
U^{11} = \hat q_i U^{11}_i & =  \hat q_i \la \df_{L1}\df_{L2} \Delta_i \ra_c
=  - \int \frac{k dk}{2 \pi^2} ~ R_{\df^2\Delta}(k) j_1( k q), \non\\
U^{20} = \hat q_i U^{20}_i &= \hat q_i \la \df_L^2 \Delta_i \ra_c
=  - \int \frac{k dk}{2 \pi^2} ~ Q_{\df^2\Delta}(k) j_1( k q), \non\\
V^{10} = \hat q_i V^{10}_i &= \hat q_i \la s^2_L \Delta_i \ra_c =  - \int \frac{k dk}{2 \pi^2} ~ Q_{s^2\Delta}(k) j_1( k q), \non\\
V^{11} = \hat q_i V^{11}_{i} &= \hat q_i \la s_{L1}^2 \df_{L1} \Delta_i \ra_c = 0, \non\\
V^{12} = \hat q_i V^{12}_{i} &= \hat q_i \la s_{L1}^2 \df_{L2} \Delta_i \ra_c =  - \int \frac{k dk}{2 \pi^2} ~ Q_{s^2\df \Delta}(k) j_1( k q),
}
Spin two terms can be split into two components, as $A_{ij} = \df^K_{ij} X + \hat q_i \hat q_j Y $. For these components we than have
\eq{
X^{10} & = \frac{1}{2}(\df^K_{ij} - \hat q_i \hat q_j) \la \df_L \Delta_i\Delta_j \ra_c
= \int \frac{k^2 dk}{2\pi^2} ~ \frac{1}{3} R_{10,0}^{(0)}(k) + \frac{1}{3} A_{10}^{(0)}(k) j_0(kq)  + \frac{1}{2}  A_{10}^{(2)}(k) j_2(kq) \non\\
&\hspace{4.5cm} = \int \frac{k^2 dk}{2\pi^2} ~ \frac{1}{3} R_{10,0}^{(0)}(k) + \frac{1}{3} \lb R_{10}^{(0)}(k) + Q_{10}^{(0)}(k) \rb j_0(kq) \non\\
&\hspace{8.1cm} + \frac{1}{2} \lb R_{10}^{(2)}(k) + Q_{10}^{(2)}(k) \rb j_2(kq), \non\\
Y^{10} & = \frac{3}{2}( \hat q_i \hat q_j - \tfrac{1}{3} \df^K_{ij}) \la \df_L \Delta_i\Delta_j \ra_c 
=  - \int \frac{k^2 dk}{2\pi^2} ~ \frac{3}{2} A_{10}^{(2)}(k) j_2(kq) \non\\
 & \hspace{4.7cm} =  - \int \frac{k^2 dk}{2\pi^2} ~ \frac{3}{2} \lb R_{10}^{(2)}(k) + Q_{10}^{(2)}(k) \rb j_2(kq), \non\\
X^{20} & = \frac{1}{2}(\df^K_{ij} - \hat q_i \hat q_j) \la s^2_L \Delta_i\Delta_j \ra_c
=  \int \frac{k^2 dk}{2\pi^2} ~ \frac{1}{3}  A_{20}^{(0)}(k) j_0(kq)  + \frac{1}{2}  A_{20}^{(2)}(k) j_2(kq) , \non\\
 & \hspace{4.5cm} = \int \frac{k^2 dk}{2\pi^2} ~ \frac{1}{3} Q_{20}^{(0)}(k) j_0(kq) 
 + \frac{1}{2} Q_{20}^{(2)}(k) j_2(kq), \non\\
Y^{20} & = \frac{3}{2}( \hat q_i \hat q_j - \tfrac{1}{3} \df^K_{ij}) \la s^2_L \Delta_i\Delta_j \ra_c
=  - \int \frac{k^2 dk}{2\pi^2} ~ \frac{3}{2} A_{20}^{(2)}(k) j_2(kq) \non\\
 & \hspace{4.8cm} =  - \int \frac{k^2 dk}{2\pi^2} ~ \frac{3}{2} Q_{20}^{(2)}(k) j_2(kq),
 }
 and we can add also the derivative terms
\eq{
\frac{\partial^2}{ k_L^2} \xi_L & =  \frac{1}{k_L^2} \int  \frac{p^2 dp}{2 \pi^2} ~ P_L(p) \partial^2 j_{0}(pq)  
= - \frac{1}{k_L^2} \int  \frac{p^2 dp}{2 \pi^2} ~ p^2 P_L(p) j_{0}(pq) ,\non\\
\frac{\partial^2}{ k_L^2}U^{10}_L &=- \frac{1}{k_L^2} \int \frac{k dk}{2 \pi^2} ~ P_{L}(k) \partial^2 j_1( k q)
= \frac{1}{k_L^2} \int \frac{p dp}{2 \pi^2} ~ p^2 P_L(p) j_{1}(pq).
} 
Note that seemingly divergent properties of the derivative terms above can be regularised by keeping only the leading $k_L^2$ contributions,
and truncating the small scales below $1/k_L$.
All the integrals above are given as spherical Bessel transforms of the integrands and thus can be easily evaluated using the FFTLog algorithm \cite{Ham00}.
For all the functions $Q_x$ and $R_x$ that come as the integrants in the expressions above, we refer the readier to the Ref.~\cite{Mat08a,Mat08b,CLPT}.
These are as well expressible as the The 1D integrals are Hankel transforms as shown in appendix of Ref.~\cite{Schmittfull2016a}.

\subsection{Halo power spectrum \& correlation function}

Final expressions, up to 1-loop, for the power spectrum and correlation
function are given in \S\ref{sec:two_points}.
Below we give explicit expressions for the individual contributions to 
Eq.~\eqref{eq:PS_real}, for the power spectrum, and Eq.~\eqref{eq:xi_real}
for the correlation function.
For additional biasing terms we have terms of the form
\begin{align}
P_{x}= 
4 \pi \int q^2 d q~ e^{ -  \frac{1}{2} k^2 (X_L + Y_L)} \Bigg( f^{(0)}_x(k,q) j_0(qk) + \sum_{n=1}^{\infty} f^{(n)}_x(k,q) \lb \frac{kY_L}{q} \rb^n j_n(qk)
\Bigg), 
\label{eq:PS_real_loop}
\end{align}
where $^{(n)}_x$ integrands are: 
\begin{center}
\begin{tabular} {  c | c c  }
$P_{x}$ & $f^{(0)}_x$  & $f^{(n)}_x$  \\ [0.2em] \hline \hline
$\text{zel} $, &
$1$, &
$1$,  \\[0.5 em]
$\text{loop}$, &
$- \frac{1}{2}k^2 \lb X^{\text{loop}} +Y^{\text{loop}} \rb $, &
\small{$- \frac{1}{2}k^2 \lb X^{\text{loop}} +Y^{\text{loop}} - q \frac{V^{\text{loop}} + \tfrac{1}{3}T^{\text{loop}}}{Y^{\text{lin}}} \rb + \frac{nY^{\text{loop}} }{Y^{\text{lin}} } - \frac{(n-1)q T^{\text{loop}} }{3 (Y^{\text{lin}})^2 }$},  \\[0.5 em]
$\df $, &
$- k^2 \big(X^{10}+ Y^{10}\big)$, &
$- k^2 \big(X^{10}+ Y^{10}\big) +2\lb nY^{10} - q U^{10} \rb/Y_L$,  \\[0.5 em]
$\df\df$, &
$\xi_L - k^2 (U^{10})^2$, &
$\xi_L - k^2 (U^{10})^2 + \lb 2n(U^{10})^2 - q U^{11} \rb /Y_L $,  \\[0.5 em]
$\df^2$, &
$- k^2 (U^{10})^2$, &
$- k^2 (U^{10})^2 + \lb 2n(U^{10})^2 - q U^{20} \rb /Y_L $,  \\[0.5 em]
$\df\df^2$, &
$0$, &
$- 2qU^{10}\xi_L/Y_L$,  \\[0.5 em]
$\df^2\df^2$, &
$\frac{1}{2}\xi_L^2$, &
$\frac{1}{2}\xi_L^2$,  \\[0.5 em]
$s^2$, &
$- k^2 \big(X^{20}+ Y^{20}\big)$, &
$- k^2 \big(X^{20}+ Y^{20}\big) +2 \lb nY^{20} - q V^{10} \rb / Y_L $,  \\[0.5 em]
$\df s^2$, &
$0$, &
$- 2 qV^{12}/Y_L$,  \\[0.5 em]
$\df^2 s^2$, &
$\chi^{12}$, &
$\chi^{12}$,  \\[0.5 em]
$s^2s^2$, &
$ \zeta_L$, &
$ \zeta_L$,  \\[0.5 em]
$\partial^2\df$, &
$ 0$, &
$ - 2q\partial^2 U^{10}/ (\Lambda_L^2Y_L)  $,  \\[0.5 em]
$\df \partial^2\df$, &
$ 2\partial^2\xi_L / \Lambda_L^2 $, &
$ 2\partial^2\xi_L / \Lambda_L^2 $,
\end{tabular}
\end{center}
The 1D integrals are Hankel transforms which can be done efficiently using
FFTs \cite{Ham00} as was shown in \cite{VlaSelBal15}.

Similar to the power spectrum it is useful to also give the explicit
expression for the 1-loop halo correlation function we have 
\eeq{
1 + \xi_{ab}(r) = \int d^3q ~ {\cal M}_{0,h}(\vec q, \vec r).
}
Using the abbreviation for the purely Gaussian part 
\eeq{
Q(\vec r -\vec q) = \frac{1}{(2\pi)^{3/2} |A_{\text{lin}}|^{1/2}} 
e^{-\tfrac{1}{2}(\vec r-\vec q)^T \vec{A}_{\text{ lin}}^{-1} (\vec r-\vec q)},
\label{eq:Q_definition}
}
we can write for individual contributions to Eq.~\eqref{eq:xi_real}
\eeq{
\xi_{x}(r) =  \int d^3q ~Q(\vec r -\vec q) F_x(\vec q), \non\\
}
where integrands $F_x$ can be tabulated as: 
\begin{center}
\begin{tabular} {  c | c | c | c  }
$\Xi_{x}$ & $F_x$  & $\Xi_{x}$ & $F_x$  \\ [0.2em] \hline \hline
$\text{zel}$, &
$1$, &
$\df^2\df^2$, &
$\frac{1}{2}\xi_L^2$,  \\[0.5 em]
$\text{loop}$, &
$ - \frac{1}{2} G_{ij} A_{ij}^{\text{loop}} + \frac{1}{6} \Gamma_{ijk}W_{ijk}^{\text{ lpt }}$, &
$s^2$, &
$ - 2 g_i V^{10}_i  - G_{ij} A^{20}_{ij} $,  \\[0.5 em]
$\text{c.t.}$, &
$  - \frac{1}{2} \text{Tr}[G_{ij}] $,  &
$\df s^2$, &
$- 2 g_i V^{12}_i$,  \\[0.5 em]
$\df $, &
$ - 2 g_i U^{10}_i  - G_{ij} A^{10}_{ij}$, &
$\df^2 s^2$, &
$\chi^{12}$,  \\[0.5 em]
$\df\df$, &
$ \xi_L - g_i U^{11}_i - G_{ij} U^{10}_i U^{10}_j  $,  &
$s^2s^2$, &
$ \zeta_L$,   \\[0.5 em]
$\df^2$, &
$ -g_i U^{20}_i - G_{ij} U^{10}_i U^{10}_j $ , &
$\partial^2\df$, &
$ - 2 g_i \partial^2 U^{10}_i / k_L^2 $,  \\[0.5 em]
$\df\df^2$, &
$-2 g_i U^{10}_i  \xi_L$, &
$2 \df \partial^2\df$, &
$2 \partial^2\xi_L / \Lambda_L^2 $,
\end{tabular}
\end{center}

\subsection{Pairwise velocity power spectrum and correlation function}

In similar way as that was used for the power spectrum, pairwise velocity 
power spectrum can be evaluated by performing the 1D Hankel transforms. 
In analogy to the Eq.~\eqref{eq:PS_real_loop} we can write
\eq{
k \widetilde \Xi^{01}_{x}(k)= 
4 \pi \int q^2 d q~ e^{ -  \frac{1}{2} k^2 (X_L + Y_L)} \Bigg( g^{(0)}_x(k,q) j_0(qk) + \sum_{n=1}^{\infty} g^{(n)}_x(k,q) \lb \frac{kY_L}{q} \rb^n j_n(qk)
\Bigg), 
}
where $g^{(n)}_x$ integrands can be tabulated as 
\begin{center}
\begin{tabular} {  c | c c  }
$\widetilde \Xi^{01}_{x}$ & $g^{(0)}_x$  & $g^{(n)}_x$  \\ [0.2em] \hline \hline
$\text{zel} $, &
$-k^2 (X^{\rm lin}+Y^{\rm lin})$, &
$k^2 (X^{\rm lin}+Y^{\rm lin}) - 2n$,  \\[0.5 em]
$\text{loop}$, &
$- 2 k^2 \lb X^{\rm loop}+Y^{\rm loop} \rb$, &
$ - 2 k^2 \lb X^{\rm loop}+Y^{\rm loop} - q \frac{V^{\text{loop}} + \tfrac{1}{3}T^{\text{loop}}}{Y^{\text{lin}}} \rb $ \\[0.5 em]
$~$ &
$~$ &
$+ 4 \lb \frac{nY^{\rm loop}}{Y_{\rm lin}} - \frac{(n-1)q T^{\text{loop}} }{3 (Y^{\text{lin}})^2} \rb $,  \\[0.5 em]
$\df$, &
$- 3 k^2 (X^{10} +Y^{10})$, &
$- 2 q\frac{\lb U^{\rm lin}_{10} + 3 U^{\rm loop}_{10} \rb}{Y_{\rm lin}} - 3 k^2 (X^{10} +Y^{10})$  \\[0.5 em]
$~$ &
$~$ &
$+2 \frac{qU^{\rm lin}_{10}}{Y_{\rm lin}} k^2 (X^{\rm lin} + Y^{\rm lin}) + 2 \frac{3n Y^{10} - 2(n-1)qU^{\rm lin}_{10}}{Y_{\rm lin}}$ ,  \\[0.5 em]
$\df\df$, &
$ - k^2 (X^{\rm lin} +Y^{\rm lin})\xi_L $ &
$- k^2 (X^{\rm lin} +Y^{\rm lin}) \xi_L - 2 q \frac{U^{11}}{Y_{\rm lin}}+ 2 n \xi_L$ \\[0.5 em]
$~$ &
$- 2k^2 \left[U^{\rm lin}_{10}\right]^2 $, &
$- 2 k^2 \left[U^{\rm lin}_{10}\right]^2 \lb 1  - \frac{2 n}{k^2 Y_{\rm lin}} \rb$,  \\[0.5 em]
$\df^2$, &
$ - 2 k^2 \left[U^{\rm lin}_{10}\right]^2 $, &
$ - 2 q U^{20} / Y_{\rm lin} - 2 k^2 \left[U^{\rm lin}_{10}\right]^2 \lb 1  - \frac{2 n}{k^2 Y_{\rm lin}} \rb $,  \\[0.5 em]
$\df\df^2$, &
$ 0 $, &
$ -2 \frac{q U^{\rm lin}_{10} \xi_L}{Y_{\rm lin}}$ .
\end{tabular}
\end{center}
Similarly the configuration space result is given as
\eeq{
\Xi^{01}_x (r) = (1+\xi(r)) v_{12,x}(r) =  \int d^3q ~Q(\vec r -\vec q) G_{x}(\vec q),
}
where we can again tabulated the configuration space integrands $G_{x}$ as
\begin{center}
\begin{tabular} {  c | c  }
$\Xi^{01}_{x}$ & $G_x$  \\ [0.2em] \hline \hline
$\text{zel}$, &
$ - \hat r_i g_j A^{\rm lin}_{ij}$,  \\[0.5 em]
$\text{loop}$, &
$ - 2 \hat r_i  g_j A^{\rm loop}_{ij}  + \frac{2}{3} \hat r_i  G_{jl} W_{ijl} $,  \\[0.5 em]
$\df $, &
$ 2 \hat r_i \lb U^{\rm lin}_{10,i} + 3 U^{\rm loop}_{10,i} \rb  - 3 \hat r_i g_j A^{10}_{ij} - 2 \hat r_i G_{jl} A^{\rm lin}_{ij} U^{\rm lin}_{10,l} $,  \\[0.5 em]
$\df\df$, &
$ 2 \hat r_i U^{11}_i -  \hat r_i g_j A^{\rm lin}_{ij} \xi_L - 2 \hat r_i g_j U^{\rm lin}_{10,i}U^{\rm lin}_{10,j}  $,  \\[0.5 em]
$\df^2$, &
$ 2 \hat r_i U^{20}_i - 2 \hat r_i g_j U^{\rm lin}_{10,i}U^{\rm lin}_{10,j}$,  \\[0.5 em]
$\df\df^2$, &
$ 2 \hat r_i U^{\rm lin}_{10,i} \xi_L$.
\end{tabular}
\end{center}
What is omitted from these tables are the shear bias terms $s^2$, as well as, all the derivative terms that appeared in the 
density to point statistics in the previous section. For full expressions in configuration space  including these terms we 
refer the reader to \cite{VlaCasWhi16}.

\subsection{Pairwise dispersion power spectrum and correlation function}

Finally we give the similar expressions, as for power spectrum and pairwise velocity, 
for pairwise dispersion power spectrum and correlation function. 
As discussed in the subsec. \ref{eq:pair_disp} we can split the 
dispersion power spectrum into two components $\widetilde \Xi_2^{(0)}$ and $\widetilde \Xi_2^{(2)}$.
Each of these can again be evaluated as the 1D Hankel transform
\eq{
\widetilde \Xi^{(\ell)}_{2,x}(k)= 
4 \pi \int q^2 d q~ e^{ -  \frac{1}{2} k^2 (X_L + Y_L)} \Bigg( h^{(0)}_{\ell,x}(k,q) j_0(qk) + \sum_{n=1}^{\infty} h^{(n)}_{\ell,x}(k,q) \lb \frac{kY_L}{q} \rb^n j_n(qk)
\Bigg),
}
where $h^{(n)}_{0,x}$ integrands can be tabulated as 
\begin{center}
\begin{tabular} {  c | c c  }
$\widetilde \Xi^{(0)}_{2x}$ & $h^{(0)}_{0,x}$  & $h^{(n)}_{0,x}$  \\ [0.2em] \hline \hline
$\text{zel} $, &
$X^{\rm lin}+\tfrac{1}{3}Y^{\rm lin}$, &
$X^{\rm lin}+\tfrac{1}{3}Y^{\rm lin} + \frac{2}{3}n \lb 2X^{\rm lin} + Y^{\rm lin} \rb$,  \\[0.5 em]
$\text{loop}$, &
\scriptsize{$4 \lb X^{(2)} + \tfrac{1}{3}Y^{(2)} \rb - \frac{1}{3} k^2 \lb X^{\rm lin} + Y^{\rm lin} \rb^2$}, &
\scriptsize{$4 \lb X^{(2)} + \tfrac{1}{3}Y^{(2)} \rb - \frac{1}{3} k^2 \lb X^{\rm lin} + Y^{\rm lin} \rb^2$} \\[0.5 em]
$~$ &
$~$ &
$- q\lb  \tfrac{5}{3} V^{(2)} + \tfrac{1}{3}T^{(2)} \rb / Y_{\rm lin}$,  \\[0.5 em]
$\df$, &
$  4 \lb X^{10}+\tfrac{1}{3}Y^{10} \rb $, &
$4 \lb X^{10}+\tfrac{1}{3}Y^{10} \rb - 2 \frac{q U^{10}}{Y_{\rm lin}} \lb \tfrac{5}{3} X_{\rm lin} + Y_{\rm lin} \rb$  \\[0.5 em]
$\df\df$, &
$  \xi_L (X^{\rm lin}+ \tfrac{1}{3}Y^{\rm lin})  +  \frac{2}{3} \lb U^{\rm lin}_{10} \rb^2 $ &
$  \xi_L (X^{\rm lin}+ \tfrac{1}{3}Y^{\rm lin})  +  \frac{2}{3} \lb U^{\rm lin}_{10} \rb^2 $ \\[0.5 em]
$\df^2$, &
$ \frac{2}{3} \left[U^{\rm lin}_{10} \right]^2 $, &
$ \frac{2}{3} \left[U^{\rm lin}_{10} \right]^2 $,  \\[0.5 em]
$\df\df^2$, &
$ 0 $, &
$ -2 \frac{q U^{\rm lin}_{10} \xi_L}{Y_{\rm lin}}$ .
\end{tabular}
\end{center}
Where $X^{(2)}$, $Y^{(2)}$ and $V^{(2)}$ and $T^{(2)}$ terms are defined by
\eq{
A^{(2)}_{ij} &= A_{ij,22}^{\rm 1-loop}+\tfrac{3}{4} A_{ij,13}^{\rm 1-loop} = \lb X_{22}+\tfrac{3}{4} X_{13} \rb \df^K_{ij} + \lb Y_{22} + \tfrac{3}{4} Y_{13} \rb \hat q_i \hat q_j, \non\\
W^{(2)}_{ijl} &= 2 W_{ijl} - W^{(112)}_{ijl}  =  (2V-V^{(112)}) \hat q_{\{i} \df^K_{jk\}} + (2T - T^{(112)}) \hat q_i \hat q_j \hat q_l.
}
Similar table holds for the $h^{(n)}_{2,x}$ 
\begin{center}
\begin{tabular} {  c | c c  }
$\widetilde \Xi^{(0)}_{2x}$ & $h^{(0)}_{0,x}$  & $h^{(n)}_{0,x}$  \\ [0.2em] \hline \hline
$\text{zel} $, &
$\tfrac{2}{3}Y^{\rm lin}$, &
$ \frac{2}{3} Y^{\rm lin} - \frac{2n(2n-1)}{k^2} + \frac{1}{3}n \lb 8 X^{\rm lin} + 10 Y^{\rm lin} \rb$,  \\[0.5 em]
$\text{loop}$, &
\scriptsize{$\frac{8}{3} Y^{(2)} - \frac{2}{3} k^2 \lb X^{\rm lin} + Y^{\rm lin}\rb^2 $}, &
\scriptsize{$8 Y^{(2)} \lb \frac{1}{3} - \frac{n}{k^2Y^{\rm lin}}\rb - \frac{2}{3} k^2 \lb X^{\rm lin} + Y^{\rm lin} \rb^2$} \\[0.5 em]
$~$ &
$~$ &
$ - \frac{q \lb 4 V^{(2)} + 2T^{(2)} \rb}{3 Y_{\rm lin}} + \frac{2 (n-1)}{k^2 Y^2}(q T)$,  \\[0.5 em]
$\df$, &
$ \frac{8}{3}Y^{10}  $, &
$ \frac{8}{3}Y^{10} - 4 \frac{q U^{10}}{Y_{\rm lin}} \lb \frac{2}{3} X_{\rm lin} + Y_{\rm lin} \rb + 4 \frac{3(n-1) q U^{10} - 2n Y^{10} }{k^2Y_{\rm lin}} $  \\[0.5 em]
$\df\df$, &
$ \frac{2}{3} \big[ \xi_L Y^{\rm lin}  + 2 \lb U^{\rm lin}_{10} \rb^2 \big] $ &
$  2\big[ \xi_L Y^{\rm lin}  + 2 \lb U^{\rm lin}_{10} \rb^2 \big] \lb \frac{1}{3} - \frac{n}{k^2 Y_{\rm lin}}\rb $ \\[0.5 em]
$\df^2$, &
$ \frac{4}{3} \left[U^{\rm lin}_{10} \right]^2 $, &
$ 4 \left[U^{\rm lin}_{10} \right]^2 \lb \frac{1}{3} - \frac{n}{k^2 Y_{\rm lin}}\rb $.
\end{tabular}
\end{center}
Similarly the configuration space result is given as
\eeq{
\Xi^{(\ell)}_{2,x} (r) = (1+\xi(r)) v_{12,x}(r) =  \int d^3q ~Q(\vec r -\vec q) H^{(\ell)}_{2,x}(\vec q),
}
where we can again tabulated the configuration space integrands $H_{x}$ as
\begin{center}
\begin{tabular} {  c | c c  }
$\Xi^{(\ell)}_{2,x}$ & $H^{(0)}_{2,x}$ & $H^{(2)}_{2,x}$  \\ [0.2em] \hline \hline
$\text{zel}$, &
$ \frac{1}{3}\df^K_{ij} A_{ij}^{\rm lin}  $, &
$ - \lb \hat r_i \hat r_j - \tfrac{1}{3} \df^K_{ij} \rb A_{ij}^{\rm lin} $,  \\[0.5 em]
$\text{loop}$, &
\scriptsize{$ \frac{1}{3}\df^K_{ij}\bigg[ 4 \lb A_{ij,22}^{\rm 1-loop}+\tfrac{3}{4} A_{ij,13}^{\rm 1-loop} \rb $} &
\scriptsize{$ - \lb \hat r_i \hat r_j - \tfrac{1}{3} \df^K_{ij} \rb
\bigg[ 4 \lb A_{ij,22}^{\rm 1-loop}+\tfrac{3}{4} A_{ij,13}^{\rm 1-loop} \rb $},  \\[0.5 em]
$~$ &
\scriptsize{$- g_l \lb 2 W_{ijl} - W^{(112)}_{ijl}  \rb  -  G_{nm} A^{\rm lin}_{in} A^{\rm lin}_{jm} \bigg] $}, &
\scriptsize{$- g_l \lb 2 W_{ijl} - W^{(112)}_{ijl}  \rb -  G_{nm} A^{\rm lin}_{in} A^{\rm lin}_{jm} \bigg] $},  \\[0.5 em]
$\df$, &
\scriptsize{$  \frac{1}{3}\df^K_{ij} \bigg[  - 2 g_l U^{\rm lin}_{10,l}  A_{ij}^{\rm lin} +  4 A_{ij}^{10} - 4 g_n  A^{\rm lin}_{in} U^{\rm lin}_{10,j} \bigg]  $},  &
\scriptsize{$  - \lb \hat r_i \hat r_j - \tfrac{1}{3} \df^K_{ij} \rb \bigg[  - 2 g_l U^{\rm lin}_{10,l}  A_{ij}^{\rm lin}  +  4 A_{ij}^{10} - 4 g_n  A^{\rm lin}_{in} U^{\rm lin}_{10,j} \bigg] $},  \\[0.5 em]
$\df\df$, &
$ \frac{1}{3} \df^K_{ij} \bigg[ \xi_L A_{ij}^{\rm lin}  + 2 U^{\rm lin}_{10,i}U^{\rm lin}_{10,j} \bigg]  $, &
$ - \lb \hat r_i \hat r_j - \tfrac{1}{3} \df^K_{ij} \rb \bigg[ \xi_L A_{ij}^{\rm lin}  +2 U^{\rm lin}_{10,i}U^{\rm lin}_{10,j} \bigg] $,  \\[0.5 em]
$\df^2$, &
$ \frac{1}{3} \df^K_{ij} \bigg[ 2 U^{\rm lin}_{10,i}U^{\rm lin}_{10,j}  \bigg] $, &
$  - \lb \hat r_i \hat r_j - \tfrac{1}{3} \df^K_{ij} \rb  \bigg[ 2 U^{\rm lin}_{10,i}U^{\rm lin}_{10,j}  \bigg] $.
\end{tabular}
\end{center}

\bibliographystyle{JHEP}
\bibliography{ms}
\end{document}